\pdfoutput=1
\documentclass[a4paper]{article}
\usepackage{jheppub}
\usepackage[utf8]{inputenc}
\usepackage[T1]{fontenc}

\usepackage{xcolor}
\definecolor{nicered}{rgb}{0.7,0.1,0.1}
\definecolor{nicegreen}{rgb}{0.1,0.5,0.1}
\definecolor{rosso}{cmyk}{0,1,1,0.4}
\definecolor{babypink}{rgb}{0.96, 0.76, 0.76}
\definecolor{babyblueeyes}{rgb}{0.63, 0.79, 0.95}
\definecolor{azure(colorwheel)}{rgb}{0.0, 0.5, 1.0}
\definecolor{MyDarkBlue}{rgb}{0,0.1,0.7}
\definecolor{amethyst}{rgb}{0.6, 0.4, 0.8}

\newcommand{\la}{\lambda_1}
\newcommand{\lb}{\lambda_2}
\newcommand{\lc}{\lambda_3}
\newcommand{\g}{\,\mathrm{GeV}}
\newcommand{\be}{\begin{equation}}
\newcommand{\ee}{\end{equation}}

\newcommand{\Vtree}{V^{(0)}}
\newcommand{\Vone}{V^{(1)}}
\newcommand{\VT}{V^{T}}
\newcommand{\f}{\varphi}
\newcommand{\ms}{\overline{\textrm{MS}}}

\newcommand{\beq}{\begin{equation}}
\newcommand{\eeq}{\end{equation}}

\newcommand{\bea}{\begin{eqnarray}}
\newcommand{\eea}{\end{eqnarray}}

\newcommand{\GeV}{\,{\rm GeV}}

\usepackage{amsmath}
\usepackage{amssymb}
\usepackage{amsthm}
\usepackage{amsfonts}
\usepackage{physics}
\usepackage[mathscr]{eucal}
\allowdisplaybreaks[3]
\usepackage{slashed}
\usepackage{float}
\usepackage{ulem}
\usepackage{enumitem}
\usepackage{siunitx} % Required for alignment
\usepackage{tikz}
\usetikzlibrary{arrows,shapes}
\usetikzlibrary{trees}
\usetikzlibrary{matrix,arrows} 				% For commutative diagram
\usetikzlibrary{positioning}				% For "above of=" commands
\usetikzlibrary{calc,through}				% For coordinates
\usetikzlibrary{decorations.pathreplacing}  % For curly braces
\usepackage{pgffor}							% For repeating patterns

\usetikzlibrary{decorations.pathmorphing}	% For Feynman Diagrams
\usetikzlibrary{decorations.markings}
\tikzset{
    vector/.style={decorate, decoration={snake}, draw},
	provector/.style={decorate, decoration={snake,amplitude=2.5pt}, draw},
	antivector/.style={decorate, decoration={snake,amplitude=-2.5pt}, draw},
    fermion/.style={draw=black, postaction={decorate},
        decoration={markings,mark=at position .55 with {\arrow[draw=black]{>}}}},
    fermionbar/.style={draw=black, postaction={decorate},
        decoration={markings,mark=at position .55 with {\arrow[draw=black]{<}}}},
    fermionnoarrow/.style={draw=black},
    gluon/.style={decorate, draw=black,
        decoration={coil,amplitude=4pt, segment length=5pt}},
    scalararrow/.style={dashed,draw=black, postaction={decorate},
        decoration={markings,mark=at position .55 with {\arrow[draw=black]{>}}}},
    scalar/.style={dashed,draw=black},
    scalarbar/.style={dashed,draw=black, postaction={decorate},
        decoration={markings,mark=at position .55 with {\arrow[draw=black]{<}}}},
    electron/.style={draw=black, postaction={decorate},
        decoration={markings,mark=at position .55 with {\arrow[draw=black]{>}}}},
	bigvector/.style={decorate, decoration={snake,amplitude=4pt}, draw},
}

\tikzstyle{block} = [draw, rectangle, 
    minimum height=3em, minimum width=6em]
\usepackage{hyperref}			% Has to be at the end.
\usepackage{orcidlink}
\hypersetup{pdfborder={0 0 0},colorlinks,breaklinks=true,
  urlcolor={MyDarkBlue},citecolor={nicegreen},linkcolor={black}
}

\usepackage{subfiles}

\title{Conformal model for gravitational waves and dark matter:\\ A status update.}

\author[a]{Maciej Kierkla \orcidlink{0000-0002-2785-5370},}  %
\author[b]{Alexandros Karam \orcidlink{0000-0002-0582-8996}} %
\author[a]{and Bogumi\l a~\'Swie\.zewska \orcidlink{0000-0003-0169-211X}} 
\affiliation[a]{\small\textit{Faculty of Physics, University of Warsaw, Pasteura~5, 02-093 Warsaw, Poland}}
\affiliation[b]{\small\textit{Laboratory of High Energy and Computational Physics, National Institute of Chemical Physics and Biophysics, R{\"a}vala pst.~10, Tallinn, 10143, Estonia}}

\emailAdd{maciej.kierkla@fuw.edu.pl}
\emailAdd{alexandros.karam@kbfi.ee}
\emailAdd{bogumila.swiezewska@fuw.edu.pl}

\abstract{
We present an updated analysis of the first-order phase transition associated with symmetry breaking in the early Universe in a classically scale-invariant model extended with a new SU(2) gauge group. Including recent developments in understanding supercooled phase transitions, we compute all of its characteristics and significantly constrain the parameter space. We then predict gravitational wave spectra generated during this phase transition and by computing the signal-to-noise ratio we conclude that this model is well-testable (and falsifiable) with LISA. We also provide predictions for the relic dark matter abundance. It is consistent with observations in a rather narrow part of the parameter space. We strongly constrain the so-called supercool dark matter scenario based on an improved description of percolation and reheating after the phase transition as well as the inclusion of the running of couplings. Finally, we devote attention to the renormalisation-scale dependence of the results. Even though our main results are obtained with the use of renormalisation-group improved effective potential, we also perform a fixed-scale analysis which proves that the dependence on the scale is not only qualitative but also quantitative.
}


\arxivnumber{2210.07075}

\begin{document}       
\maketitle

%\newpage\pagebreak
%\noindent\rule{\textwidth}{1pt}
%\vspace{-1cm}
%\setcounter{tocdepth}{2}
%\tableofcontents %JHEP demands commenting this
%\noindent\rule{\textwidth}{1pt}
%\newpage

%\subfile{sections/TODOs}
\section{Introduction\label{sec:intro}}

With the recent direct detection of gravitational waves (GW) by the LIGO and Virgo Collaborations~\cite{Abbott:2016-2, Abbott:2016, Abbott:2017, Abbott:2017-2, LIGOScientific:2017ycc, LIGOScientific:2017vox} and the prospect of the scheduled Laser Interferometer Space Antenna (LISA)~\cite{Bartolo:2016ami, Caprini:2019pxz, Gowling:2021gcy, LISACosWG:2022jok, Boileau:2022ter, Gowling:2022pzb} and other future experiments -- such as AION~\cite{Badurina:2019hst}, MAGIS~\cite{Graham:2016plp, Graham:2017pmn}, AEDGE~\cite{AEDGE:2019nxb}, and the Einstein Telescope (ET)~\cite{Punturo:2010zz, Hild:2010id} (along with the ongoing aLIGO/aVirgo~\cite{Harry:2010zz, VIRGO:2014yos, LIGOScientific:2014pky, LIGOScientific:2019lzm}) 
-- collecting gravitational-wave data across a wide frequency range, it is prudent to seek ways of using GW to probe fundamental physics. One of the promising phenomena that could leave their imprint in primordial GW background is a first-order phase transition (PT) in the early Universe~\cite{Caprini:2015, Caprini:2019pxz, LISACosWG:2022jok, Gowling:2021gcy, Boileau:2022ter, Gowling:2022pzb}. For a signal produced during a PT proceeding around the temperatures characteristic for the electroweak (EW) PT, $T\sim 100\g$, frequencies within the LISA sensitivity window are expected. However, it turned out that in many models the signal is not strong enough to be observable. The class of models with classical scale invariance is a counterexample -- such models typically predict strong GW signal within the reach of LISA. This is due to a logarithmic potential, which provides conditions for large supercooling and huge latent heat release during the transition. This kind of behaviour has been studied in models with strong dynamics or extra dimensions, see e.g.~\cite{Randall:2006, Konstandin:2010, Konstandin:2011, vonHarling:2017, Bruggisser:2018,Kubo:2016, Baldes:2021aph} and in models with perturbative classically scale-invariant potentials~\cite{Hambye:2013, Jaeckel:2016, Hashino:2016, Jinno:2016, Marzola:2017, Ghorbani:2017lyk, Baldes:2018, Prokopec:2018, Marzo:2018, Mohamadnejad:2019vzg, Ghoshal:2020vud, Kang:2020jeg, Mohamadnejad:2021tke, Dasgupta:2022isg}.
 
Among the variety of classically conformal models, the ones with an extra gauge group are very promising -- being highly predictive and perturbative. The minimal extensions of the conformal Standard Model (SM) with a gauge group are models with an extra U(1) symmetry~\cite{Hempfling:1996, Sher:1996ib, Chang:2007, Iso:2009, Iso:2012jn, Khoze:2013-1, Khoze:2013-2, Khoze:2013-3, Hashimoto:2013hta, Hashimoto:2014ela, Benic:2014xho, Khoze:2014, Benic:2014aga, Okada:2014nea, Guo:2015, Humbert:2015epa, Oda:2015gna, Humbert:2015yva, Plascencia:2015, Haba:2015lka, Das:2015nwk, Haba:2015nwl, Wang:2015sxe, Jinno:2016, Das:2016zue, Oda:2017kwl, Hambye:2018, Loebbert:2018, Marzo:2018, YaserAyazi:2019caf, Kim:2019ogz, Mohamadnejad:2019vzg, Kang:2020jeg, Gialamas:2021enw, Barman:2021lot, Barman:2203, Mohamadnejad:2021tke, Dasgupta:2022isg} and SU(2) symmetry~\cite{Hambye:2013, Carone:2013, Khoze:2014, Pelaggi:2014wba, Karam:2015, Plascencia:2016, Chataignier:2018, Hambye:2018, Baldes:2018, Prokopec:2018, Marfatia:2020}. Another option of realising the classical scaling symmetry are models with extended scalar sectors~\cite{Sher:1996ib, Meissner:2006, Foot:2007s, Foot:2007-3, Foot:2007, Foot:2010av, AlexanderNunneley:2010, Foot:2010et, Lee:2012, Farzinnia:2013, Gabrielli:2013, Steele:2013fka, Guo:2014, Salvio:2014soa, Khoze:2014, Davoudiasl:2014, Farzinnia:2014xia, Lindner:2014oea, Kang:2014cia, Kannike:2015apa, Endo:2015ifa, Kang:2015aqa, Endo:2015nba, Ahriche:2015loa, Wang:2015cda, Ghorbani:2015xvz, Farzinnia:2015fka, Helmboldt:2016, Ahriche:2016cio, Ahriche:2016ixu, Wu:2016jdo, Marzola:2017, Ghorbani:2017lyk, YaserAyazi:2018lrv, Oda:2018zth, Brdar:2018, Brdar:2018num, Mohamadnejad:2019wqb, Kannike:2019upf, Jung:2019dog, Brdar:2019qut, Braathen:2020vwo, Kannike:2020ppf, Kubo:2020fdd, Ahriche:2021frb, Soualah:2021xbn}. Also, models with larger gauge groups, extra fermions or more baroque architecture have been considered~\cite{Dias:2005jk, Holthausen:2009uc, Heikinheimo:2013, Dermisek:2013, Holthausen:2013ota, Kubo:2014ida, Altmannshofer:2014, Antipin:2014qva, Giudice:2014tma, Ametani:2015jla, Carone:2015jra, Kubo:2015joa, Latosinski:2015pba, Haba:2015qbz, Karam:2016, Kubo:2016, Ishida:2019gri, Dias:2020ryz, Aoki:2020mlo, Dias:2022hbu}. In the present work, we focus on the PT in the classically scale-invariant model with an extra SU(2)$_X$ symmetry and a scalar which transforms as a doublet under this group while being a singlet of the SM. Apart from undergoing a strong first-order PT it provides a candidate for dark matter (DM) particle stabilised by a residual symmetry remaining after the SU(2)$_X$ breaking~\cite{Gross:2015, Hambye:2008}. 

The exciting prospect of detecting GW from a phase transition and probing processes taking place in the very early Universe is overshadowed by the pessimistic estimates of the precision of theoretical predictions~\cite{Croon:2020, Athron:2022}. The renormalisation-scale dependence amounts to one of the main sources of theoretical uncertainties. Classically scale-invariant models, due to the logarithmic nature of their potential, cover a wide energy range, thus being specially prone to scale dependence issues.

The aim of the present article is threefold. First, we present an update of the predictions of the stochastic GW background within the classically scale-invariant model with SU(2)$_X$ symmetry, implementing recent progress in understanding supercooled PTs~\cite{Ellis:2018, Ellis:2019, Lewicki:2019, Ellis:2020-2, Ellis:2020}. In particular, we pay special attention to an accurate formulation of the nucleation condition and we check whether the PT successfully ends with percolation which should not be taken for granted in models with strong supercooling. To our knowledge, the condition for percolation has not been included in previous studies of the SU(2)$_X$ model\footnote{With the exception of ref.~\cite{Baldes:2018}, however, there this condition was evaluated numerically for a single point in the parameter space and a general discussion followed.}, while we prove that it significantly affects the parameter space. We also evaluate the GW spectra using updated simulations~\cite{Caprini:2019pxz, Lewicki:2020, Lewicki:2022pdb} and determine the dominant source (sound waves vs bubble collisions) using recent developments~\cite{Ellis:2019, Ellis:2020, Hoche:2020ysm, Gouttenoire:2021kjv}. Second, we pay special attention to the renormalisation-scale dependence of the results. To minimise this dependence we employ a renormalisation-group (RG) improved effective potential. We perform an expansion in powers of couplings consistent with the conditions from conformal symmetry breaking as well as the radiative nature of the transition, and include all the leading terms. Moreover, we perform separate scans of the parameter space at different fixed renormalisation scales to study the dependence of the results on the arbitrary scale. Last, we study the DM phenomenology in light of the updated picture of the phase transition.

The structure of the article is as follows. In section~\ref{sec:zero-temp}, we introduce the model, discuss the generation of masses and analyse the parameter space. In section~\ref{sec:finite-temp}, we discuss the finite-temperature effective potential, the expansion in powers of couplings and the RG improvement. Then, a discussion of the supercooled PT and GW production follows in section~\ref{sec:PT-and-GW}, and of DM production and phenomenology in section~\ref{sec:DM}. Section~\ref{sec:results} is devoted to the presentation of results on GW observability prospects combined with the predictions on DM abundance and constraints from direct detection experiments. Contrary to the main body of the paper, in section~\ref{sec:scale-dep} we present results obtained at fixed renormalisation scale and discuss the difference in predictions with respect to the RG-improved case. Finally, in section~\ref{sec:summary} we summarise the obtained results. The main part of the article is followed by appendices in which scalar contributions to the effective potential are discussed (appendix~\ref{app:scalars}); the expressions for scalar self-energies are listed (appendix~\ref{app:self-energies}); the numerical procedure used for scanning the parameter space is described in detail (appendix~\ref{app:numerical-procedure}); various approximations to the energy transfer rate present in the literature are discussed (appendix~\ref{app:reheating}) and spectra of GW computed according to recent results of ref.~\cite{Lewicki:2022pdb} are presented.
\section{The model \label{sec:zero-temp}}
%XXXXXXXXXXXXXXXXXXXXXXXXXXXXXXXXXXXXXXXXXXXXXXXXX

In %the remaining part of 
this section we introduce the potential and analyse the symmetry breaking, paying special attention to a consistent expansion in powers of couplings. We derive the expressions for the physical masses of the scalar particles and the mixing between them, and we present a scan of the parameter space at zero temperature.

%~~~~~~~~~~~~~~~~~~~~~~~~~~~~~~~~~~~~~~~~~~
\subsection{Field content and potential}
%~~~~~~~~~~~~~~~~~~~~~~~~~~~~~~~~~~~~~~~~~~

In this work, we analyse a model called SU(2)cSM~\cite{Hambye:2013, Carone:2013} which is invariant under classical scale symmetry. It consists of the Standard Model (SM) constrained by the scaling symmetry, i.e.\ with the Higgs mass term excluded from the Lagrangian, and an additional sector consisting of a new SU(2)$_X$ gauge group and a scalar doublet of this group, which is a singlet under the SM symmetries. The two sectors communicate via a standard Higgs-portal coupling. The tree-level potential of the model reads
\be
\Vtree(\Phi, \Psi)=\la \left(\Phi^{\dagger}\Phi\right)^2 + \lb \left(\Phi^{\dagger}\Phi\right) \left(\Psi^{\dagger}\Psi \right)+ \lc \left(\Psi^{\dagger}\Psi\right)^2,\notag
\ee
where $\Phi$ is the SM scalar field, while $\Psi$ is the new scalar field. The gauge bosons of SU(2)$_X$ will be referred to as $X$ bosons. They couple to $\Psi$ via the covariant derivative and to the SM sector only via the mixing between $\Psi$ and $\Phi$.

As can be seen, at tree level the potential is scale invariant, there are no dimensionful parameters. This symmetry is broken when one-loop corrections are included and the masses of all particles are generated via the so-called Coleman--Weinberg mechanism~\cite{Coleman:1973} (also referred to as radiative symmetry breaking or dimensional transmutation). 

%~~~~~~~~~~~~~~~~~~~~~~~~~~~~~~~~~~~~~~~~~
\subsection{Effective potential and symmetry breaking}
%~~~~~~~~~~~~~~~~~~~~~~~~~~~~~~~~~~~~~~~~~
Due to the SU(2) symmetry of the SM and the new SU(2)$_X$ symmetry one can write the effective potential in terms of two real scalar fields $h$ and $\f$ which correspond to the radial components of the scalar doublets,
\be
\Phi^{\dagger}\Phi=\frac{1}{2} h^2, \quad \Psi^{\dagger}\Psi=\frac{1}{2} \f^2.\notag
\ee
The one-loop effective potential  written in terms of $h$ and $\f$ reads
\be
V(h,\f)=\Vtree(h,\f) + \Vone(h,\f),\label{eq:one-loop-potential}
\ee
with
\be
\Vtree(h,\f)=\frac{1}{4}\left(\la h^4 + \lb h^2 \f^2 + \lc \f^4\right)\label{eq:Vtree}
\ee
and the one-loop potential is given by the standard Coleman--Weinberg formula (in $\ms$ scheme and Landau gauge) 
\be
\Vone(h,\f)=\frac{1}{64 \pi^2}\sum_{a}n_a M_a^4(h,\f)\left(\log\frac{M_a^2(h,\f)}{\mu^2}-C_a\right), \label{eq:one-loop}
\ee
where the sum runs over all particle species. For simplicity of notation, we do not distinguish between the radial components of the quantum fields and the classical fields which are arguments of the effective potential. In eq.~\eqref{eq:one-loop}, $M_a(h,\f)$ denotes the field-dependent mass of a particle (for a scalar particle it is an eigenvalue of the second derivative of the tree-level potential), $n_a$ counts the number of degrees of freedom associated with each species and $C_a=\frac{5}{6}$ for vector bosons and $C_a=\frac{3}{2}$ for other particles.\footnote{These constants depend on the regularisation and renormalisation schemes chosen. In the present work, we use dimensional regularisation and $\ms$. If dimensional reduction was used, the $C_a$ would equal $\frac{3}{2}$ for all particle species.} For a particle of spin $s_a$ the factor $n_a$ is given by
\be
n_a=(-1)^{2s_a} Q_a N_a (2s_a+1),\notag%\label{eq:dof}
\ee
where $Q_a=1$ for uncharged particles, and $Q_a=2$ for charged particles, $N_a=1,\,3$ for uncoloured and coloured particles, respectively. The particles that contribute are the top quark (we neglect lighter quarks), the gauge $W^{\pm}$, $Z$ and $X$ bosons and, in principle, the scalar particles -- the physical scalars and the Goldstones. In section~\ref{sec:scaling-of-couplings} below and in appendix~\ref{app:scalars} we argue that the scalar contributions can and should be dropped for consistency.

The tree-level field-dependent mass matrix for the scalar sector reads
\be
 M^2(h,\f)=\left(\begin{array}{cc}
3\la h^2 + \frac{\lb}{2}\f^2		 & \lb h \f\\
\lb h \f 			& 3 \lc \f^2 + \frac{\lb}{2} h^2\\
 \end{array}
 \right), \label{eq:tree-level-mass-matrix}
 \ee
which gives tree-level mass eigenvalues as
\begin{align}
M^2_{\pm}(h,\f)& = \frac{1}{2}\left(3 \lambda_1 + \frac{\lambda_2}{2}\right)h^2 + \frac{1}{2}\left(\frac{\lambda_2}{2} + 3\lambda_3\right)\varphi^2  \notag\\[2pt]
& \ \ \ \pm\frac{1}{2}\sqrt{\left[\left(3\lambda_1-\frac{\lambda_2}{2}\right)h^2-\left(3\lambda_3-\frac{\lambda_2}{2}\right)\varphi^2\right]^2+4\lambda_2^2h^2\varphi^2}.\label{eq:M-plus-minus}
\end{align}
The field-dependent Goldstone masses read
\begin{align}
    M_G^2(h,\f)&=M_{G^{\pm}}^2(h,\f)=\la h^2 + \frac{1}{2}\lb \f^2,\\
    M_{{G_X}}^2(h,\f)&=M_{G_X^{\pm}}^2(h,\f)=\lc \f^2 + \frac{1}{2}\lb h^2.\label{eq:X-Goldstone-masses}
\end{align}
The tree-level vector boson and top masses are given by the standard formulas
\be
M_{W^{\pm}}(h,\f)=\frac{1}{2}g_2 h,\quad M_Z(h,\f)=\frac{1}{2} \sqrt{g_2^2+g_Y^2} h,\quad M_X(h,\f)=\frac{1}{2}g_X \f,\quad M_t(h,\f)=\frac{1}{\sqrt{2}} y_t h,
\ee
where $g_X$ is the gauge coupling of SU(2)$_X$.

%%%%%%%%%%%%%%%%%%%%%%%%%%%%%%%%%%%%%%%%%%%%%%%%%%%%%%%%
\subsubsection{Radiative symmetry breaking and scaling of couplings \label{sec:scaling-of-couplings}}
%%%%%%%%%%%%%%%%%%%%%%%%%%%%%%%%%%%%%%%%%%%%%%%%%%%%%%%%

Symmetry breaking in the considered model is introduced via the Coleman--Weinberg mechanism~\cite{Coleman:1973}. It relies on the interplay of the tree-level and one-loop terms and results in a non-vanishing VEV of the scalar field. This nontrivial interplay indicates that perturbative computations should not be organized in terms of loops -- for radiative symmetry breaking to work, the one-loop corrections need to be comparable to the tree-level terms. This can be done without violating perturbativity if a scaling relation between couplings arises. This relation is straightforward to define in an archetypal model of RSB -- the massless scalar QED -- reading $\lambda\sim e^4$, where $e$ is the U(1) charge~\cite{Coleman:1973}. Then, all the quantities should be expanded to a fixed order in $e$ and the contributions up to order $\mathcal{O}(e^4)$ constitute the leading order. This includes the tree-level potential, and the one-loop contribution from the gauge boson (but excludes one-loop corrections from the scalars). However, with richer field content the relation between couplings becomes obscured, therefore, before proceeding with the analysis of symmetry breaking we discuss the scaling of couplings, see also ref.~\cite{Chataignier:2018RSB}.

We will consider two types of scaling for the couplings: a coupling that is relevant at tree level (i.e.\ it dominates over loop contributions) scales as $g^2$, whereas a coupling comparable to loop contributions from e.g.\ SM gauge bosons scales as $g^4$. The $g^2$ or $g^4$ scaling should be understood formally as a bookkeeping device, which allows to organise the perturbative computations via a systematic expansion in powers of $g$ -- as in the massless scalar electrodynamics. In particular, we will assume that the SM couplings $g_2^2$, $g_Y^2$ and $y_t^2$ scale as $g^2$.

Earlier results~\cite{Carone:2013, Hambye:2013, Khoze:2014, Plascencia:2016, Hambye:2018, Baldes:2018, Chataignier:2018RSB, Marfatia:2020}  indicate that $\lb$ should be small, typically $\lb<10^{-3}$ (and it can acquire much smaller values), therefore we assume that $\lb\sim g^4$. The $\lc$ coupling, as will be shown shortly, scales as $g_X^4$. On the other hand, $\la$ acquires SM-like values and thus scales as $g^2$. Moreover, in conformal models, typically the VEV of the new scalar field, coupled to the new gauge group is much larger than the SM Higgs VEV, $w/v\gg 10$. Since in minimisation conditions, eqs.~\eqref{eq:min1}--\eqref{eq:min2} below, apart from powers of couplings, powers of the VEVs appear, terms proportional to couplings scaling as $g^4$ can still be brought to the leading order if they are multiplied by a large ratio $w/v$. 

The scalar contributions to the effective potential are typically neglected. Indeed they scale as $g^8$ so formally they belong to higher orders. However, large ratios of the VEVs can appear making the omission of scalars less obvious. The discussion of these subtleties is relegated to appendix~\ref{app:scalars}. Consequently, in the zero-temperature one-loop effective potential, we will only consider the contributions from the gauge bosons ($W^{\pm}$, $Z$, $X$) and the top quark. A bonus of this choice is that it makes the effective potential gauge-independent since at one-loop level all the dependence on the gauge-fixing parameters resides in the Goldstone-boson field-dependent masses~\cite{Loebbert:2018}.\footnote{Discussion of the gauge dependence of the vacuum decay rate is beyond the scope of the present work, see refs.~\cite{Metaxas:1995, Endo:2017gal, Croon:2020, Lofgren:2021ogg, Hirvonen:2021zej, Schicho:2022wty} for more details. Typically the uncertainty related to gauge dependence is much smaller than the one associated with renormalisation-scale dependence~\cite{Croon:2020}.}

%%%%%%%%%%%%%%%%%%%%%%%%%%%%%%%%%%%%%%%%%%%%%%%%%%%%%%%%
\subsubsection{Minimisation of the one-loop potential}
%%%%%%%%%%%%%%%%%%%%%%%%%%%%%%%%%%%%%%%%%%%%%%%%%%%%%%%%

Equipped with the knowledge about the scaling of various couplings, we now turn to studying the symmetry breaking in the SU(2)cSM. Let us look at the stationary point equations divided by the VEVs, $
v=\langle h \rangle$, $w=\langle\f\rangle$,
\begin{align}
\frac{1}{v^3}\frac{\partial V}{\partial h}=\la&  +\frac{1}{2} \lb \left(\frac{w}{v}\right)^2 +\frac{1}{v^3}\left.\frac{\partial \Vone}{\partial h}\right|_{h=v, \f=w}=0,\label{eq:min1}\\
\frac{1}{w^3}\frac{\partial V}{\partial \f}=\lc& +\frac{1}{2} \lb \left(\frac{v}{w}\right)^2 +\frac{1}{w^3}\left.\frac{\partial \Vone}{\partial \f}\right|_{h=v, \f=w}=0.\label{eq:min2}
\end{align}
Starting from eq.~\eqref{eq:min2}, the term proportional to $\lb$ is suppressed by $(v/w)^2$ which means that it can be neglected, being subleading with respect to $\lc$. As explained above, in the one-loop term we keep only the contribution from the $X$ gauge bosons.  Therefore, writing explicitly the Coleman--Weinberg term, the relevant equation reads
\be
\lc= - \frac{9}{256\pi^2}g_X^4\left[2\log\left(\frac{g_X}{2}\frac{w}{\mu}\right) -\frac{1}{3}\right].\label{eq:min-phi}
\ee
This is a typical Coleman--Weinberg relation between the scalar coupling and the gauge coupling which ensures that the loop term is of the order of the tree-level term such that radiative symmetry breaking is possible. This confirms our earlier assumption that $\lc\sim g_X^4\sim g^4$.

In the other minimisation condition, eq.~\eqref{eq:min1}, the leading order corresponds to the tree-level terms. The term proportional to $\lb$ is enhanced by the ratio $\left(\frac{w}{v}\right)^2$ so should not be neglected. We then obtain a relation  as follows:
\be
\la +\frac{1}{2} \lb \left(\frac{w}{v}\right)^2=0.\label{eq:tree-level-min}
\ee
This is an SM-like relation, where the second term can be identified with the mass term for the Higgs boson. We can include the NLO corrections, which correspond to the loop contributions from the SM particles. Thus, the loop-corrected minimisation condition reads
\be
\la +\frac{1}{2} \lb \left(\frac{w}{v}\right)^2 +\frac{1}{16\pi^2}\sum_{W^{\pm},Z,t}n_a \frac{M_a^4(h,\f)}{v^4}\left(\log\frac{M_a^2(h,\f)}{\mu^2}-C_a+\frac{1}{2}\right)=0. \label{eq:condition-lambda2}
\ee

The considerations above show that the symmetry breaking in the $\f$ direction is truly of Coleman--Weinberg nature, while along $h$ its character is SM-like, with the ``tree-level mass term'' being generated by the portal coupling.

%XXXXXXXXXXXXXXXXXXXXX
%~~~~~~~~~~~~~~~~~~~~~~~~~~~~~~~~~~~~~~~~~
\subsection{Masses and mixing of the scalars \label{sec:masses-and-mixing}}
%~~~~~~~~~~~~~~~~~~~~~~~~~~~~~~~~~~~~~~~~~
%XXXXXXXXXXXXXXXXXXXXX

The most common method used to compute masses of scalar particles is to use the eigenvalues of the matrix of second derivatives of the effective potential -- the so-called running masses. Since the effective potential corresponds to the momentum-independent part of the effective action (zeroth order in expansion in momenta), running masses are evaluated at momentum equal to zero. On the other hand, the physical mass corresponds to the pole of the propagator, i.e.\ is evaluated away from $p^2=0$, and is given by the following equation:
\be
M^2_{\mathrm{pole}}=m^2_{\textrm{tree-level}}+\Re[\Sigma(p^2=M^2_{\mathrm{pole}})].\label{eq:pole-mass}
\ee
In the present paper, we improve upon the existing studies by computing the pole masses for the scalars, instead of the usual running masses (see also ref.~\cite{Pelaggi:2014wba}). Since in the studied model loop corrections are equally relevant as tree-level contributions, we claim that it is worthwhile checking how the momentum-dependent contributions to one-loop corrections modify the result.\footnote{In fact, to compute the physical pole mass, one should perform renormalisation in the on-shell scheme. In this work, we compute the pole masses of eq.~\eqref{eq:pole-mass} in the $\ms$ scheme in order to facilitate comparisons with earlier results as well as not to overly complicate the computations related to the phase transition (tunnelling, nucleation, percolation). The pole masses that we obtain are thus still scale dependent and should be considered as the running masses corrected by momentum dependence.}

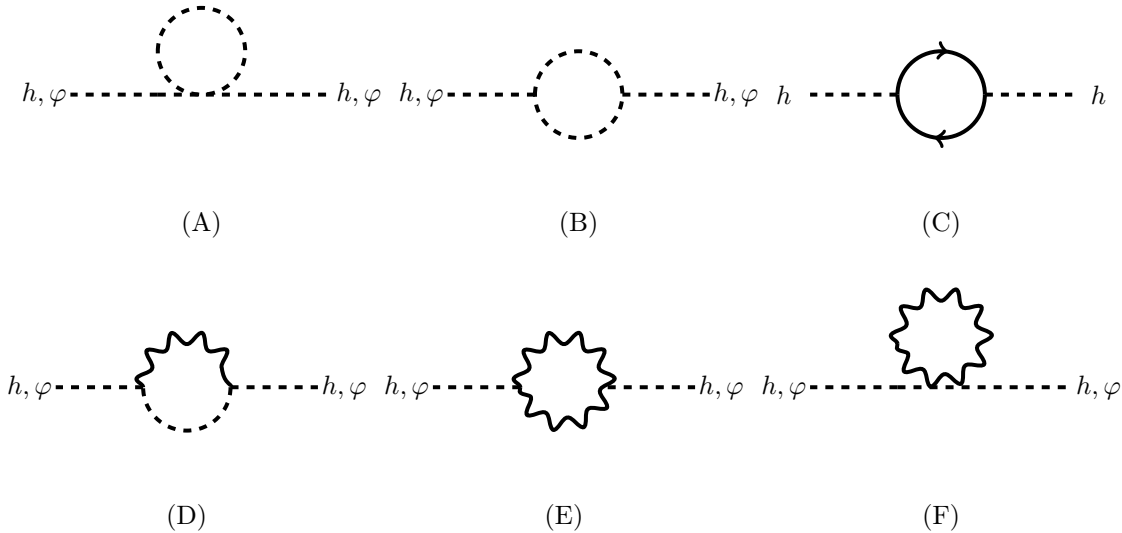
\begin{figure}[ht]
\center
\begin{tikzpicture}[line width=1.5 pt, scale=1.15]
	\draw[scalar] (0:1)--(0,0);
	\draw[scalar] (1,0.5) arc (180:-180:.5);
	%\draw[vector] (2,0) arc (0:-180:.5);
	\draw[scalar] (1,0) --(3,0);
	\node at (-.3,0) {$h,\f$};
	\node at (3.3,0) {$h,\f$};
	%\node at (40:2) {$$};
	\node at (1.5,-1.5) {(A)};
\end{tikzpicture}%\hspace{.5cm}
\begin{tikzpicture}[line width=1.5 pt, scale=1.15]
	\draw[scalar] (0:1)--(0,0);
	\draw[scalar] (1,0) arc (180:-180:.5);
	%\draw[vector] (2,0) arc (0:-180:.5);
	\draw[scalar] (2,0) --(3,0);
	\node at (-.3,0) {$h,\f$};
	\node at (3.3,0) {$h,\f$};
	%\node at (30:1.7) {$$};
	\node at (1.5,-1.5) {(B)};
\end{tikzpicture}%\hspace{.5cm}
\begin{tikzpicture}[line width=1.5 pt, scale=1.15]
	\draw[scalar] (0:1)--(0,0);
	\draw[fermion] (1,0) arc (180:0:.5);
	\draw[fermion] (2,0) arc (0:-180:.5);
	\draw[scalar] (2,0) --(3,0);
	\node at (-.3,0) {$h$};
	\node at (3.3,0) {$h$};
	%\node at (30:1.7) {$t, b$};
	\node at (1.5,-1.5) {(C)};
\end{tikzpicture}\\
\vspace{.5cm}
\begin{tikzpicture}[line width=1.5 pt, scale=1.15]
	\draw[scalar] (0:1)--(0,0);
	\draw[vector] (1,0) arc (180:0:.5);
	\draw[scalar] (2,0) arc (0:-180:.5);
	\draw[scalar] (2,0) --(3,0);
	\node at (-.3,0) {$h,\f$};
	\node at (3.3,0) {$h,\f$};
	%\node at (30:1.7) {$$};
	%\node at (-30:1.7) {$$};
	\node at (1.5,-1.5) {(D)};
\end{tikzpicture}%\hspace{.1cm}
\begin{tikzpicture}[line width=1.5 pt, scale=1.15]
	\draw[scalar] (0:1)--(0,0);
	\draw[vector] (1,0) arc (180:-180:.5);
	%\draw[vector] (2,0) arc (0:-180:.5);
	\draw[scalar] (2,0) --(3,0);
	\node at (-.3,0) {$h,\f$};
	\node at (3.3,0) {$h,\f$};
	%\node at (30:1.7) {$$};
	%\node at (-30:1.7) {$$};
	\node at (1.5,-1.5) {(E)};
\end{tikzpicture}%\hspace{.1cm}
\begin{tikzpicture}[line width=1.5 pt, scale=1.15]
	\draw[scalar] (0:1)--(0,0);
	\draw[vector] (1,0.5) arc (180:-180:.5);
	%\draw[vector] (2,0) arc (0:-180:.5);
	\draw[scalar] (1,0) --(3,0);
	\node at (-.3,0) {$h,\f$};
	\node at (3.3,0) {$h,\f$};
	%\node at (40:2) {$$};
	\node at (1.5,-1.5) {(F)};
\end{tikzpicture}
\hspace{.5cm}
\caption{Diagrams contributing to the scalar self energy.\label{fig:self-energy}}
\end{figure}

We define a mass matrix which contains the tree-level contributions of eq.~\eqref{eq:tree-level-mass-matrix} and loop corrections from self energies which introduce momentum dependence,
\be
 M^2(p)=\left(\begin{array}{cc}
3\la v^2 + \frac{\lb}{2}w^2		 & \lb v w\\
\lb v w 			& 3 \lc w^2 + \frac{\lb}{2} v^2\\
 \end{array}
 \right)
 +
\left(\begin{array}{cc}
\Sigma_{hh}(p)		 & \Sigma_{h\f}(p)\\
\Sigma_{h\f}(p)			& \Sigma_{\f\f}(p)\\
 \end{array}
 \right).
\ee
The $\Sigma$ matrix contains also the zero-momentum corrections given by the effective potential computed in the previous section, $\Sigma(p=0)=\frac{\partial^2 \Vone}{\partial \f_i\partial\phi_j}|_{h=v,\f=w}$. The diagrams contributing to the self-energy matrix are depicted in figure~\ref{fig:self-energy}. The first two diagrams correspond to purely scalar diagrams, the third one is the fermionic contribution, while in the second line gauge bosons and Goldstone bosons can propagate in the loop.  In what follows we will compute the self-energy corrections to the masses keeping the approximation assumed in the previous section -- neglecting purely scalar contributions in the loop terms. Moreover, based on arguments from appendix~\ref{app:scalars}, we fix the Goldstone masses in the loops of the type (D) to zero. Since the mixing terms in the self-energy $\Sigma_{h\f}$ are purely scalar (see the first line of figure~\ref{fig:self-energy}) they can be neglected and the only source of mixing remains the tree-level term. Even though the mixing will not affect the masses significantly, we do not neglect it since it is the only source of coupling between the dark sector and the SM sector. For this reason, it will be crucial in determining how reheating after the phase transition proceeds. 

The momentum-dependent mass eigenvalues are given by
\begin{align}
M^2_{\pm}(p^2)& = \frac{1}{2}\Biggl\{\left(3 \lambda_1 + \frac{\lambda_2}{2}\right)v^2 + \frac{1}{2}\left(\frac{\lambda_2}{2} + 3\lambda_3\right)w^2 +\Sigma_{hh}(p^2)+\Sigma_{\f\f}(p^2) \notag\\[2pt]
& \ \ \ \pm\sqrt{\left[\left(3\lambda_1-\frac{\lambda_2}{2}\right)v^2-\left(3\lambda_3-\frac{\lambda_2}{2}\right)w^2+\Sigma_{hh}(p^2)-\Sigma_{\f\f}(p^2)\right]^2+4\lambda_2^2v^2w^2}\,\Biggr\}.\label{eq:momentum-dependent-masses}
\end{align}
$\Sigma_{hh}$ and $\Sigma_{\f\f}$ can be found by computing the expressions corresponding to diagrams~(C)--(F) from figure~\ref{fig:self-energy}. The results are expressed in terms of the well-known Passarino-Veltman functions $a$ and $b$~\cite{Passarino:1978}, and can be found in Appendix~\ref{app:self-energies}. We have checked both analytically and numerically that in the limit of vanishing momenta the masses of eq.~\eqref{eq:momentum-dependent-masses} reduce to the running masses obtained from the effective potential.

In order to determine which of the mass eigenvalues corresponds to the Higgs particle we can apply an approximation neglecting terms suppressed by a product of a small coupling, $\lb$ or $\lc$ and the Higgs VEV. (In numerical computations we use the full expression for $M_{\pm}$ as it appears in eq.~\eqref{eq:momentum-dependent-masses}, we do not impose the approximations discussed here). We then find the following expressions:
\begin{align}
    &M_{+}^{2}(h,\f) =3 \lc \f^2+ \Sigma_{\f\f}(p^2), \\
    &M_{-}^{2}(h,\f) = 3\la h^2 +\frac{1}{2}\lb \f^2+\Sigma_{hh}(p^2),
\end{align}
for $3\la h^2-3\lc \f^2+\frac{1}{2}\lb \f^2+\Sigma_{hh}(p^2)-\Sigma_{\f\f}(p^2)<0$. For the opposite sign, $M_{+}$ and $M_{-}$ are interchanged. We always want to identify the mass eigenstate $H$ with the state that is SM-like, i.e.\ is mostly composed of the SM-like scalar $h$. Since $M_+$ is always greater than $M_-$, it is clear that in different parameter space regions we need to consider different mass orderings of $H$ and $S$. 

To obtain the momentum-corrected masses we solve the gap equations
\begin{align}
    M_H^2&=M_{\mp}^2(p^2=M_H^2),\label{eq:gap-Higgs}\\
    M_S^2&=M_{\pm}^2(p^2=M_S^2).\label{eq:gap-scalar}
\end{align}
The first one is used to fix the $\la$ coupling by demanding $M_H=125\g$~\cite{ParticleDataGroup:2020ssz, Workman:2022ynf} (the details of the numerical procedure are described in appendix~\ref{app:numerical-procedure}), while the other gives the mass of the new scalar $S$. The possible range of the mass of the scalar $S$, as well as differences in masses computed from the effective potential approximation and with the inclusion of self energies, are discussed in section~\ref{sec:parameter-space} below.

 The mass eigenstates are obtained from the gauge eigenstates by a rotation matrix as follows:\footnote{The fields $h$ and $\f$ in eq.~\eqref{eq:mixing} should be understood as translated by the respective VEVs, $h\to h-v$, $\f\to\f-w$ so that the physical fields (mass eigenstates) have zero VEVs.}
 \be
\left(\begin{array}{c}\phi_-\\ \phi_+\end{array}\right)=
\left(\begin{array}{rl}
\cos\theta & \sin\theta\\
-\sin\theta & \cos\theta
\end{array}\right)\left(\begin{array}{c}h \\ \f\end{array}\right),\label{eq:mixing}
\ee
where $\phi_-$ corresponds to the lower mass $M_{-}$ and $\phi_+$ to $M_{+}$. The mixing angle $\theta$ is in the range between $-\frac{\pi}{2}$ and $\frac{\pi}{2}$. To avoid confusion in the mixing parameters related to the change of mass ordering we define the mixing parameters $\xi_H$ and $\xi_S$ which represent rescalings of respective scalars' couplings with respect to the SM as
\begin{equation}
\xi_H=\left\{\begin{array}{rrl}
    \cos\theta & \textrm{ for }& M_H\leqslant M_S  \\
    -\sin\theta & \textrm{ for }& M_H>M_S
\end{array}\right.,\quad
\xi_S=\left\{\begin{array}{lrl}
    -\sin\theta & \textrm{ for }& M_H\leqslant M_S  \\
    \phantom{-}\cos\theta &\textrm{ for }& M_H>M_S
\end{array}\right..
\label{eq:xi}
\end{equation}

The inclusion of momentum dependence via the self-energies results in a difficulty in computing $\theta$, since depending on whether we use $p^2=M_H^2$ or $p^2=M_S^2$ the result will be different. This could be resolved by applying carefully chosen renormalisation conditions for the mixing angle, see e.g.~\cite{Krause:2016, Krause:2017}. Here, we simply use the value obtained at $p^2=M_H^2$. We checked that the results obtained at $p^2=0$ (in the effective potential approximation) are very close. However, going to $p^2=M_S^2$ would lead to changes up to $30\%$ in the value of $\xi_H$ when $M_S$ becomes large. We leave the careful analysis of possible renormalisation conditions for the mixing angle for future work.

%XXXXXXXXXXXXXXXXXXXXXXXXXXXXXXXXXXXXXXXXXXXXXXXXX
\subsection{Parameter space \label{sec:parameter-space}}
%XXXXXXXXXXXXXXXXXXXXXXXXXXXXXXXXXXXXXXXXXXXXXXXXX

The SU(2)cSM lagrangian contains 4 parameters apart from the SM ones -- $\la$, $\lb$, $\lc$ and $g_X$. Using the measured values of the Higgs VEV and Higgs mass we can eliminate two of them and be left with two free parameters. We choose the free parameters to be $g_X$ and $M_X$. In this section, we present the results of the scan of this parameter space, showing the region available for DM and GW studies. A detailed description of the numerical procedure can be found in appendix~\ref{app:numerical-procedure}.

One of the aims of this article is to emphasize the effect of scale dependence on the results obtained within classically scale-invariant models. In these models, typically the VEVs of the new scalar and of the SM scalar are separated by orders of magnitude, therefore a choice of scale for the computations of different quantities can matter significantly. In our analysis, we choose the free parameters at the renormalisation scale $\mu=M_X$. The masses of the scalars are computed at the electroweak scale ($\mu=M_Z$), the couplings and field renormalisation factors are evolved between these scales using their one-loop running (see ref.~\cite{Chataignier:2018RSB} for the $\beta$ functions), the running of the fields is also included. In this way, we perform the computations at scales relevant for the quantities considered. In the phase-transition computations we will use one-dimensional RG-improved effective potential along the $\f$ field direction. For it to be well defined down to low values of the field, we need the couplings not to hit Landau poles in the IR. Because of that, we set a constraint on $g_X$:
\be
g_X(M_Z)\leqslant 1.15, \label{eq:pert-gx}
\ee
which allows the RG-improved potential to be well-behaved throughout the scales considered. It should be noted that in some previous works the running of $g_X$ was not considered, see e.g.~refs~\cite{Baldes:2018, Hambye:2018, Marfatia:2020}, in some cases allowing large values of $g_X\sim 10$~\cite{Marfatia:2020}, which we disregard here as they would lead to the breakdown of perturbativity at the electroweak scale.

\begin{figure}[h!t]
\center
\includegraphics[width=.4\textwidth]{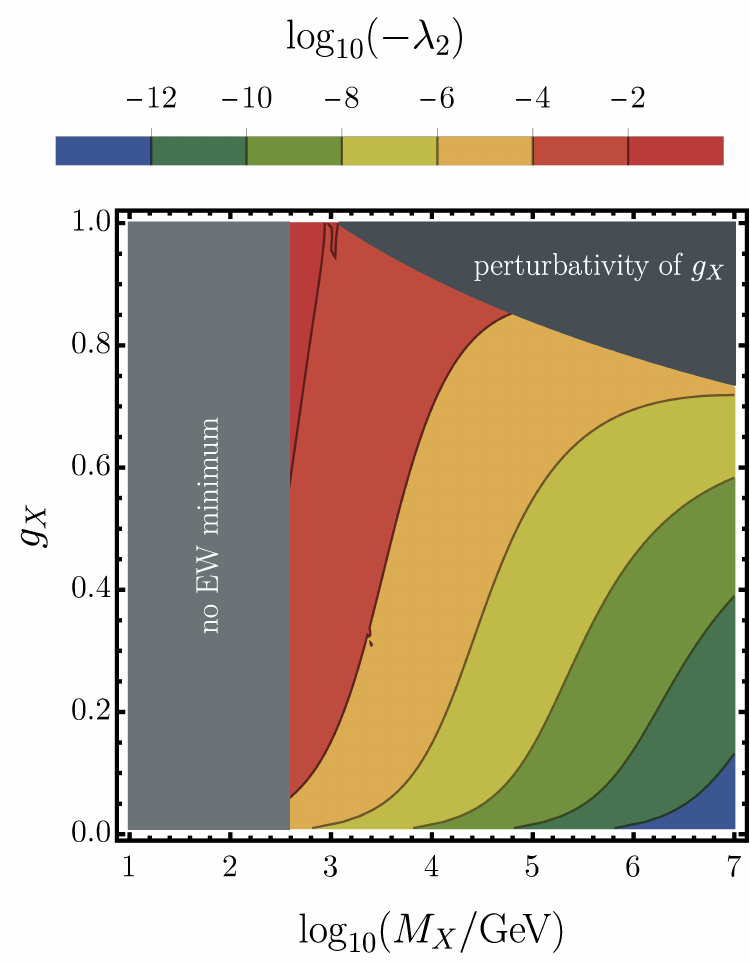}\hspace{20pt}
\includegraphics[width=.4\textwidth]{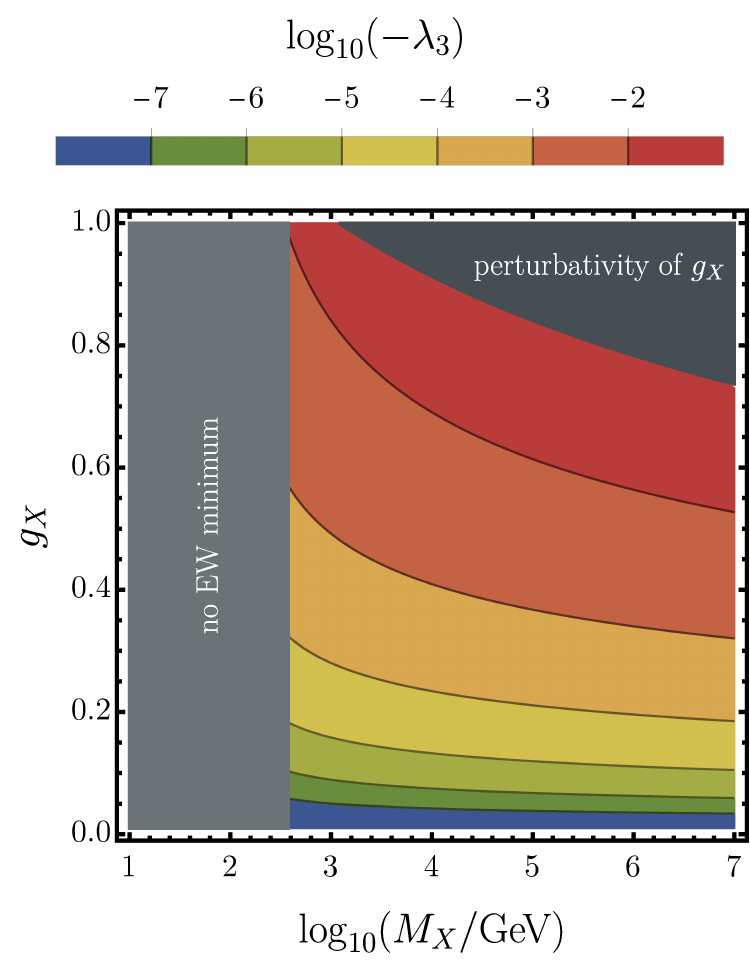}
\caption{Values of the scalar couplings $\lb$ (left panel) and $\lc$ (right panel) evaluated at the electroweak scale. Gray shaded regions are excluded, from left to right: no electroweak minimum with correct mass and VEV of the Higgs exists, perturbativity of $g_X$ (see eq.~\eqref{eq:pert-gx}).\label{fig:scalar-couplings}}
\end{figure}

Figure~\ref{fig:scalar-couplings} shows the values of the scalar couplings $\lb$ (left panel) and $\lc$ (right panel) (evaluated at the electroweak scale). The results confirm that these couplings are very small compared to the SM couplings and the formal scaling $\lb, \lc \sim g^4$ is justified. We do not display the values of the $\la$ coupling as they are very close to 0.14, which is the SM value at $\mu=M_Z$, especially in the region where $M_S>M_H$. Small deviations are allowed in the region where $M_S<M_H$ and around the $M_S=M_H$ line.

Let us also discuss the constraints in the $(M_X,\,g_X)$ plane. First, the region of low $X$ masses is excluded because it is not possible the reproduce a stable minimum with the correct Higgs VEV and mass in this regime. The upper right corner is cut off by the condition of eq.~\eqref{eq:pert-gx}. %\comment{what follows can be CROSSED OUT: \sout{ The right-most region of high $M_X$ is excluded because there $\lambda_1$ becomes negative already at the scale $\mu=M_X$. This is the same problem as with the stability of the SM vacuum, see e.g.~\cite{Buttazzo:2013} and references therein, where the Higgs self-coupling is driven to negative values by the top corrections. The running of $\la$ in SU(2)cSM is the same as in the SM up to very small corrections from the $\lambda_2$ coupling.  As in the SM, the scale at which $\lambda_1$ turns negative depends strongly on the value of the top quark mass. It can therefore be changed by including two-loop running. Detailed study of this issue is beyond the scope of the present work, but we note that the constraint on large $M_X$ can be shifted to some extent by carefully studying the running of SM couplings. However, we also do not expect $\lb$ to be able to stabilise the potential up to the Planck scale\footnote{In ref.~\cite{Prokopec:2018} an optimistically low value of the top Yukawa coupling at low scale was assumed which led to $\la$ being positive up to the Planck scale.}, and therefore there will always be a constraint on large values of $M_X$.}}

\begin{figure}[ht]
\center
\includegraphics[width=.4\textwidth]{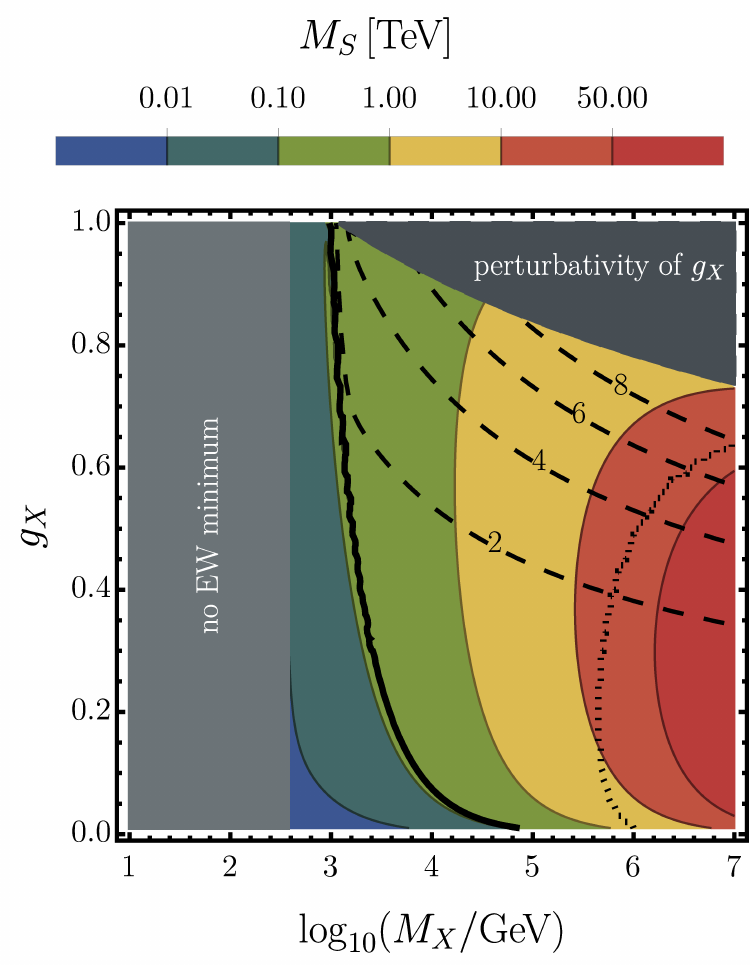}\hspace{20pt}
\includegraphics[width=.4\textwidth]{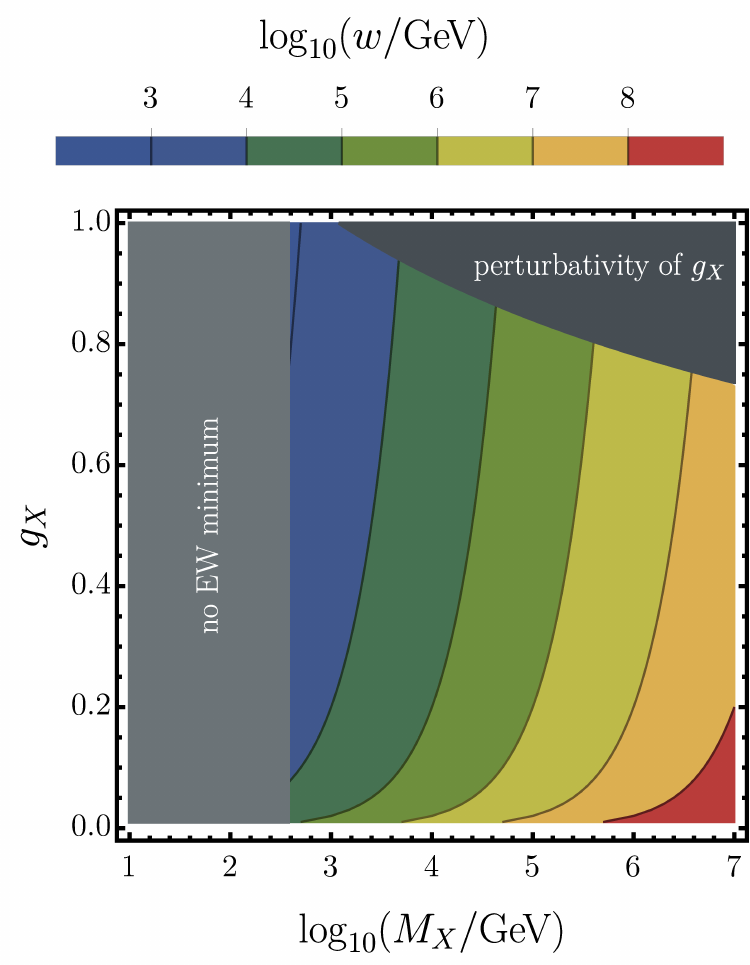}
\caption{Values of the new scalar mass $M_S$ (left panel) and the VEV $w$ (evaluated at $\mu=M_X$) (right panel). In the left panel the thick black line indicates where $M_S=M_H=125\g$ and across this line mass ordering between $S$ and $H$ changes (to the left of the line $M_S<M_H$, and to the right $M_H<M_S$). To the right of the dotted line $\xi_H$ becomes numerically equal to 1. The dashed lines indicate a discrepancy between the running and the pole mass (in percent). Grey shaded regions are excluded, see caption of figure~\ref{fig:scalar-couplings}.\label{fig:MS-w}}
\end{figure}

In figure~\ref{fig:MS-w}, the result of the scan for $M_S$ and $w$ (the VEV of $\f$) is shown.
From the left panel of figure~\ref{fig:MS-w} it is clear that the new scalar $S$ in most of the parameter space is heavier than the Higgs boson ($M_S>M_H$ to the right of the thick black line) and it is lighter than the DM candidate $X$. The mixing between the mass eigenstates $H$ and $S$ parameterised by $\xi_H$, see eq.~\eqref{eq:xi}, is very weak, in most of the parameter space $\xi_H>0.99$ (with the exception of the close vicinity of the $M_H=M_S$ line), therefore, the experimental constraint on the mixing $\xi_H>0.95$ does not reduce the parameter space (see e.g. ref.~\cite{Robens:2021} for recent constraints on the mixing angle in the singlet-extended SM, which are applicable also to our model). The dotted line indicates where $\xi_H$ becomes numerically equal to 1 (to the right of the line). Dashed lines encode the difference between the pole mass computed (iteratively) by solving eq.~\eqref{eq:gap-scalar} and the mass computed from the effective potential approximation. The differences are non-vanishing but not too large, going up to 10\% in the upper right part of the parameter space. Analogous difference for the Higgs mass is much smaller, not exceeding 2\% except for the $M_H=M_S$ line.

The right panel of figure~\ref{fig:MS-w} shows the VEV of the $\f$ field, which is orders of magnitude above the SM Higgs VEV, as anticipated before. This is due to the logarithmic nature of the scale-invariant potential and is crucial for the generation of GW.

%XXXXXXXXXXXXXXXXXXXXXXXXXXXXXXXXXXXXXXXXXXXXXXXXX
\section{Finite temperature \label{sec:finite-temp}}
%XXXXXXXXXXXXXXXXXXXXXXXXXXXXXXXXXXXXXXXXXXXXXXXXX
In this section, we introduce the finite-temperature effective potential, discuss applied approximations (including the expansion in powers of couplings) and the method of renormalisation-group improvement. We compare our approach to approximate schemes of evaluating the effective potential present in the literature and show that including the terms which are often omitted changes the results substantially.

%XXXXXXXXXXXXXXXXXXXXXXXXXXXXXXXXXXXXXXXXXXXXXXXXX
\subsection{Scalar potential at finite temperature \label{sec:finite-temp-potential}}
%XXXXXXXXXXXXXXXXXXXXXXXXXXXXXXXXXXXXXXXXXXXXXXXXX

The temperature-dependent effective potential is  obtained at one-loop order by adding a correction to the zero-temperature one-loop effective potential,
\be
V(h,\f,T)=\Vtree(h,\f)+\Vone(h,\f)+\VT(h,\f,T)+V_{\textrm{daisy}}(h,\f,T).\label{eq:V-eff}
\ee
The finite-temperature correction is given by the following formula:
\be
\VT(h,\f,T)=\frac{T^4}{2\pi^2}\sum_{a} n_a J_a\left(\frac{M_a(h,\f)^2}{T^2}\right),\label{eq:thermal-pot}
\ee
where the sum runs over particle species. $J_a$ denotes the thermal function, which is given by
\be
J_{F,B}(y^2) =  \int_0^\infty \dd{x} x^2 \log(1\pm e^{-\sqrt{x^2 + y^2}}),\label{eq:thermal-functions}
\ee
where the ``$+$'' sign is used for fermions ($J_F$), while the ``$-$'' for bosons ($J_B$).

It has been shown~\cite{Carrington:1992, Parwani:1992, Arnold:1992} that this basic formula for the effective potential is not enough since in the high-temperature limit higher-loop contributions can grow as large as the tree-level and one-loop terms. This means that the perturbative expansion in terms of  loops fails and one has to improve the computation scheme by resumming a class of leading contributions. This is commonly attained by a resummation of the so-called daisy diagrams in the high-temperature limit. 

We choose to use the Arnold-Espinosa~\cite{Arnold:1992} approach to daisy resummation. Therefore, we define $V_{\textrm{daisy}}$ as
\be
V_{\textrm{daisy}}(h,\f,T)=-\frac{T}{12\pi}\sum_i n_i \left[(M_{i,\textrm{th}}^2(h,\f,T))^{3/2}-(M_i^2(h,\f))^{3/2}\right],
\ee
where $n_i$ denotes the number of degrees of freedom (we only sum the scalar and longitudinal bosonic degrees of freedom), $M_{i,\textrm{th}}$ denotes thermally corrected mass, and $M_i$ the usual field-dependent mass. For the thermal masses see ref.~\cite{Prokopec:2018}.

%XXXXXXXXXXXXXXXXXXXXX
%~~~~~~~~~~~~~~~~~~~~~~~~~~~~~~~~~~~~~~~~~
\subsection{Renormalisation-group improvement\label{sec:RG-improvement}}
%~~~~~~~~~~~~~~~~~~~~~~~~~~~~~~~~~~~~~~~~~
%XXXXXXXXXXXXXXXXXXXXX

\begin{figure}[h!t]
    \centering
    \includegraphics[height=.2\textheight]{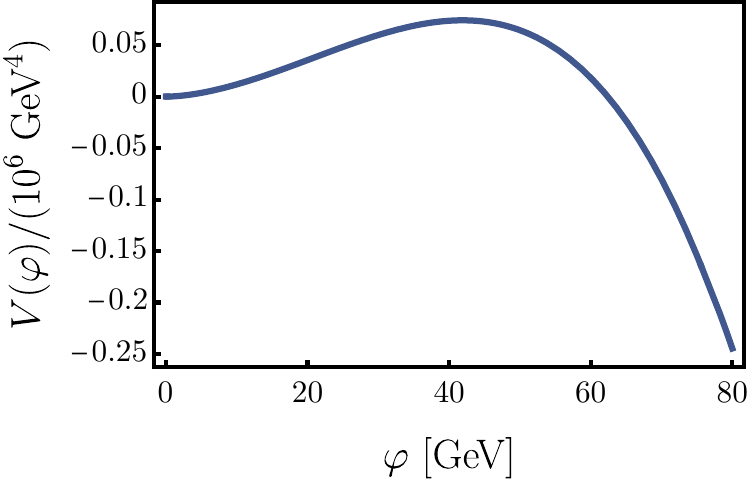}\hspace{6pt}
    \includegraphics[height=.2\textheight]{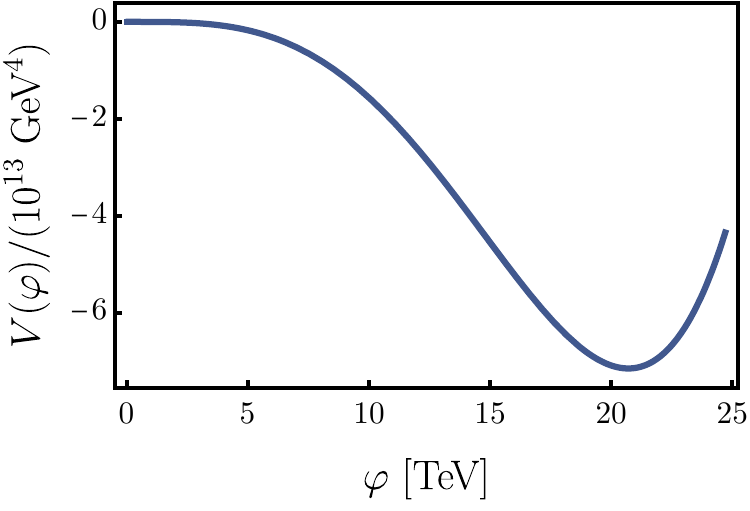}
    \caption{RG-improved effective potential at nucleation temperature along the $\f$ direction for a benchmark point with $g_X=0.9$, $M_X=10^4\g$ in two ranges of the field values -- around the barrier, where the tunnelling takes place (left panel), and around the minimum (right panel).}
    \label{fig:VatTnuc}
\end{figure}

The dependence on the renormalisation scale is a source of significant uncertainty in computations of gravitational wave spectra resulting from phase transitions in the early Universe, see e.g.~\cite{Croon:2020, Gould:2021}. The problem becomes even more severe in models with classical scale symmetry, where vastly different physically relevant scales are present. As has been shown in figure~\ref{fig:MS-w}, the VEV of the new scalar field is orders of magnitude above the EW scale. Moreover, the value of the field around which the thermal barrier forms, which is where the tunnelling takes place, is typically below the EW scale. Figure~\ref{fig:VatTnuc} shows the potential for a representative benchmark point (at nucleation temperature). For this point, the barrier is located around $\f=40\g$, whereas the minimum forms around $\f=20$\,TeV. Therefore, we cannot choose a single physically distinguished scale for the computations. Thus, we resort to renormalisation-group (RG) improvement of the effective potential. In this way, the renormalisation scale traces the value of the field.

We use the full one-loop finite-temperature effective potential defined in eq.~\eqref{eq:V-eff}, with the couplings replaced by the running couplings as $g\to g(t)$, $t=\log(\mu/\mu_0)$, and the field-renormalisation factors $Z$ included (with the replacement $\f^2\to Z_{\f}(t)\f^2$, $h^2 \to Z_{h}(t)h^2$).\footnote{See eq.~\eqref{eq:V-one-loop-along-phi} below for an explicit formula for the temperature independent part of the potential.} For the PT-related computations we restrict the effective potential to the $\f$ direction (using \texttt{CosmoTransitions}~\cite{Wainwright:2011} we have checked for a sample of parameter-space points that the $h$ component remains zero in the critical bubble solution, see also ref.~\cite{Prokopec:2018}). To obtain a potential that resums the leading logarithms we set the renormalisation scale to
\be
\mu=\textrm{max}\left(\frac{1}{2}g_X(M_X)\f,\  0.1\g\right)\equiv \textrm{max}\left(\overline{M}_X(\f),\ 0.1\g\right).\label{eq:choice-of-mu}
\ee
We introduced $\overline{M}_X$ which differs from the full $M_X$ in that it does not include the running of $g_X$ and the field $\f$ (otherwise the condition for $\mu$ would be implicit). 
We will show below that this difference is negligible from the point of view of the RG-improved potential. In the infrared, for small values of $\mu$, the SM couplings --- the strong coupling and the top Yukawa coupling --- and the SU(2)$_X$ coupling start growing rapidly, approaching their Landau poles.
This influences the effective potential at small field values. Therefore, we freeze the running of the couplings and fields below $M_X(\f)=0.1\g$ in eq.~\eqref{eq:choice-of-mu} in order not to transfer the bad behaviour of the SM couplings to the potential.\footnote{The SM couplings enter the potential restricted to the $\f$ axis via the thermal mass for the $h$ field, which is $h$ independent but depends on $\f$ via the RG improvement.}
Therefore, the running couplings and fields are evaluated at $t=\log\frac{\textrm{max}(M_X(\f),\ 0.1\g)}{M_Z}$~\footnote{In the temperature-dependent part of the potential we add the temperature to the cut-off function as $t=\log\frac{\textrm{max}(M_X(\f),\ T,\ 0.1\g)}{M_Z}$.}, where $M_Z$ is the reference scale. 
The dependence on the reference scale is negligible (see e.g.\ ref.~\cite{Chataignier:2018RSB} for an explicit check). For the $\beta$ and $\gamma$ functions of the model, see ref.~\cite{Chataignier:2018RSB}.

In section~\ref{sec:scale-dep}, we show explicitly how the physically relevant results would change if we did not use the RG-improved potential and discuss the implications.

In order to compute quantities that require non-vanishing values of the SM scalar field we use the potential with running couplings and fields to evolve between different scales. In this case, we do not fix $\mu=\overline{M}_X(\f)$. This allows us to compute the scalar masses with $\mu$ fixed to the electroweak scale and the decay width of the scalar $S$, which is essential for reheating (see section~\ref{sec:PT}), at $\mu=M_S$. It is important because, since the splitting of various scales is so large, the running of couplings between these scales can also be significant. %\comment{To be crossed out: \sout{Moreover, we can check whether the electroweak minimum even exists at the scale corresponding to the minimum along the $\f$ direction, which leads to the constraint presented in figures~\ref{fig:scalar-couplings} and~\ref{fig:MS-w}.}}

To emphasize the importance of the approximations applied to the effective potential, we now compare our approach described above with the prescription for the RG-improved effective potential used in the literature concerning the PT in the SU(2)cSM model~\cite{Hambye:2018, Baldes:2018} as well as another common treatment (see e.g.\ ref.~\cite{Ellis:2020}). We will focus here on the zero-temperature part of the effective potential along the $\f$ direction. The potential used in the present work reads
\be
V(\f)=\frac{1}{4}\lc(t) Z_{\f}(t)^2\f^4+\frac{9M_X(\f,t)^4}{64\pi^2}\left(\log\frac{M_X(\f,t)^2}{\mu^2}-\frac{5}{6}\right),\label{eq:V-one-loop-along-phi}
\ee
where $t=\log\frac{\mu}{\mu_0}$, $\mu_0=M_Z$, $M_X(\f,t)=\frac{1}{2}g_X(t)\sqrt{Z_{\f}(t)}\f$ and for $\mu$ we make the choice of eq.~\eqref{eq:choice-of-mu}.

Both of the approaches used in the literature mentioned above amount to approximating the one-loop zero-temperature effective potential of eq.~\eqref{eq:V-one-loop-along-phi} by the tree-level part with a running coupling.  
\begin{enumerate}
\item The approach of refs.~\cite{Hambye:2018,Baldes:2018} approximates the running quartic coupling via its $\beta$ function, relates the renormalisation scale with the field and uses as a reference scale the scale at which $\lambda_3$ changes sign,
\be
V_1\approx \frac{1}{4}\lc(t) \f^4 \approx \frac{1}{4}\frac{9 g_X^4}{128\pi^2}\log(\frac{\f}{\f_0}),\label{eq:potential-Strumia}
\ee
where $t=\log\frac{\mu}{\f_0}$, $\lc(0)=0$ and $g_X$ is interpreted as evaluated at $\mu=\f_0$ (the running of $g_X$ is not included).
\item The approach of ref.~\cite{Ellis:2020} also approximates the one-loop potential by the tree-level potential with running coupling but uses $\mu=\f$ and some fixed reference scale $\mu_0$, e.g.\ the top mass,
\be
V_2\approx \frac{1}{4}\lc(t) \f^4,\label{eq:potential-updated}
\ee
where $t=\log(\frac{\f}{\mu_0})$.
\end{enumerate}

It should be noted that with the choice of scale $t=\log(\frac{\f}{\mu_0})$ neither the logarithm of the one-loop potential of eq.~\eqref{eq:V-one-loop-along-phi} nor the term proportional to the constant $-\frac{5}{6}$ is strictly cancelled. It is common lore to claim that the remaining contributions are subleading, as well as the contribution from the running of the couplings in the one-loop correction to the effective potential. One should, however, remember that in the case of classically scale-invariant models, the one-loop contribution and the tree-level term constitute together the leading order contribution to the potential and should be treated on equal footing (at fixed order expansion in the couplings). Therefore, in this work, we include all the terms and simplify the result by neglecting contributions of order $g^8$ and higher (including Goldstone bosons). 

\begin{figure}[ht]
    \centering
    \includegraphics[height=.23\textheight]{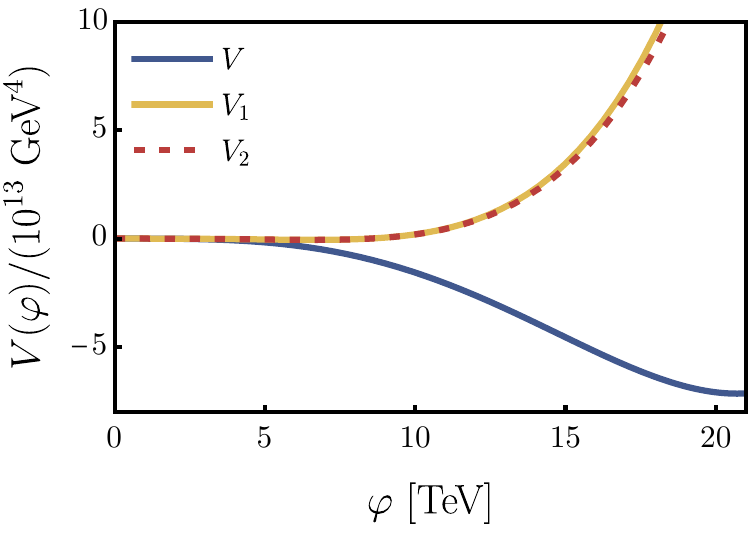}\hspace{2pt}
    \includegraphics[height=.23\textheight]{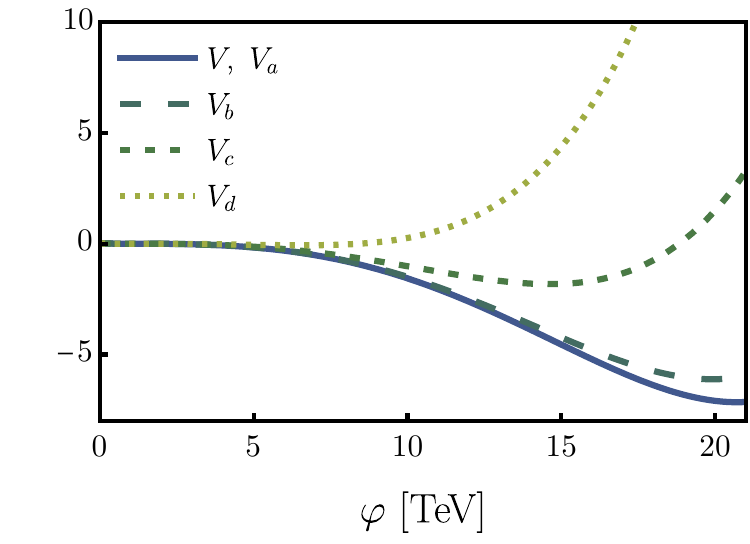}
    \caption{Effective potential at zero temperature along the $\f$ direction for a benchmark point with $g_X=0.9$, $M_X=10^4\g$ (defined at $\mu=M_X$). Left panel: Comparison of different approaches used in the literature, $V_1$ of eq.~\eqref{eq:potential-Strumia} (yellow solid), $V_2$ of eq.~\eqref{eq:potential-updated} (dashed red) and the full potential $V$ of eq.~\eqref{eq:V-one-loop-along-phi} used in this work (solid blue). Right panel: Comparison of different approximations imposed on the full potential $V$ of eq.~\eqref{eq:V-one-loop-along-phi} used in this work (solid blue) discussed in points~\ref{point-log}--\ref{point-tree} in the main text: $V_b$ (long-dashed darkest green), $V_c$ (medium-dashed dark green), $V_d$ (short-dashed light green).}
    \label{fig:potential-comparison}
\end{figure}

The comparison of the approaches described above can be seen in figure~\ref{fig:potential-comparison} (left panel). The two approaches of eqs.~\eqref{eq:potential-Strumia} and~\eqref{eq:potential-updated} prove, as expected, to be very close to each other. However, they differ strikingly from our result, which proves that the omitted contributions are non-negligible.

To better understand which contributions are crucial we perform a series of approximations or modifications on our approach, the results of which are presented in the right panel of figure~\ref{fig:potential-comparison}. Namely:
\begin{enumerate}[label=(\alph*)]
    \item\label{point-log} $V_a$ corresponds to the potential of eq.~\eqref{eq:V-one-loop-along-phi} with the part proportional to the logarithm neglected. $V_a$ exactly overlaps with the full potential (solid blue line).
    \item $V_b$ corresponds to the potential of eq.~\eqref{eq:V-one-loop-along-phi} with the choice of $\mu=\f$ (darkest green, long-dashed line). This choice alone does not modify the potential significantly with respect to our choice (solid blue line).
    \item $V_c$ corresponds to the potential of eq.~\eqref{eq:V-one-loop-along-phi} with the constant $-\frac{5}{6}$ neglected (dark green, medium-dashed curve). Since, as explained in point~\ref{point-log} above, the omission of the logarithm (with our choice of the scale) does not visibly modify the result, $V_c$  is equivalent to using the tree-level part of eq.~\eqref{eq:V-one-loop-along-phi}. Here the difference with respect to the full potential is significant. It is understandable since the choice of the scale was such as to annihilate the logarithmic term but not the $\frac{5}{6}$ constant.
    \item\label{point-tree} $V_d$ corresponds to the tree-level part of eq.~\eqref{eq:V-one-loop-along-phi} but with the choice $\mu=\f$ (light green, short-dashed line), which makes this choice very close to $V_1$ and $V_2$ discussed above. This can be also seen by comparing the left and the right panel of figure~\ref{fig:potential-comparison}. Clearly, $V_d$ differs significantly from the full potential.
\end{enumerate}

This simple check proves that the approximations widely used in the literature modify the potential significantly, affecting e.g.\ the location of the minimum, which influences the GW spectra. Admittedly, shifting the VEV by a factor of 3 may not change the GW spectrum significantly as in the discussed model the values of $\alpha$ (see section~\ref{sec:PT-and-GW}) are huge and consequently drop out from the final expressions. It is, however, harder to assess the influence of the modified shape of the potential on the computation of the length or time scale of the transition. With the use of the full (up to order $g^7$) one-loop potential, we improve on the reliability of the existing results.

%~~~~~~~~~~~~~~~~~~~~~~~~~~~~~~~~~~~~~~~~~
\subsubsection{Consistent expansion in powers of couplings\label{sec:expansion-in-couplings-finite-T}}
%~~~~~~~~~~~~~~~~~~~~~~~~~~~~~~~~~~~~~~~~~
In section~\ref{sec:scaling-of-couplings}, we discussed the expansion in powers of couplings by assigning formal scaling properties to various couplings. The same should be done in the temperature-dependent part of the potential. The necessity of consistent expansion in powers of couplings in order to provide the scale- and gauge-independence of the results has been emphasized in the literature, see e.g.\ \cite{Arnold:1992, Ekstedt:2020, Gould:2021, Ekstedt:2022}. Commonly, in models with mass terms present already at tree level, it is assumed that $\lambda\sim g^2$ and then from the requirement that in the vicinity of a first-order phase transition, different terms in the potential are of similar order yields $\f\sim gT$. These assumptions are modified if the barrier is thermally induced, then $\lambda\sim g^3$ (see refs.~\cite{Arnold:1992, Ekstedt:2022}). In the case of radiatively generated minimum, this relation is further modified because $\lambda\sim g^4$ and comparison of different terms in the potential amounts to 
\be
g\f\sim T.\label{eq:finite-T-scaling}
\ee
Using this scaling we estimate orders of different contributions to the potential and use an expansion to order $g^7$. We could have used the first non-trivial order of $g^4$, however, then we would not include contributions from thermal masses and the potential would be substantially modified. Therefore, we choose to work at $\mathcal{O}(g^7)$, which allows us to ignore the scalar contributions to the zero-temperature one-loop effective potential. 

In a recent article~\cite{Gould:2021} it has been shown that, in order to maintain RG-scale invariance of the finite-temperature effective potential, one should also consider some two-loop diagrams as they are needed to cancel the scale dependence of lower-loop-order diagrams at fixed order in $g$. The additional terms that should be added according to ref.~\cite{Gould:2021} scale as $g^4 T^2 \f^2$, which would amount to $g^6 \f^4$ with our scaling of eq.~\eqref{eq:finite-T-scaling} coming from scale invariance. This suggests that indeed, also in our case, in order to preserve renormalisation scale invariance we should study two-loop diagrams. One should note, however, that the reasons behind the scaling issues discussed in this section rely on the high-temperature expansion of the effective potential, which does not hold everywhere in the case of conformal potentials (which is clear from eq.~\eqref{eq:finite-T-scaling}). Therefore, a detailed study taking into account the assumption of classical scale invariance would be needed, which we leave for future work.  
%XXXXXXXXXXXXXXXXXXXXXXXXXXXXXXXXXXXXXXXXXXXXXXXXX
\section{Phase transition and gravitational-wave signal\label{sec:PT-and-GW}}
%XXXXXXXXXXXXXXXXXXXXXXXXXXXXXXXXXXXXXXXXXXXXXXXXX

In this section, we review the parameters that characterise a first-order phase transition and are needed to compute the predictions for the gravitational-wave signal and outline our computation methods. Presenting the results of the scan of the parameter space, we discuss the character of the phase transition. Finally, we present a sample of GW spectra produced in a first-order PT in the SU(2)cSM model.

%~~~~~~~~~~~~~~~~~~~~~~~~~~~~~~~~~~~~~~~~~~~~~~~~~
\subsection{Phase transition\label{sec:PT}}
%~~~~~~~~~~~~~~~~~~~~~~~~~~~~~~~~~~~~~~~~~~~~~~~~~
A first-order phase transition proceeds through nucleation, growth and percolation of bubbles filled with the broken-symmetry phase in the sea of the symmetric phase. This corresponds to the fields tunnelling through a potential barrier. In principle, the tunnelling proceeds in the full field space of $h$ and $\f$, i.e.\ we should solve for the critical bubble in the two-dimensional field space. For the model at hand, we have checked with the use of \texttt{CosmoTransitions}~\cite{Wainwright:2011} that for a sample of benchmark points $h$ remains zero in the solution. This means that the tunnelling proceeds along the $\f$ direction, while the transition in the $h$ direction is smooth. Assuming that this holds for the full parameter space, i.e.\ that only the $\f$ direction is relevant for the tunnelling, allows us to obtain better accuracy and simplifies the issue of RG improvement of the potential (see section~\ref{sec:RG-improvement}). This was also a common assumption in earlier works~\cite{Hambye:2018, Baldes:2018} and was checked in ref.~\cite{Prokopec:2018}. Note that, as was pointed out in ref.~\cite{Prokopec:2018}, this is incompatible with the Gildener--Weinberg approach, which only analyses the potential along the direction from the origin of the field space to the true minimum.

In models with classical scale invariance the onset of the phase transition can be delayed so much that QCD phase transition at $T_{\mathrm{QCD}}\approx0.1\g$ proceeds first~\cite{Iso:2017uuu, vonHarling:2017, Hambye:2018}. Then the quark condensate forms and by coupling to the Higgs boson generates an effective mass term in the potential, thus changing the mechanism and dynamics of the phase transition. This kind of behaviour was analysed e.g.\ in refs.~\cite{Hambye:2018, Baldes:2018,Ellis:2020}. In the present work, we do not consider the QCD-sourced PT and focus on the phase transition caused by the tunnelling, because we can model it with more accuracy. Therefore, we do not present results for points with percolation temperature below 0.1\,GeV, which corresponds to the region in the lower left part of the plots shown in the following in the darkest grey.

In what follows we review the temperature evolution of the potential, defining the parameters characterising the phase transition. For all the computations beyond solving the bounce equation, we use self-developed code, optimized for the case of a scale-invariant potential. We perform a scan of the parameter space, taking as a starting point the result of the zero-temperature scan described in section~\ref{sec:zero-temp}.

\begin{figure}[h!t]
\center
\includegraphics[width=.8\textwidth]{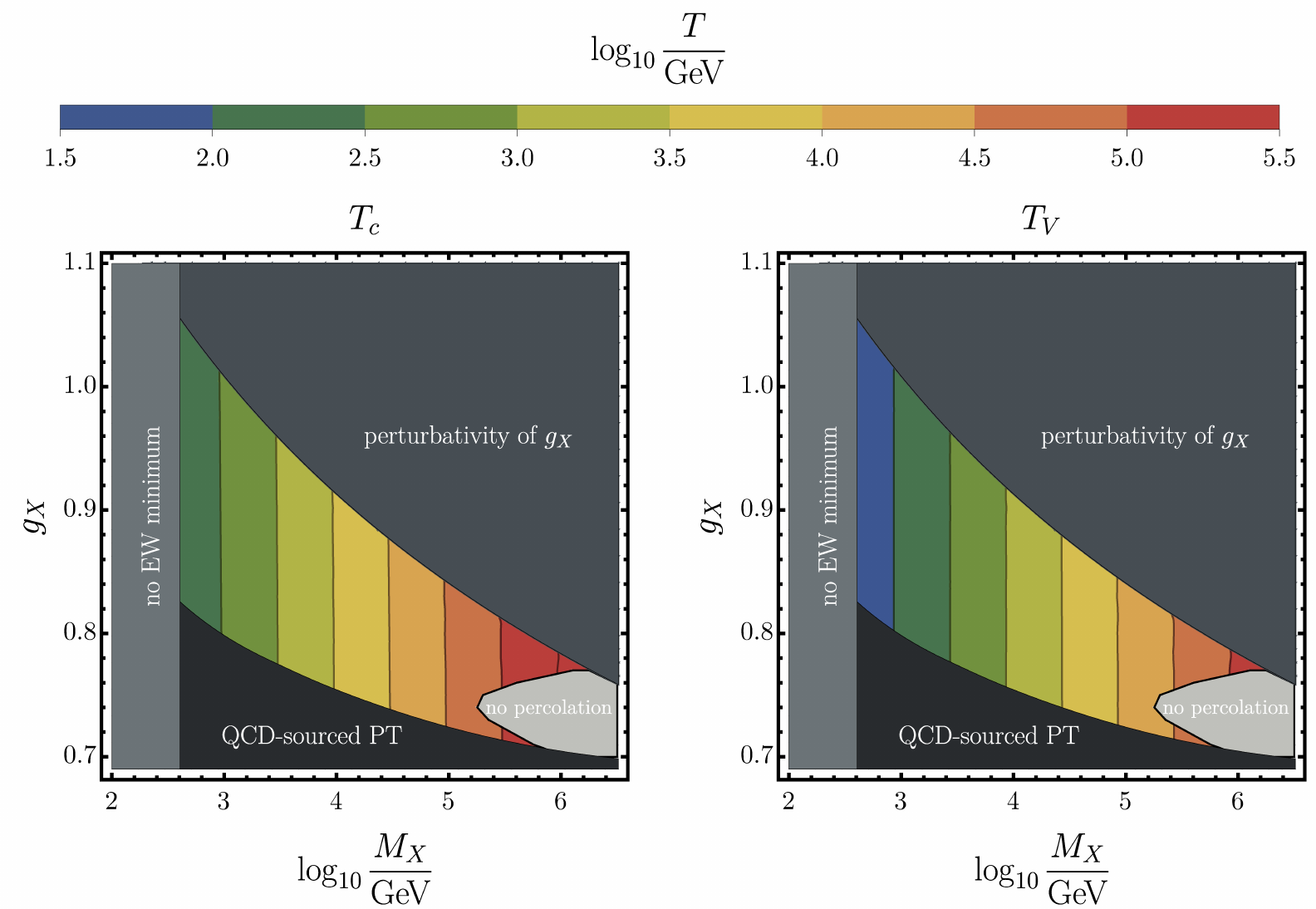}
\caption{The values of the critical temperature $T_c$ (left panel) and the temperature at which thermal inflation starts $T_V$ (right panel). \label{fig:Tc-TV}}
\end{figure}

\paragraph{Critical temperature $T_c$} At high temperatures the symmetry is restored and the effective potential has a single minimum at the origin of the field space. As the Universe cools down, a second minimum is formed. At the critical temperature, the two minima are degenerate, and for lower temperatures, the minimum with broken symmetry becomes the true vacuum. This is the temperature at which the tunnelling becomes possible. The values of the critical temperature obtained for the SU(2)cSM are shown in the left panel of figure~\ref{fig:Tc-TV}. The lower the mass of $X$, the closer the critical temperature is to the electroweak values, $T_c\approx 100\g$. For higher $M_X$ the critical temperature becomes significantly higher than in the standard picture of the EW phase transition.

In figure~\ref{fig:Tc-TV} there are the same excluded areas as before (see the discussion of figure~\ref{fig:scalar-couplings}), and two new shaded regions. The lower left corner (darkest grey) is not analysed because there the PT is sourced by the QCD phase transition, which is beyond the scope of the present work. The light-grey region around $M_X\approx 10^6\g$ is where the percolation criterion of eq.~\eqref{eq:percolation-crit} is violated and is discussed in more detail below.

\paragraph{Temperature $T_V$ at which thermal inflation starts}
As pointed out in ref.~\cite{Hambye:2018, Ellis:2018}, if there is large supercooling, i.e.\ the phase transition is delayed to low temperatures, much below the critical temperature, it is possible that a period of thermal inflation due to the false vacuum energy appears before the phase transition completes. The Hubble parameter can be written as
\be
H^2=\frac{1}{3\bar{M}_{\rm{Pl}}^2}(\rho_R+\rho_V)=\frac{1}{3\bar{M}_{\rm{Pl}}^2}\left(\frac{T^4}{\xi^2_g}+\Delta V\right),
\ee
where $\xi_g=\sqrt{30/(\pi^2 g_*)}$ and $g_*$ is the number of degrees of freedom in the plasma, while $\Delta V$ is the difference of the values of the effective potential at false and true vacuum, $\bar{M}_{\textrm{Pl}}$ denotes the reduced Planck mass, $\bar{M}_{\textrm{Pl}}=2.435 \cdot 10^{18}\g$. The onset of the period of thermal inflation can be approximately attributed to the temperature at which vacuum and radiation contribute to the energy density equally, 
\be \label{eq:T-thermal-inflation}
 %\frac{T_V^4}{\xi^2_g} = \Delta V
 T_V \equiv \qty(\xi^2_g \Delta V )^\frac{1}{4}.
\ee
For supercooled transitions, it is a good approximation to assume that $\Delta V$ is independent of the temperature below $T_V$.

By using the temperature $T_V$, the Hubble constant can be rewritten in the following way:
\be
H^2 \simeq \frac{1}{3\bar{M}_{\rm{Pl}}^2 \xi^2_g}\left(T^4+T_V^4\right).
\ee
This approximation works very well and we implement it in our calculations. Moreover, in the case of large supercooling, the contribution to the Hubble parameter from radiation energy can be neglected leaving
\be
H^2 \simeq H^2_V=\frac{1}{3\bar{M}_{\rm{Pl}}^2}\Delta V.
\ee
The values of $T_V$ obtained for the parameter space of SU(2)cSM are presented in the right panel of figure~\ref{fig:Tc-TV}. The vacuum domination begins not much below the critical temperature.

\paragraph{Nucleation temperature $T_n$} Below the critical temperature nucleation of bubbles of true vacuum becomes possible. In order to compute the decay rate of the false vacuum we start by solving the bounce equation,
\be
\frac{\mathrm{d}^2\f}{\mathrm{d}r^2}+\frac{2}{r}\frac{\mathrm{d}\f}{\mathrm{d}r}=\frac{\mathrm{d}V(\f,T)}{\mathrm{d}\f},
\ee
with the following boundary conditions: $\frac{\mathrm{d}\f}{\mathrm{d}r}=0$ for $r=0$ and $\f\to0$ for $r\to\infty$. To this end, we use the dedicated code \texttt{CosmoTransitions}~\cite{Wainwright:2011}. Above, $V(\f,T)$ denotes the RG-improved finite-temperature potential computed along the $\f$ direction, as discussed in section~\ref{sec:finite-temp}. Once the bubble profile is known we can compute the Euclidean action along the tunneling path
\be
S_3(T)=4\pi\int \mathrm{d}r ~r^2 \left[ \frac{1}{2}\left(\frac{\mathrm{d}\f}{\mathrm{d}r}\right)^2 + V(\f,T) \right].
\ee
Then the decay rate of the false vacuum due to the thermal fluctuations\footnote{In the considered parameter space, the zero-temperature tunnelling rate is always smaller than the thermal one, thus the tunnelling proceeds via the thermal critical bubble.} is given by
\be
\Gamma(T)\approx T^4\left(\frac{S_3(T)}{2\pi T}\right)^{3/2} e^{-S_3(T)/T}.
\ee

\begin{figure}[h!t]
\center
\includegraphics[width=.8\textwidth]{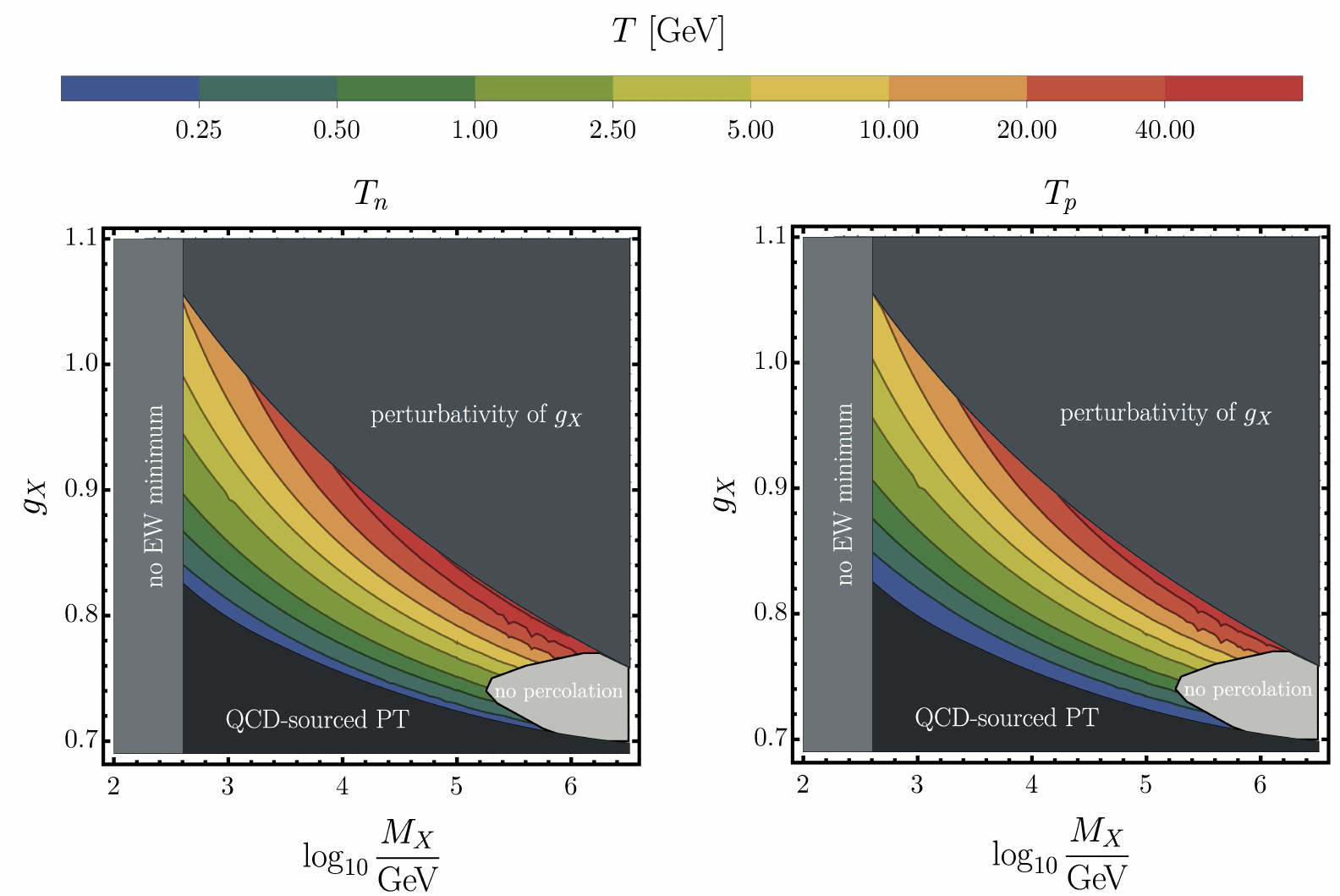}
\caption{The values of the nucleation temperature $T_n$ (left panel) and the percolation temperature $T_p$ (right panel).\label{fig:Tn-Tp}}
\end{figure}

The nucleation temperature is defined as the temperature at which at least one bubble is nucleated per Hubble volume, which can be interpreted as the onset of the PT. It can be defined as follows~\cite{LINDE198137, LINDE1983421}: 
\be
N(T_n)=1=\int_{t_c}^{t_n}dt\frac{\Gamma(t)}{H(t)^3}=\int_{T_n}^{T_c}\frac{dT}{T}\frac{\Gamma(T)}{H(T)^4}.
\ee
It should be noted that the common criterion for evaluating $T_n$ as $S_3/T_n\approx 140$ is not reliable in the case of strongly supercooled transitions as it relies on the assumption of radiation domination at the moment of the phase transition. The values of the nucleation temperature $T_n$ obtained for the parameter space of the SU(2)cSM are presented in figure~\ref{fig:Tn-Tp} (left panel). One can appreciate the degree of supercooling present in the studied model by comparing the nucleation temperature with the temperature at which thermal inflation starts -- the former is orders of magnitude below the latter, see figure~\ref{fig:Tc-TV}. That is a common feature of (nearly) conformal models, and it results in the production of a strong GW signal, as we will show shortly.
\paragraph{Percolation temperature $T_p$} Percolation of bubbles can be considered as the moment of finalising the phase transition. In strongly supercooled transitions, it is important to assure percolation, as in principle the thermal inflation could prevent the bubbles of true vacuum from filling the entire space.  It also needs not be very close to the nucleation temperature, as is often assumed in models with polynomial potentials. We choose the percolation temperature as the characteristic temperature of the PT, at which we will evaluate the parameters relevant for the GW generation. In computing the percolation temperature we follow ref.~\cite{Ellis:2018}.

The probability of finding a point still in the false vacuum at a certain temperature is given by $P(T) = e^{-I(T)}$, where $I(T)$ is the amount of true vacuum volume per unit comoving volume and reads as follows~\cite{Guth:1981}: 
\begin{align}
		I(T) = \frac{4\pi}{3} \int_{T}^{T_c} \dd{T^\prime} \frac{\Gamma(T^\prime)}{T^{\prime 4} H(T')}\qty(\int_{T}^{T'} \frac{\dd{\tilde T} }{ H(\tilde T)})^3.
\end{align}
In order to simplify this expression we can distinguish between the vacuum and radiation domination period which leads to the Hubble parameter in the following form:
\begin{align} \label{hubble_approx}
	H(T) \simeq 
\left\{\begin{array}{ll}
	H_{\mathrm{R}}(T)=\frac{T^{2}}{\sqrt{3} \bar{M}_{\mathrm{pl}} \xi_{g}}, &\quad \mathrm{for}\quad T > T_{V}, \\
	H_{\mathrm{V}}=\frac{T_{V}^{2}}{\sqrt{3} \bar{M}_{\mathrm{pl}} \xi_{g}}, &\quad \mathrm{for}\quad T < T_{V}.
\end{array}\right.
\end{align}
We can thus write a simplified version of $I(T)$ valid in the region where $T<T_V$:
\begin{align}
\begin{split}
I_{\mathrm{RV}}(T) =\frac{4 \pi}{3 H_{\mathrm{V}}^4}\left(\int_{T_V}^{T_c} \frac{d T^{\prime} \Gamma\left(T^{\prime}\right)}{T^{\prime 6}} T_V^2\left(2 T_V-T-\frac{T_V^2}{T^{\prime}}\right)^3+\int_T^{T_V} \frac{d T^{\prime} \Gamma\left(T^{\prime}\right)}{T^{\prime}}\left(1-\frac{T}{T^{\prime}}\right)^3\right) .
\end{split}
\end{align}
Commonly, the moment of percolation is estimated by the criterion from ref.~\cite{Vinod:1971}, $I(T_p) \gtrsim 0.34$. It corresponds to the ratio of the volume in equal-size and randomly-distributed spheres (including overlapping regions) to the total volume of space being equal to 0.34. A concrete percolation criterion is not known for de Sitter space~\cite{Guth:1981}, thus we resort to the aforementioned criterion assuming it should be an approximation to the correct value. 
Thus, we approximate the value of percolation temperature using
\be
I_{\rm RV}(T_p) = 0.34.\label{eq:percolation-temp}
\ee

The results of the scan of the parameter space for $T_p$ can be found in figure~\ref{fig:Tn-Tp} (right panel). Comparing to the values of $T_n$ one can see that these two temperatures are of the same order, yet they differ, hence we will not use $T_n$ as a proxy for the temperature at which the PT proceeds (see~\cite{Ellis:2018}).

Moreover, the simple criterion above may not be sufficient for the completion of the transition in a vacuum domination scenario. One also needs to make sure that the volume of the false vacuum $V_f \sim a^3(T) P(T)$  \cite{Turner:1992} is  decreasing around the percolation temperature. This condition is especially constraining in models featuring strong supercooling, as thermal inflation can prevent bubbles from percolating. It can be expressed as follows:
\begin{align}
\frac{1}{V_f} \dv{V_f}{t} = 3 H(t) - \dv{I(t)}{t} = H(T)\qty(3+T\dv{I(T)}{T}) < 0.\label{eq:percolation-crit}
\end{align}
%In older literature \cite{Enqvist:1992} one can find a slightly different percolation criterion $I(\tilde T_p) > 1 $, which leads to lower value of percolation temperature. In that case the condition of false vacuum shrinking can be satisfied easier.

\begin{figure}
    \centering
    \includegraphics[width=.4\textwidth]{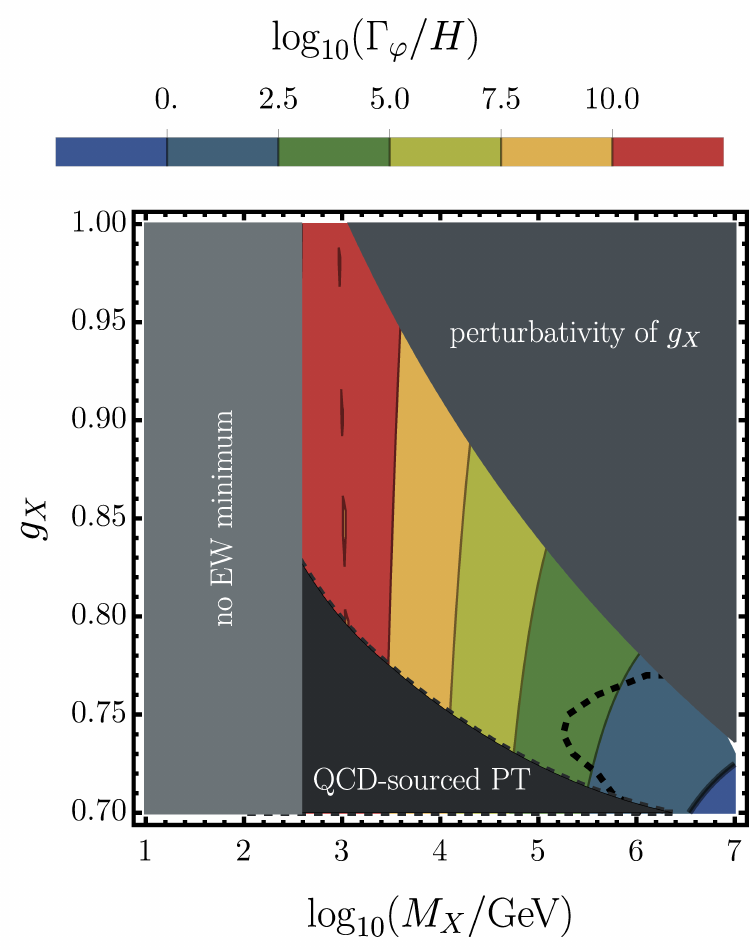}%\hspace{2pt}
    \caption{Contour plot of the decimal logarithm of the ratio of the energy transfer rate $\Gamma_{\f}$ to the Hubble parameter $H$. The equality $H=\Gamma_{\f}$ is indicated as a thick black solid line in the lower right corner. The percolation bound is shown as a black dashed line (in other plots it is shown as a light-grey region).
    \label{fig:reheating}}
\end{figure}

\begin{figure}[h!t]
\center
\includegraphics[width=.4\textwidth]{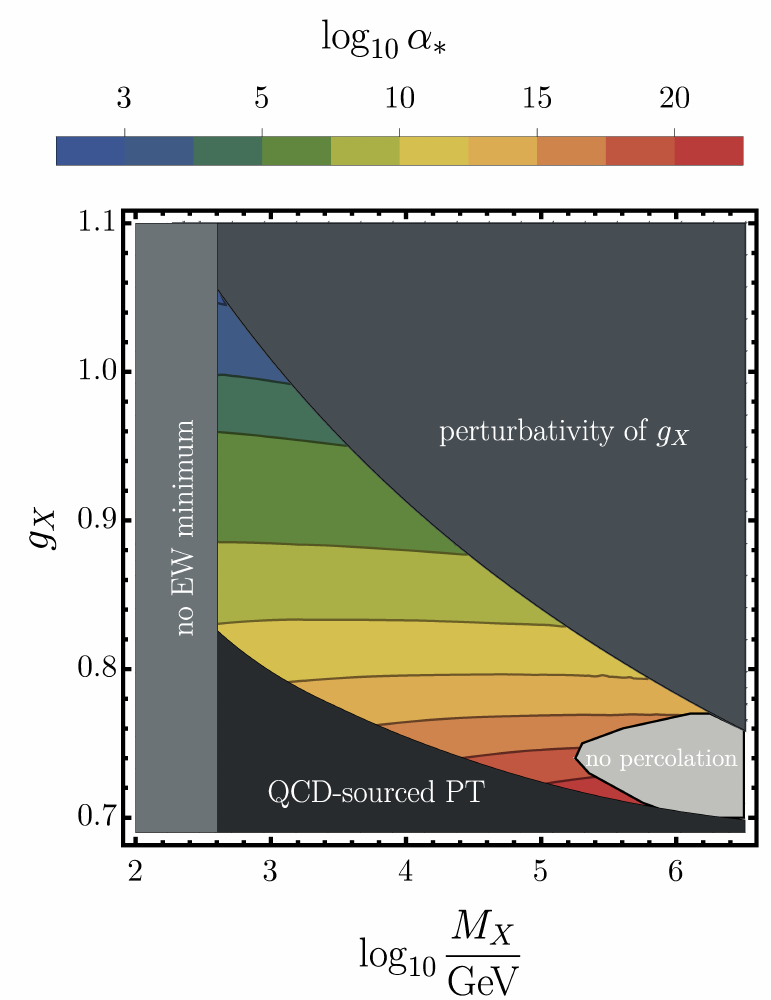}
\caption{The values of transition strength parameter $\alpha_*$.\label{fig:alpha}}
\end{figure}

\begin{figure}[h!t]
\center
\includegraphics[width=.4\textwidth]{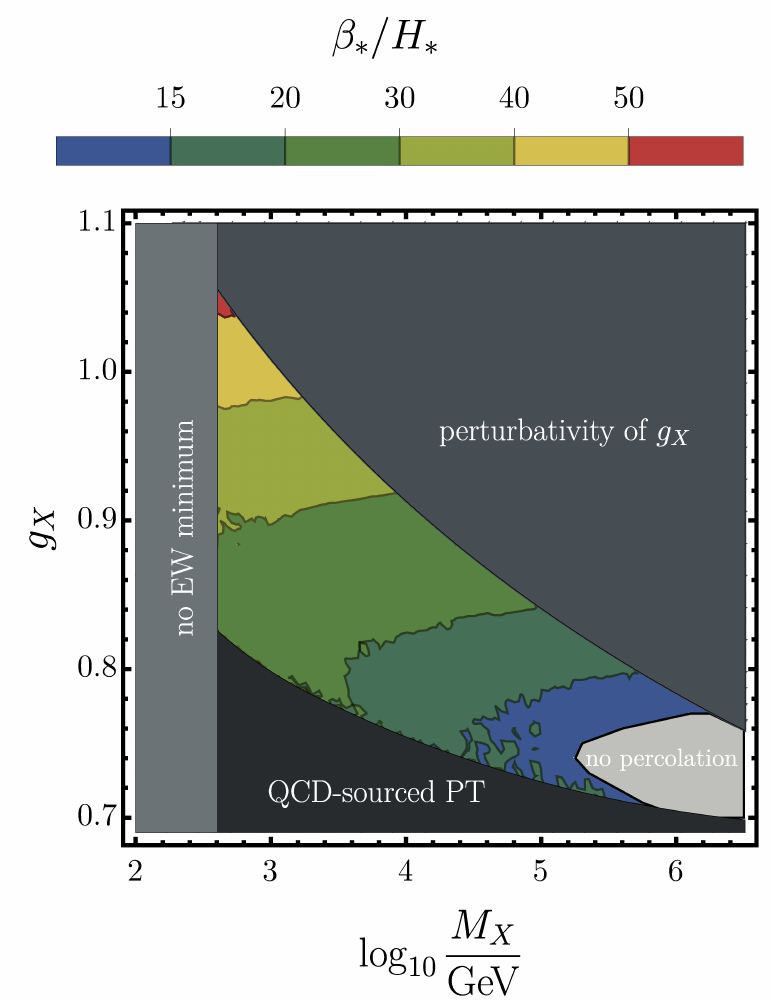}\hspace{20pt}
\includegraphics[width=.4\textwidth]{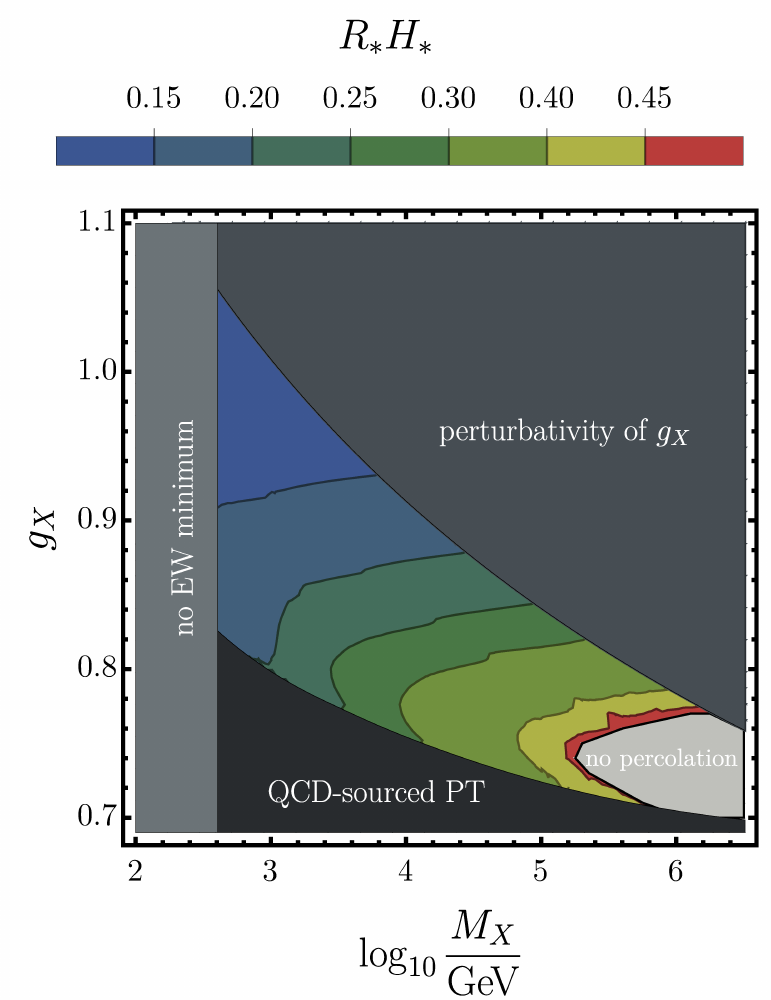}
\caption{The values of inverse time scale $\beta / H_*$ (left panel) and the length scale of the transition $R_*H_*$ (right panel).\label{fig:beta-RH}}
\end{figure}

\begin{figure}[h!t]
\center
\includegraphics[width=.8\textwidth]{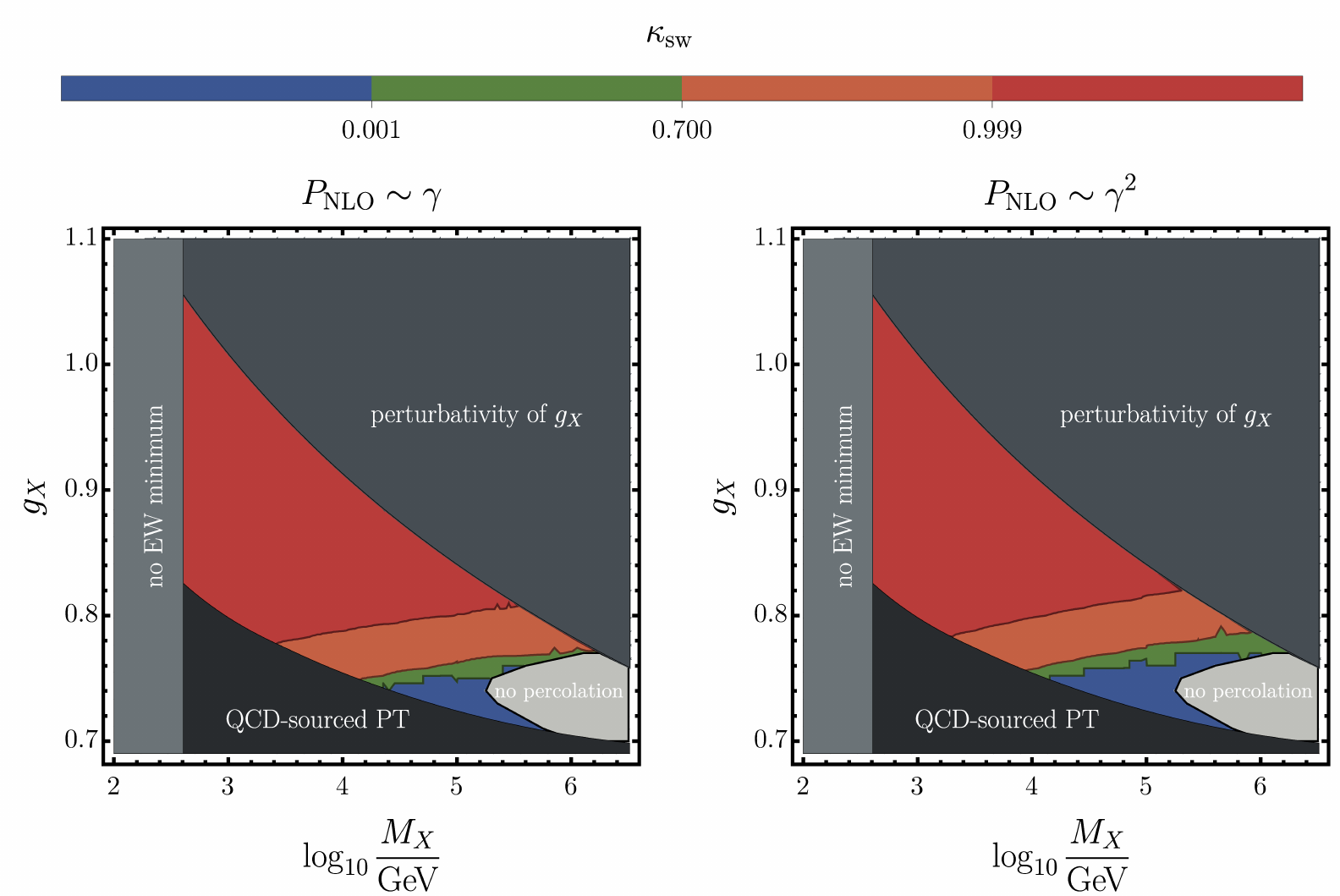}\hspace{20pt}
\caption{The values of the efficiency factor $\kappa_{\mathrm{sw}}$ of transferring energy of the PT into sound waves. The results were obtained by using the friction force of ref.~\cite{Gouttenoire:2021kjv} (left panel) and of ref.~\cite{Hoche:2020ysm} (right panel). \label{fig:kappa-sw}}
\end{figure}

The region where the percolation criterion of eq.~\eqref{eq:percolation-crit} is not satisfied is shown in figures~\ref{fig:Tc-TV}--\ref{fig:kappa-sw} in light-grey (lower-right part of the plots). We know that in this region the phase transition does not complete via bubble percolation we thus do not analyse this region further, as our aim in this paper is to analyse the phase transition which is sourced by tunnelling under the potential barrier and proceeding via nucleation and percolation of bubbles. Nonetheless, we will now comment on the possible ways of completing the PT in this region. First, it is possible that the Universe keeps inflating until the onset of the QCD phase transition and then the PT completes because of the appearance of the quark condensate. Then, this region would belong to the same class as the region in the lower part of our parameter space (dark grey). If the thermal inflation lasts long enough, it may become possible to end the PT via quantum fluctuations, see ref.~\cite{Lewicki:2021xku}\footnote{We thank the referee for pointing out this option.}. In ref.~\cite{Lewicki:2021xku}, it was shown that before a phase transition is ended by quantum fluctuations a substantially long period of thermal inflation takes place and there is a discontinuous change in the number $N$ of $e$-folds of inflation between the region where percolation of bubbles ends the phase transition and the region where quantum fluctuations become significant. The first scenario was found to be realised for $N\sim\mathcal{O}(1)$--$\mathcal{O}(10)$, whereas the latter for $N\approx 20$--$50$. We can evaluate the number of $e$-folds until QCD PT by computing $N=\log\frac{T_V}{T_{\mathrm{QCD}}}$~\cite{Hambye:2018}. We find $N\approx 15$ at the right edge of the $M_X$ axis, therefore we expect that throughout the presented parameter space the phase transition is QCD induced. Extrapolating the results of our scan we expect that the bound from perturbativity of $g_X$ and the region of QCD sourced PT should meet around $M_X=10^8\g$, which would result in $N\approx 18$ so we do not expect the scenario of quantum fluctuations ending the PT to be realised (or possibly it may be realised in a very small part of the parameter space). The detailed study of the boundary between the two scenarios is beyond the scope of the present article.
%The region where the percolation criterion, eq.~\eqref{eq:percolation-crit} is not satisfied is shown in figures~\ref{fig:Tc-TV}--\ref{fig:kappa-sw} as the light-grey excluded region. It constrains significantly the region of the strongest possible transitions (compare with figure~\ref{fig:alpha}), in particular the region of the parameter space where GW could be sourced by bubble-wall collisions (see figure~\ref{fig:kappa-sw} below). Moreover, it also excludes the possibility of inefficient reheating at lower values of $g_X$, which in turn prohibits the supercool DM scenario (for details see below, section~\ref{sec:DM}). This shows that this criterion is important, also in models with classical scale invariance.

\paragraph{Reheating temperature $T_r$} The total energy released in the phase transition corresponds to the energy stored (before the PT) in the true vacuum and is given by $\Delta V(T_p)\approx \Delta V (T=0)\equiv \Delta V$. For reheating the Universe, the energy has to be transferred from the scalar field $\f$ to the relativistic plasma. If reheating is instantaneous, this whole energy is turned into the energy of radiation,
\be
\Delta V = \rho_R(T_r)=\rho_R(T_V),\label{eq:condition-Tr-1}
\ee
where we assume that the number of relativistic degrees of freedom remains constant and use eq.~\eqref{eq:T-thermal-inflation}. This gives  (see e.g.\ refs.~\cite{Ellis:2018, Hambye:2018})
\be
T_r=T_V.\label{eq:Treh=TV}
\ee

On the other hand, if at $T_p$ the rate of energy transfer from the $\f$ field to the plasma, $\Gamma_{\f}$, is smaller than the Hubble parameter, $\Gamma_{\f} < H(T_p)$, then the energy will be stored in the scalar field oscillating about the true vacuum and redshift as matter until %the temperature reaches $T_d$, when $\Gamma=H(T_d)$, thus
%$$
%\rho_R(T_r)=\left(\frac{a(T_p)}{a(T_d)}\right)^3 \Delta V.
%$$
%The ratio of the scale factors can be rewritten in terms of the Hubble rates at respective times, where $H(T_d)=\Gamma$ and $H(T_p)=H_*$.  With this we obtain the formula for the reheating temperature
%\be
%T_r=\left(\Gamma H_*\right)^{1/2} \left(\frac{30 \Delta V}{\pi^2 g_*}\right)^{1/4}=\left(\Gamma H_*\right)^{1/2} T_V^4 .
%\ee
$\Gamma_{\f}$ becomes comparable to the Hubble parameter. In this case, the reheating temperature will read~\cite{Ellis:2019, Hambye:2018}
\be
T_r=T_V\sqrt{\frac{\Gamma_{\f}}{H_*}}  .
\label{eq:Treh-general}
\ee
For a more refined treatment of the reheating temperature see e.g.\ ref.~\cite{Ellis:2020}. %In our numerical study we find that for all the points in the parameter space reheating is instantaneous, thus we do not discuss this issue further.

In order to determine which of these scenarios takes place one has to evaluate $\Gamma_{\f}$. Before the phase transition, the energy is stored in the $\f$ field. Thus, for assessing the efficiency of reheating we should know the decay rate of the $\f$ field which quantifies the energy transfer rate from $\f$ to the plasma. The $\f$ field can be understood as a mixture of the mass eigenstates $H$ and $S$ (by the inverse of eq.~\eqref{eq:mixing}) for which the decay widths are well defined. The decay width of $H$ is equal to the SM Higgs decay rate rescaled due to the mixing by $\xi_H^2$ (see eq.~\eqref{eq:xi}). Since the SM Higgs decay width is fairly large, we can safely assume that the $H$ component of $\f$ decays quickly. Due to the mixing between $h$ and $\f$ the decay width of $S$ is a sum of SM-like decays with couplings rescaled by $\xi_S$ and, in case $M_S>2M_H$ the scalar decay $S\to HH$. The decay to a pair of gauge bosons $S\to XX$ is kinematically forbidden. Therefore, the rate of energy transfer from $\f$ to the plasma reads
\begin{align}
    \Gamma_{\f}=& (1-\xi_S^2)\Gamma(S)=\xi_S^2(1-\xi_S^2)\Gamma_{\mathrm{SM}}(S)+(1-\xi_S^2)\Gamma(S\to HH),\label{eq:gamma-phi}
\end{align}
where $\Gamma_{\mathrm{SM}}$ denotes a decay width computed as in the SM, i.e. with the same couplings and decay channels, but for a particle of mass $M_S$.\footnote{One should note that as $M_S$ increases new decay channels (as compared to the SM Higgs case) open up: $S\to ZZ$, $S\to W^+W^-$, $S\to t\bar{t}$.} From the above formula it is clear that if there is no mixing between the scalars the only available decay channel is the scalar one, $S\to HH$. It could be suspected that, since the mixing between the two scalars is small, the approximation of no mixing should hold. However, it turns out that the mixing enhances the decay width twofold, first, it amplifies the coupling $SHH$ as compared to $\f h h$ and, moreover, it allows a contribution from the SM sector, which is especially important when the $S\to HH$ decay is kinematically forbidden. %The $SHH$ coupling reads
%\be
%\frac{1}{4}\left(6 \cos\alpha\sin\alpha(\lambda_3 w\sin\alpha-\lambda_1 v\cos\alpha)+\lambda_2 w \cos\alpha(\cos^2\alpha-2\sin^2\alpha)+\lambda_2 v \sin\alpha(2\cos^2\alpha-\sin^2\alpha)\right). \label{eq:SHH-coupling}
%\ee
To compute $\Gamma_{\rm SM}$ we use the formulas for the SM Higgs decay rates following refs.~\cite{Djouadi:1995, Djouadi:2005, Djouadi:2005-2, Spira:1997}, summarised e.g.\ in the appendix of ref.~\cite{Swiezewska:2016} and cross-checked with \texttt{hdecay}~\cite{Djouadi:1997, Djouadi:2018xqq}\footnote{We thank J.\ Viana for his help with \texttt{hdecay}.}.

It should be noted that we compute the decay rate at the renormalisation scale $\mu=M_S$, which is the correct scale for considering decays of a particle of mass $M_S$, and can be widely separated both from the EW scale and the scale of $M_X$. Running between $\mu=M_X$ and $\mu=M_S$ modifies $\Gamma_{\f}$, this includes also running of the VEVs needed for the computation of the $SSH$ coupling. The running of the VEVs is significant and will be discussed in section~\ref{sec:scale-dep} (see figure~\ref{fig:scale-dep}).

We compute the energy transfer rate $\Gamma_{\f}$ and the Hubble parameter $H$ and compare them in order to determine whether reheating is instantaneous. This comparison is presented in figure~\ref{fig:reheating}, which shows the logarithm of the ratio of $\Gamma_{\f}$ to $H$ throughout the parameter space. The two parameters, $\Gamma_{\f}$ and $H$, only become equal for large values of the $X$ mass (along the thick black line), in the region where the percolation condition of eq.~\eqref{eq:percolation-crit} is violated (to the right of the black dashed line). This means that in the region where the phase transition proceeds due to nucleation and percolation of bubbles, which is the focus of the present paper, the reheating is always instantaneous and $T_r=T_V$ as stated in eq.~\eqref{eq:Treh=TV}. Apparently, the inclusion of the percolation criterion changes the picture of the phase transition, as compared to earlier works that studied the SU(2)cSM model in the context of the PT~\cite{Hambye:2018, Marfatia:2020, Baldes:2018} but did not investigate percolation (with the exception of ref.~\cite{Baldes:2018} but there it has been numerically evaluated only for a single point in the parameter space).

The numerical results for $T_V$ can be found in figure~\ref{fig:Tc-TV} (right panel). As explained in the next section, the region with inefficient reheating is essential for the supercool DM scenario, which thus will be eliminated at least in the region where the PT is sourced by the tunnelling, not by the QCD phase transition.

In the literature~\cite{Hambye:2018, Ellis:2020, Marfatia:2020} one can find various approaches to the computation of $\Gamma_{\f}$, which result in varying predictions for the reheating temperature. In appendix~\ref{app:reheating} we discuss different approximations and compare them to our approach.

\paragraph{Transition strength} The latent heat released during the transition consists of the difference in free energy (or effective potential) and additionally an entropy variation. In the limit of large supercooling, $T_p\ll T_c$, the entropy contribution can be neglected~\cite{Espinosa:2010, Marzola:2017}. For the gravitational wave signal the relevant quantity, called the transition strength, is the ratio of $\Delta V$ to the energy density of radiation at the time of the transition, $\rho_R(T_p)$,
\begin{align}
\alpha_* = \frac{\Delta V}{\rho_R(T_p)}.
\label{eq:alpha}
\end{align}
The strength of the PT in the studied model is very large, see figure~\ref{fig:alpha}. This is due to the large VEV of the new scalar field $\f$ and low percolation temperature.
\paragraph{The bubble-wall speed $v_w$} Since the PT in the studied model is extremely strong as can be appreciated in figure~\ref{fig:alpha} we can safely assume that the wall velocity is equal to the speed of light $v_w=1$.

\paragraph{Inverse time scale of the transition $\beta$} One of the parameters characterising the phase transition, important for the GW computations, is the (inverse) time scale of the transition. It is given by 
\be
\Gamma(t) \sim e^{-\beta t},
\ee
which translates to
\be
\frac{\beta}{H_*} = T_p \eval{ \frac{\dd(S_3/T)}{\dd T}}_{T=T_p}.\label{eq:beta}
\ee
Since $\beta/H_*$ is computed as the derivative, it can become numerically unstable. In our scan, we paid special attention to smoothing the numerical approximation of $S_3/T$ but still the results for $\beta/H_*$ computed based on eq.~\eqref{eq:beta} are less reliable than for the other parameters, see figure~\ref{fig:beta-RH} (left panel). Therefore, as an alternative, we will characterise the PT by its length scale, see below.

The plot showing the values of $\beta_*$, however, can help us understand the shape of the region excluded by the percolation criterion of eq.~\eqref{eq:percolation-crit} as it follows the shape of lines of constant $\beta_*$. $\beta_*$ is given as the derivative of the Euclidean action divided by the temperature. The larger $\beta_*$, the faster $S_3/T$ decreases with decreasing temperature, which means that the tunnelling rate increases quickly as the temperature decreases. This works in favour of percolation since a growing tunnelling rate can balance the expansion of the Universe due to thermal inflation. In contrast, if $\beta_*$ is smaller -- as it is near the region excluded by the percolation condition -- the decay rate of the false vacuum does not increase as fast with decreasing temperature. This, combined with thermal inflation leads to the phase transition not completing.

\paragraph{Length scale of the transition} The length scale of the transition is approximately given by the average bubble radius $R_*$ at the time of percolation. It is approximately given by the cubic root of the inverse of the bubble number density $n_B$ \cite{Turner:1992, Enqvist:1992}:
\be
R_* \equiv n_B^{-1/3} =  \left( T_p^3 \int^{T_c}_{T_p} \frac{\dd{T^\prime}}{T^{\prime 4}} \frac{\Gamma(T^\prime)}{H(T^\prime)} e^{- I(T^\prime)} \right)^{-1/3}.
\ee
There is a relation between the average radius $R_*$ and the inverse time scale $\beta$ given by $R_* \simeq \frac{(8\pi)^{1/3}}{\beta}$ which holds at $\tilde T_p$ such that $I(\tilde T_p)=1$~\cite{Enqvist:1992}.\footnote{In case of weaker transition one have to also take wall velocity $v_w$ into consideration.} Using the results presented in figure~\ref{fig:beta-RH} one can check that it does not hold exactly, this is partially due to the percolation condition used, see eq.~\eqref{eq:percolation-temp}.\footnote{In a recent article~\cite{Lewicki:2022pdb} it is pointed out that a more adequate relation would be $R_*=5/{\beta}$ and this is also in reasonable agreement with our numerical results. }

%XXXXXXXXXXXXXXXXXXXXXXXXXXXXXXXXXXXXXXXXXXXXXXXXX
\paragraph{Lorentz factor of the bubble wall\label{sec:lorentz_factor}}
%XXXXXXXXXXXXXXXXXXXXXXXXXXXXXXXXXXXXXXXXXXXXXXXXX

Another question crucial for GW production is whether the bubble wall reaches the terminal velocity before the collision or it keeps accelerating. Depending on that, the main source of GW are sound waves in the plasma or bubble wall collisions, respectively. To determine the behaviour of the bubble wall before the collision we have to examine the pressure exerted on it by the surrounding plasma.

The pressure difference across the wall can be expressed as
\be
\Delta P = \Delta V - P_{\mathrm{LO}} - P_{\mathrm{NLO}},
\ee
where $P_{\mathrm{LO}}$ is the leading-order (LO) pressure accounting for $1\rightarrow1$ scattering~\cite{Bodeker:2009}, while $P_{\mathrm{NLO}}$ is the next-to-leading-order (NLO) contribution associated with $1\rightarrow N$ splittings in the vicinity of the bubble wall~\cite{Bodeker:2017,Hoche:2020ysm,Gouttenoire:2021kjv}. The first quickly reaches a constant value
\be
P_{\mathrm{LO}} = \sum_a k_a c_a \frac{\Delta m_a^2 T_{p}^2}{24}, \quad c_a = \mbox{$1$ $(1/2)$ for bosons (fermions)},
\ee
where the sum runs over particle species, $k_a$ denotes the number of degrees of freedom of a given particle, $\Delta m_a^2$ is the change in mass across the wall. 

To compute the latter term, $P_{\mathrm{NLO}}$, it is necessary to sum contributions from $1\to N$ splitting processes, for arbitrary $N$. That is a difficult task and the outcome is a subject of debate in the literature. According to ref.~\cite{Hoche:2020ysm} the NLO friction can be expressed as
\be
P_{\mathrm{NLO}}^{(2)}  \sim \gamma^2 \sum_i k_i g_i   T^4_p,\label{eq:friction-Jessica}
\ee
where $g_i$ is the $i$-th boson's gauge coupling. On the other hand, ref.~\cite{Gouttenoire:2021kjv} reports the following expression for the NLO friction:
\be
P_{\mathrm{NLO}}^{(1)}  \sim \gamma \sum_i g_i  m_i T^3_p \log{\left( \frac{m_i}{\mu_{\mathrm{ref}}}\right)},\label{eq:friction-Yann}
\ee
where $m_i$ is the mass in the broken phase and $\mu_{\mathrm{ref}}$ is an IR cutoff proportional to $T_p$ (for a detailed discussion see~\cite{Gouttenoire:2021kjv}). The crucial difference between the two results lies in the power of the Lorentz $\gamma$ factor. The $\gamma^2$-scaling of eq.~\eqref{eq:friction-Jessica} suggests a faster damping of walls' velocity and thus a constrained possibility of GW production via bubble collisions. It is beyond the scope of the present work to determine which of the results quoted above is correct, instead, we will compare the results obtained using both of them.

To determine the behaviour of the wall at the moment of the collision we consider the evolution of $\gamma$ with the bubble radius $R$~\cite{Ellis:2020, Lewicki:2022pdb}. At the initial stage of expansion, when the $\gamma$ parameter is not large yet, but\footnote{We have checked that the initial radius is indeed much smaller than the final one for the whole parameter space.} $R_0\ll R$, the constant LO term in the friction dominates, leading to the Lorentz factor of the bubble wall increasing linearly $\gamma_{\textrm{run-away}} \simeq R/3R_0$~\cite{Gouttenoire:2021kjv}, where $R_0$ is the initial bubble radius. It is given by~\cite{Ellis:2019}
\be
R_0 \equiv\left[\frac{3 E_{0, V}}{4 \pi \Delta V(t=0)}\right]^{1/3},
\ee
where $\Delta V(t=0)$ is the energy difference between the center of the initial bubble and the outside, while $E_{0, V}$ is the potential energy contribution to the energy of the initial bubble. It is a good assumption to identify the initial size with the critical radius when the kinetic and potential energy are equal which gives $E_{0, V} \simeq S_3(\phi=\phi_{\textrm{bounce}})/2$. If bubbles collide in this first step of accelerated expansion, practically it resembles the run-away scenario.

If the wall expands further, the NLO terms become important and finally the wall reaches a stationary state where $\Delta P = 0$, which leads to a Lorentz factor of the form
\be
\gamma_{\textrm{eq}} \equiv \left( \frac{\Delta V - P_{\mathrm{LO}}}{P_{\mathrm{NLO}}^{(n)}} \right)^{\frac{1}{n}},
\ee
with $n$ being the power of $\gamma$ in the NLO pressure term. 

Finally, the $\gamma$ factor of the wall at the moment of the collision reads
\be
\gamma_* = \min{(\gamma_{\textrm{eq}}, \gamma_{\textrm{run-away}})}.
\ee

%XXXXXXXXXXXXXXXXXXXXXXXXXXXXXXXXXXXXXXXXXXXXXXXXX
\paragraph{Energy transfer\label{sec:energy_transfer}}
%XXXXXXXXXXXXXXXXXXXXXXXXXXXXXXXXXXXXXXXXXXXXXXXXX
Knowing the terminal Lorentz factor, it is possible to investigate the transfer of the released energy to the plasma. The efficiency factor $\kappa_{\text{col}}$ at the end of the transition is given as a ratio of the energy stored in the bubble wall to the total released energy. We use the form derived in~\cite{Lewicki:2022pdb}
\be
\kappa_{\textrm{col}} =
\left( 1-\frac{\alpha_{\infty}}{\alpha} \right)
\left( 1-\frac{1}{\gamma_{\text {eq }}^c}\right)
\frac{R_{\text {eq}}}{R_*} \frac{\gamma_*}{\gamma_{\text {eq}}},
\ee
where $\alpha_{\infty} = P_{\mathrm{LO}}/\rho_R$ \cite{Ellis:2020} and $R_{\text {eq}}=3R_0\gamma_{\text{eq}}$ \cite{Lewicki:2022pdb}. The energy that goes into sound waves in the plasma is parameterised by the efficiency factor $\kappa_{\text{sw}}$ \cite{Espinosa:2010, Hindmarsh:2013, Hindmarsh:2015}\footnote{Note that the recent results of refs.~\cite{Giese:2020rtr,Giese:2020znk} do not apply to strongly supercooled phase transitions with $\alpha_*>1$.}:
\be
\kappa_{\mathrm{sw}}=\frac{\alpha_{\mathrm{eff}}}{\alpha_*} \frac{\alpha_{\mathrm{eff}}}{0.73+0.083 \sqrt{\alpha_{\mathrm{eff}}}+\alpha_{\mathrm{eff}}}, \quad \text{ with } \quad \alpha_{\mathrm{eff}}=\alpha_*\left(1-\kappa_{\mathrm{col}}\right),
\ee
which for strong supercooled transitions is simply $\kappa_{\mathrm{sw}} \simeq (1-\kappa_{\text{col}})$. We show the resulting values of the efficiency factor for sound waves in figure~\ref{fig:kappa-sw} considering both forms of $P_{\mathrm{NLO}}$.  As expected, the model under consideration allows generation of GW signal both by the sound waves in the plasma and the bubble collisions (recall that $\kappa_{\mathrm{col}}=1-\kappa_{\mathrm{sw}}$). There is also a small region where both sources contribute significantly leading to ``mixed'' spectra. The difference between the two approaches to the friction force numerically is not large -- with the $\gamma^2$ scaling the onset of bubble collisions as the main source of GW is slightly shifted towards the region of stronger transitions, and consequently the available parameter space where GW from bubble collisions can be observed is reduced. In the following evaluation of GW spectra, we will use the $P_{\mathrm{NLO}} \sim \gamma$ scaling, remembering that with the $P_{\mathrm{NLO}} \sim \gamma^2$ the results would be very similar.
%

%XXXXXXXXXXXXXXXXXXXXXXXXXXXXXXXXXXXXXXXXXXXXXXXXX
\subsection{Gravitational waves\label{sec:GW}}
%XXXXXXXXXXXXXXXXXXXXXXXXXXXXXXXXXXXXXXXXXXXXXXXXX

Stochastic gravitational wave background produced by the first-order phase transition can be associated with three sources: bubble collisions, sound waves and turbulence in the plasma. The signal produced by turbulence remains a subject of ongoing discussion in the community, and its contribution is still burdened with large uncertainties. Therefore, we neglect it in this work.

%XXXXXXXXXXXXXXXXXXXXXXXXXXXXXXXXXXXXXXXXXXXXXXXXX
\paragraph{Bubble collisions\label{sec:collisions}}
%XXXXXXXXXXXXXXXXXXXXXXXXXXXXXXXXXXXXXXXXXXXXXXXXX
% As discussed in the previous section, the GW signal in the SU(2)cSM model can be sourced by bubble collisions. 
The spectrum generated during the transition at temperature $T_*=T_p$ can be evaluated using the following formulae resulting from simulations~\cite{Lewicki:2020azd}:
% \be %2.93 is the (8pi)^1/3
% \Omega_{\text{col}}(f) = \qty(\frac{R_*H_*}{2.93})^2 \qty( \frac{\kappa_{\text{col}} \alpha }{1+\alpha} )^2 S_{\textrm{col}}(f)\,. \label{eq:spectra-old}
% \ee
\be %2.93 is the (8pi)^1/3
\Omega_{\text{col},*}\qty(\frac{f}{f_{\textrm{col},*}}) = \qty(\frac{R_*H_*}{ \sqrt[3]{8\pi} })^2 \qty( \frac{\kappa_{\text{col}} \alpha }{1+\alpha} )^2 S_{\textrm{col}}\qty(\frac{f}{f_{\textrm{col},*}})\,, \label{eq:spectra-old}
\ee
where the spectral shape $S_{\textrm{col}}$ is defined using broken power-law:
%a = 2.34; b=2.41; c=4.2; A = 3.61/100 
%\be
%S_{\textrm{col}} = \bar{S} (a+b)^c  \qty[ b \qty(\frac{f}{ f_{\textrm{col}} })^{-\frac{a}{c}} + a \qty(\frac{f}{ f_{\textrm{col}} })^{\frac{b}{c}} ]^{-c}, 
%\ee
%with fitted values $a = 2.34$, $b=2.41$, $c=4.2$, $100\bar{S} = 3.61$ from the lattice simulations. \comment{or just like that in shorter way}
\be
S_{\textrm{col}}\qty(\frac{f}{f_{\textrm{col}}}) = 25.09  \qty[ 2.41 \qty(\frac{f}{ f_{\textrm{col}} })^{-0.56} + 2.34 \qty(\frac{f}{ f_{\textrm{col}} })^{0.57} ]^{-4.2}. 
\ee
The peak frequency $f_{\textrm{col}}$ of collisions spectrum, at $T_*$ is given by
% \be
% f_{\textrm{col}} \simeq 0.13 \qty(\frac{5}{R_*H_*}). %\qty(\frac{f_p}{\beta})
% \ee
\be 
%omega_bar/beta = 0.82, f_bar/beta = omega/2pi/beta = 0.13
f_{\textrm{col},*} \simeq 0.13 \beta \simeq 0.13 \qty(\frac{\sqrt[3]{8\pi}}{R_*}) .
\ee
In order to obtain present-day values, we have to redshift (assuming radiation domination) the amplitude and the peak frequency. One can do that by applying the following rescalings (as reviewed e.g.\ in \cite{Ellis:2020}):
\be
\Omega_{\text{col},0} \qty(\frac{f}{f_{\textrm{col},0}})
= \qty(\frac{a_r}{a_0})^4 \qty(\frac{H_r}{H_0})^2 \Omega_{\text{col},*}\qty(\frac{f}{f_{\textrm{col},0}})
= \frac{1.67 \cross 10^{-5}}{h^2} \qty[\frac{100}{g_*}]^\frac{1}{3} \Omega_{\text{col},*}\qty(\frac{f}{f_{\textrm{col},0}}),
\ee
\be
f_{\textrm{col},0} 
= h_* \qty(\frac{\beta}{H_*}) \qty(\frac{f_{\textrm{col},*}}{\beta})
\simeq h_* \qty(\frac{\sqrt[3]{8\pi}}{R_* H_*}) \qty(\frac{f_{\textrm{col},*}}{R_*}),
\ee
where $a_r $, $H_r$ are scale factor and Hubble parameter evaluated at $T_r$, then $h=0.674$ denotes the dimensionless Hubble parameter and 
\be
h_*  = 1.67 \cross 10^{-5} \textrm{ Hz} \qty(\frac{T_r}{100 \textrm{ GeV}}) \qty(\frac{g_*}{100})^\frac{1}{6} ,
\ee
is the inverse Hubble time at the transition redshifted to today.
%kierkla: starting from "...where [...]" is basically copied from marek, so rephrase it a little  

%XXXXXXXXXXXXXXXXXXXXXXXXXXXXXXXXXXXXXXXXXXXXXXXXX
\paragraph{Sound waves\label{sec:sound-waves}}
%XXXXXXXXXXXXXXXXXXXXXXXXXXXXXXXXXXXXXXXXXXXXXXXXX
The spectra of the sound-wave-sourced GW were obtained using lattice simulations and derived in a series of articles \cite{Hindmarsh:2013, Hindmarsh:2015, Hindmarsh:2017}. Using them we can write the spectrum redshifted to today, which has the following form:
% \be
% \Omega_\textrm{sw}(f) = \qty(\frac{R_*H_*}{5}) \qty( 1 - \frac{1}{\sqrt{1+2\tau_\textrm{sw}H_* }} ) 
% \qty( \frac{\kappa_{\text{sw}} \alpha }{1+\alpha} )^2 S_{\textrm{sw}}(f),
% \ee
\be
\Omega_{\textrm{sw},0}\qty(\frac{f}{f_{\textrm{sw},0}}) = \frac{4.011 \cross 10^{-7} }{h^2}\qty(R_*H_*) \qty( 1 - \frac{1}{\sqrt{1+2\tau_\textrm{sw}H_* }} ) 
\qty( \frac{\kappa_{\text{sw}} \alpha }{1+\alpha} )^2 S_{\textrm{sw}}\qty(\frac{f}{f_{\textrm{sw},0}}), \label{eq:spectrum-sw}
\ee
with the spectrum shape given as
\be
S_\textrm{sw}\qty(\frac{f}{f_{\textrm{sw},0}}) =\qty(\frac{f}{f_{\textrm{sw},0}})^3 \qty[\frac{4}{7}+\frac{3}{7}\qty(\frac{f}{f_{\textrm{sw},0}})^2]^{-7/2}\,,
\ee
and the peak frequency today  
% \be
% f_\textrm{sw} \simeq 0.54 \qty(\frac{5}{R_*H_*}).\label{eq:spectra-old-2}
% \ee
\be \label{eq:spectra-old-2}
f_{\textrm{sw},0} \simeq 2.62 \cross 10^{-5} \textrm{ Hz} \qty(\frac{1}{R_*H_*}) \qty(\frac{T_r}{100 \textrm{ GeV}}) \qty(\frac{g_*}{100})^\frac{1}{6} .
\ee
Equation~\eqref{eq:spectrum-sw} includes the duration of the sound wave period, which normalised to Hubble can be expressed as
\be
\tau_\textrm{sw} H_* = \frac{R_*H_*}{U_f}, \quad U_f \simeq \sqrt{\frac{3}{4}\frac{\alpha}{1+\alpha}\kappa_\textrm{sw}}.
\ee

At the final stage of completion of the present work a new article appeared~\cite{Lewicki:2022pdb} in which the authors attempt to numerically model the GW production in strongly supercooled transitions. The energy transferred into fluid is assumed to be localised in thin shells following behind the bubble wall before the collision and propagating in the same direction after the collision.  Under these assumptions, it is shown there that the spectrum sourced by the sound waves in the plasma depends on the behaviour of the fluid shell following the bubble after the collision. Its velocity can either quickly relax to the sound velocity and then we should see the typical sound-wave spectra; or, if the transition is very strong, it can continue propagating at the speed of light, and then the sound-wave spectra are the same as the spectra from bubble collisions. In the latter case, one does not see a transition in the shape of the spectra as one passes from $\kappa_{\mathrm{sw}}=1$ to $\kappa_{\mathrm{col}}=1$. Unfortunately, there is no simple criterion to determine to which category a given point in the parameter space belongs -- a simulation would be needed to this end. Since the well-established formulae for the GW spectra that we use in this work, eqs.~\eqref{eq:spectra-old}--\eqref{eq:spectra-old-2}, are based on simulations performed for rather weak transitions, in studying strong transitions we need to extrapolate their results to the region where we cannot ascertain their validity. Therefore, in the absence of full PT simulations of strong transitions, we find it interesting to study the implications of the numerical solutions of ref.~\cite{Lewicki:2022pdb}, intended for such cases. We devote appendix~\ref{app:new-spactra} to the discussion of the spectra computed according to ref.~\cite{Lewicki:2022pdb}.

\begin{figure}[h!t]
\center
\includegraphics[width=.48\textwidth]{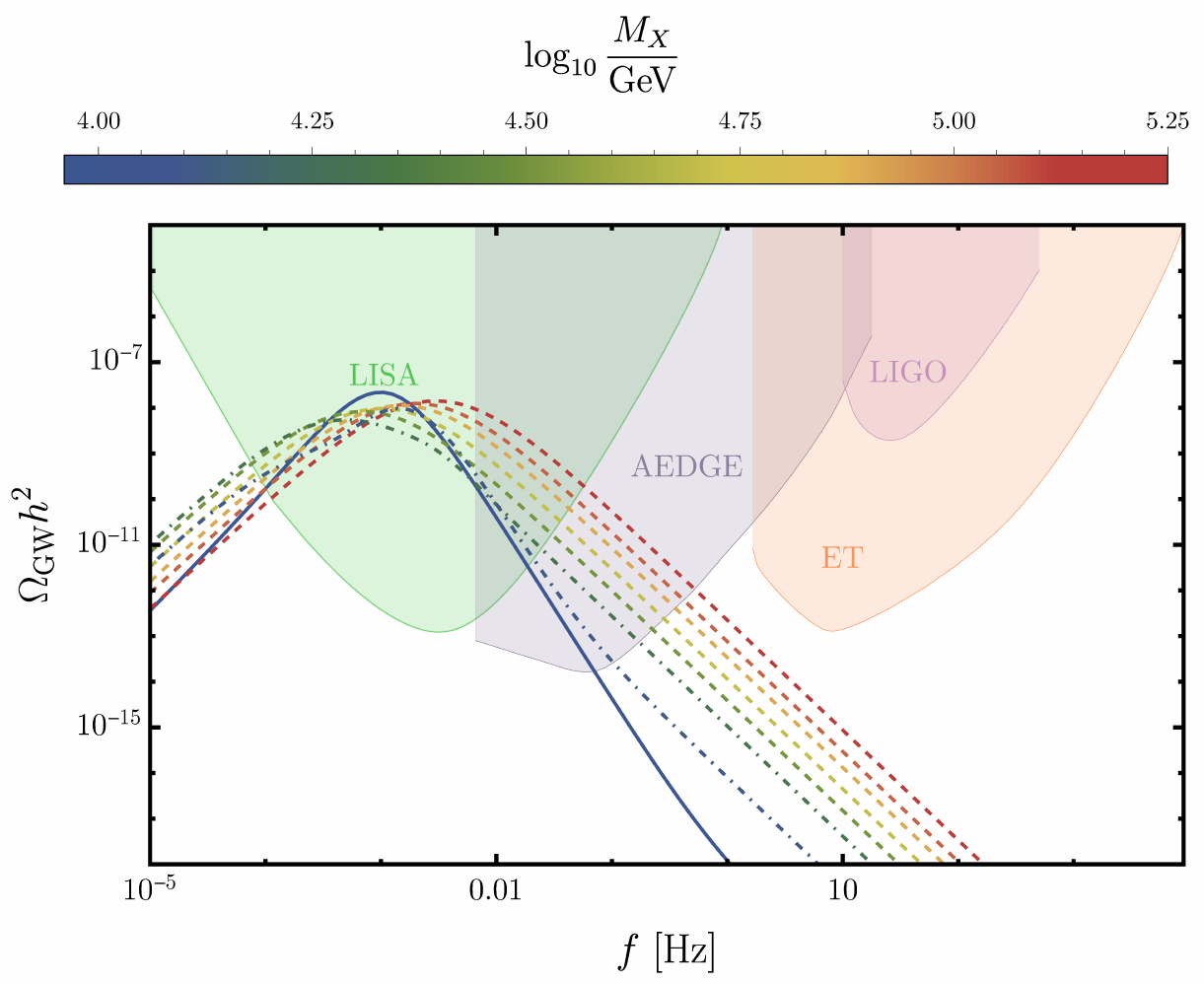}\hspace{2pt}
\includegraphics[width=.48\textwidth]{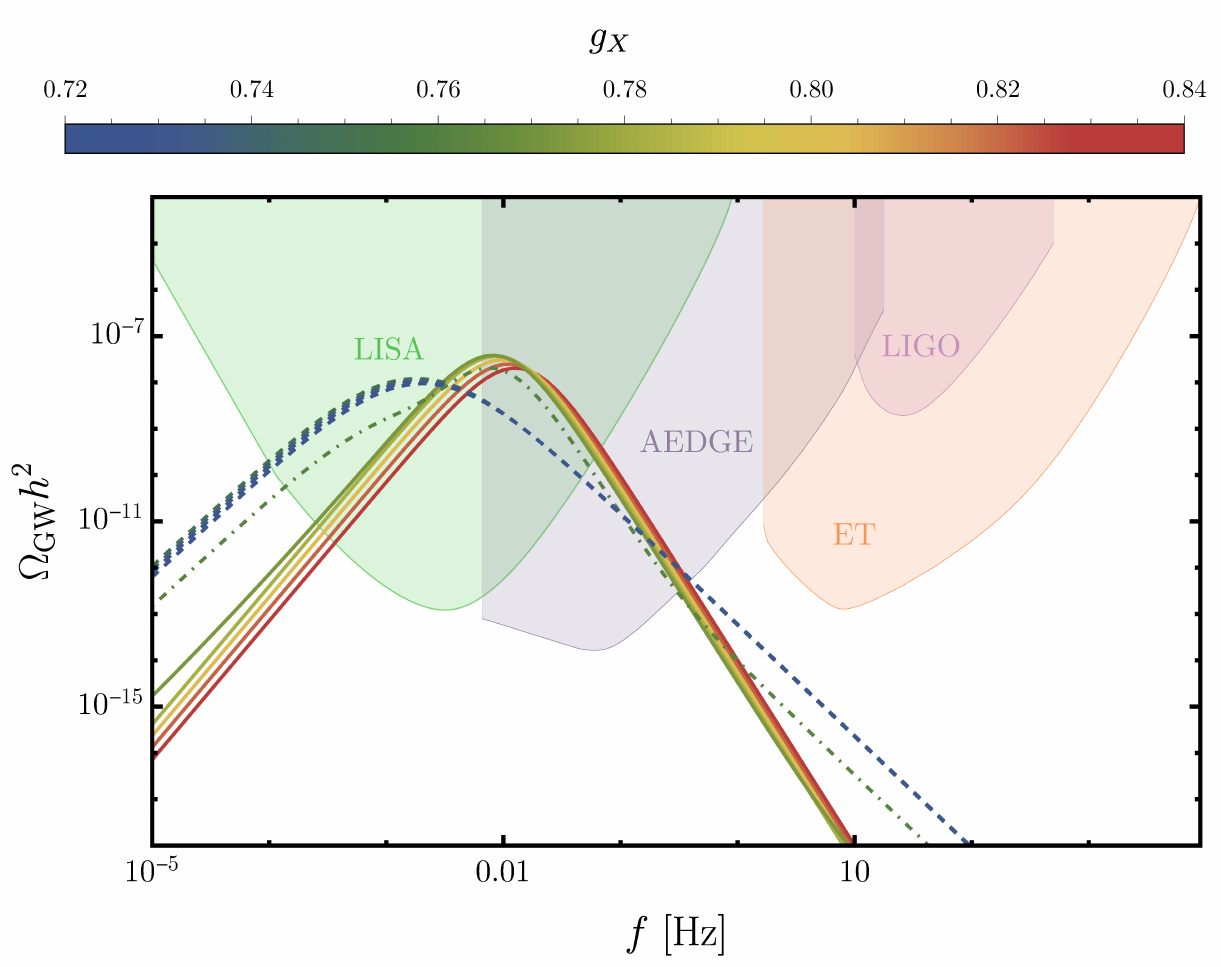}
\caption{Predictions for spectra of gravitational waves together with integrated sensitivity curves for LISA, AEDGE, ET and LIGO. Left: for a fixed value of $g_X = 0.76$ and varying value of $M_X$ (colour-coded). Right: for a fixed value of $M_X = 100$\,TeV and varying value of $g_X$ (colour-coded).  \label{fig:spectra-gx-MX-caprini}}
\end{figure}

We now present sample spectra of GW generated during a phase transition within the SU(2)cSM model. Left panel of figure~\ref{fig:spectra-gx-MX-caprini} shows the GW spectra for a sample of points in the parameter space with $g_X=0.76$ and the $X$ mass varying (see the colour coding). The right panel presents the spectra computed for fixed $M_X=100$\,TeV and varying $g_X$. First and foremost, it should be noted that the predicted signal is strong and well within the reach of LISA.  Second, looking at figure~\ref{fig:kappa-sw} we can see that as $M_X$ or $g_X$ changes, we should pass from sound-wave sourced GW to bubble collisions being the dominant source. Indeed, the spectra change their shape as $M_X$ or $g_X$ are changed -- solid lines denote spectra sourced by sound waves, dashed by bubble collisions and the dashed-dotted indicate a mixed case, where both production mechanisms contribute. It has the characteristic feature of a flatter central part, following from superimposing the two peaks. 

To better evaluate the observability potential of the signal generated within SU(2)cSM we will compute the signal-to-noise ratio for the predicted spectra. We relegate this to section~\ref{sec:results} to combine the information about observability of GW with the predictions for DM relic abundance and the constraints from DM direct detection.

%XXXXXXXXXXXXXXXXXXXXXXXXXXXXXXXXXXXXXXXXXXXXXXXXX
\section{Dark Matter\label{sec:DM}}
%XXXXXXXXXXXXXXXXXXXXXXXXXXXXXXXXXXXXXXXXXXXXXXXXX
Our DM candidates are the three vector bosons $X_\mu^a$ (where $a = 1,\,2,\,3$) of the hidden sector gauge group $SU(2)$. As discussed in~\cite{Gross:2015}, the gauge bosons are stable due to an intrinsic $\mathbb{Z}_2 \cross \mathbb{Z}_2'$ symmetry associated with complex conjugation of the group elements and discrete gauge transformations. This discrete symmetry actually generalizes to a custodial $SO(3)$~\cite{Hambye:2008} and the dark gauge bosons are degenerate in mass. Vector DM from various gauge groups has been studied e.g.\ in~\cite{Hambye:2013, Carone:2013, Khoze:2014, Pelaggi:2014wba, Karam:2015, Plascencia:2016, Karam:2016, Arcadi:2016kmk, Arcadi:2016qoz, Heikinheimo:2017ofk, Choi:2017zww, Duch:2017khv, Heikinheimo:2018duk, Hambye:2018, Baldes:2018, YaserAyazi:2019caf, Mohamadnejad:2019vzg, Baouche:2021wwa, Mohamadnejad:2021tke}.

Before the first-order phase transition, the dark $SU(2)_X$ gauge symmetry is unbroken and thus the DM particles are massless. After the phase transition to the true vacuum, the DM particles (as well as the other particles of the model) become massive and the energy stored in $\Delta V$ reheats the Universe to the temperature $T_r$. As shown in figure~\ref{fig:reheating}, for all the parameter space under consideration we have $\Gamma_\varphi > H_*$, therefore $T_r = T_V$. Furthermore, the thermal inflation temperature can be approximately related to the DM mass as $T_V \simeq M_X / 8.5$~\cite{Hambye:2018}, which implies that $T_r \simeq M_X / 8.5$. On the other hand, the usual freezeout or decoupling temperature is $T_{\rm dec} \simeq M_X/25$. If $T_r < T_{\rm dec}$, then the DM abundance receives in principle two contributions: supercool and sub-thermal production via scattering, as discussed in~\cite{Hambye:2018, Baldes:2018, Marfatia:2020}. The supercool DM population is what remains after the late time inflation. The gauge bosons are originally massless and their abundance gets suppressed by the dilution due to thermal inflation. In addition, after reheating, there is a sub-thermal population which can be produced through the thermal bath via scattering effects.\footnote{Note that during supercooling any preexisting baryon asymmetry gets washed out due to the period of thermal inflation. We thus need some mechanism to regenerate the baryon asymmetry after supercooling, around the electroweak scale. This can be achieved through leptogenesis. A minimal way to achieve leptogenesis would require right-handed neutrinos (RHN) $N$ that couple to the SM neutrinos through a term of the form $Y_N N L H$, plus an extra real scalar singlet $S'$ with Yukawa coupling $Y_S S' N^2 / 2$ that gets a VEV and dynamically generates the mass of the RHN as in e.g.~\cite{Karam:2015}. Leptogenesis in this model has been studied in~\cite{Plascencia:2016}, but a full analysis of this model goes beyond the scope of our paper. See also~\cite{Hambye:2018} for a more detailed discussion of leptogenesis in supercool scenarios.}

Nevertheless, since we find $T_r > T_{\rm dec}$ for all parameter points of interest, the supercool DM population gets diluted away, the sub-thermal population reaches thermal equilibrium again, and thus the relic abundance is produced as in the standard freezeout scenario.\footnote{The relation between $T_r$ and $T_{\mathrm{dec}}$ is relaxed when the gauge interactions are nonperturbative and in these scenarios supercool DM can be present, see e.g.~\cite{Baldes:2020kam, Baldes:2021aph}. See Ref.~\cite{Bernal:2015} for other production mechanisms in this model.} This discrepancy between our results and those of refs.~\cite{Hambye:2018, Baldes:2018, Marfatia:2020} is attributed to the inclusion of the percolation criterion in our work and to differences in computing the energy transfer rate $\Gamma_{\varphi}$ (see appendix~\ref{app:reheating} for a detailed discussion), which e.g.\ in the case of ref.~\cite{Marfatia:2020} led to the underestimation of $\Gamma_{\f}$.
%the fact that the authors of those papers considered simplified expressions for the decay rate $\Gamma_\varphi$, which resulted in $\Gamma_\varphi < H_*$ for a large part of their parameter space, leading to supercool DM. In our case, indeed $\Gamma_\varphi < H_*$ occurs but only when $M_X \gtrsim 2 \times 10^6$, and DM masses above this limit are excluded from the percolation criterion (eq.~\eqref{eq:percolation-crit}), which was not included in refs.~\cite{Hambye:2018, Baldes:2018, Marfatia:2020}. 

%%%%%%%%%%%%%%%%%%%%%%%%%%%%%%%%%%
\subsection{DM Abundance}
\begin{figure}[H]
	\centering
	\includegraphics[height=0.17\textwidth]{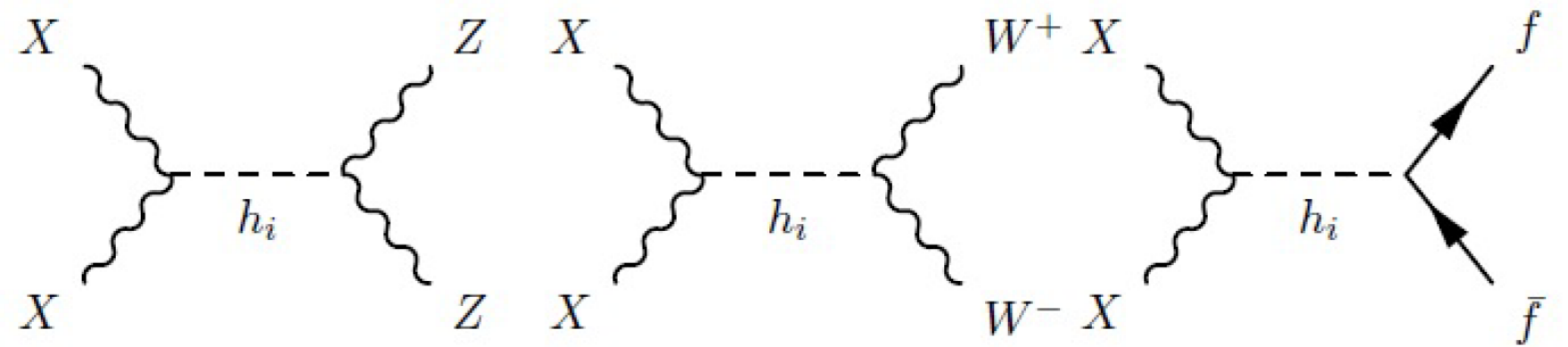}
	\caption{Feynman diagrams for DM annihilation to SM gauge bosons and fermions, with $h_i = \left\lbrace H , S \right\rbrace$.}
	\label{fig:DMann1}
%\end{flushleft}
\end{figure}

\begin{figure}[H]
	\centering
	\includegraphics[height=0.17\textwidth]{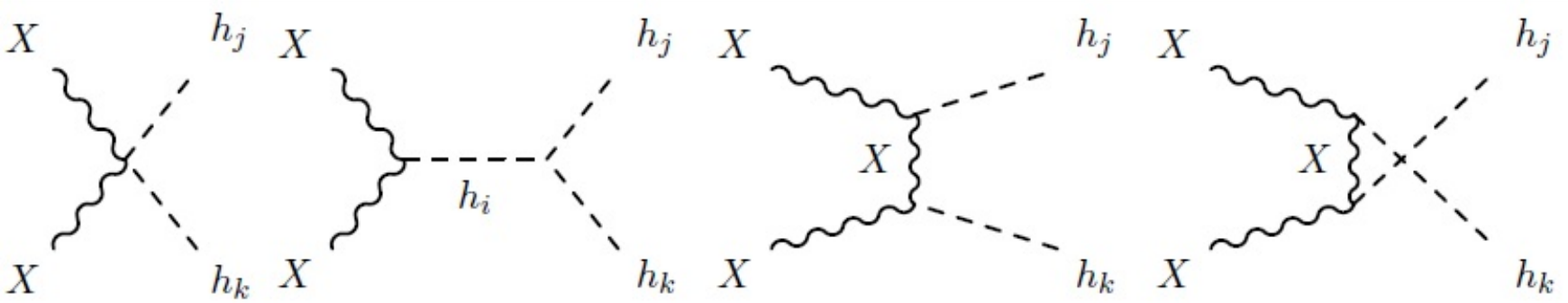}
	\caption{Feynman diagrams for DM annihilation to scalars, with $\left\lbrace h_j , h_k \right\rbrace = \left\lbrace H , S \right\rbrace$.}
	\label{fig:DMann2}
\end{figure}

\begin{figure}[H]
	\centering
	\includegraphics[height=0.17\textwidth]{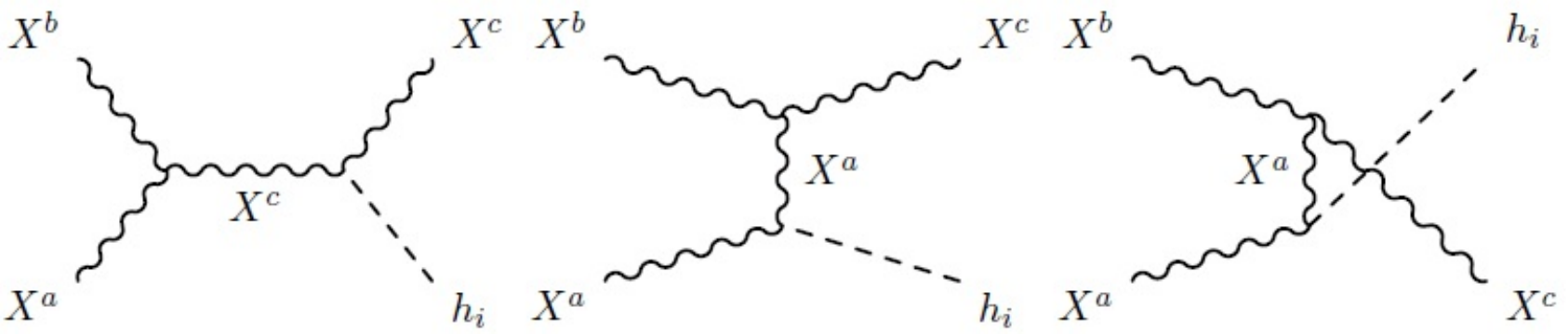}
	\caption{Feynman diagrams for DM semi-annihilation, with $h_i = \left\lbrace H , S \right\rbrace$.}
	\label{fig:DMsemi}
\end{figure}

%%%%%%%%%%%%%%%%%%%%%%%%%%%%%%%%%%
In order to compute the dark matter relic abundance we have to solve the Boltzmann equation which describes the evolution of the number density $n$ of a given particle species with time. An interesting feature of the gauge nature of the dark vector bosons $X^a$ is that they can both annihilate and semi-annihilate~\cite{DEramo:2010keq, Belanger:2012vp}. We present the Feynman diagrams for the relevant processes in figures~\ref{fig:DMann1}-\ref{fig:DMsemi}.

%%%%%%%%%%%%%%%%%%%%%%%%
\vspace{0.1cm}
%%%%%%%%%%%%%%%%%%%%%%%%%%
%\vspace*{0.4 cm}
%%%%%%%%%%%%%%%%%%%%%%%%%%
\vspace{0.1cm}
%%%%%%%%%%%%%%%%%%%%%%%%%%
%\vspace*{0.4 cm}
%%%%%%%%%%%%%%%%%%%%%%%%%%
%%%%%%%%%%%%%%%%%%%%%%%

The Boltzmann equation has the following form~\cite{Khoze:2014}:
%%%%%%%%%%%%%%%%%%%%%%%
\be 
\frac{\dd n_X}{\dd t} + 3\,H\,n_X = - \, \frac{\left< \sigma v \right>_{\rm ann}}{3}  \left( n_X^2 - n^2_{X,eq} \right) - \frac{2\left< \sigma v \right>_{\rm semi}}{3} \, n_X \left( n_X - n_{X,eq} \right), 
\label{eq:BoltzmannEquation}
\ee
%%%%%%%%%%%%%%%%%%%%%%%
where $\left< \sigma v \right>$ is the thermally averaged cross section of the dark gauge bosons times their relative velocity, and the subscripts $\rm ann$ and $\rm semi$ denote annihilation and semi-annihilation respectively. In the nonrelativistic approximation, the thermally averaged cross section times velocity is given by~\cite{Srednicki:1988ce}
%%%%%%%%%%%%%%%%%%%%%%%
\be 
\left< \sigma v \right> \simeq \frac{1}{M^2_X} \left[ W(s) - \frac{3}{2x} \left( 2 W(s) - W'(s) \right)  \right] \Bigg\vert_{s = 4 M^2_X} \,.
\ee
%%%%%%%%%%%%%%%%%%%%%%%
The quantity $W(s)$ is defined as
%%%%%%%%%%%%%%%%%%%%%%%
\be 
W(s) = \frac{1}{4} \left( 1 - \frac{\delta_{j k}}{2} \right) \beta\left( s, m_j, m_k \right) \int \frac{\dd( \cos\omega)}{2}\,\sum\, \lvert \mathcal{M} \left( X\,X \rightarrow \text{all} \right) \rvert^2,
\ee
%%%%%%%%%%%%%%%%%%%%%%%
where $\sum\lvert \mathcal{M} \rvert^2$ denotes the matrix element squared of all possible annihilation and semi-annihilation channels, averaging over initial polarizations and summing over final spins, $\beta\left( s, m_j, m_k \right) = \frac{1}{8\pi} [ 1 - ( m_j + m_k )^2/{s}  ]^{1/2} [ 1 - ( m_j - m_k)^2/{s} ]^{1/2}$ is the final-state Lorentz invariant phase space (with $m_j$ and $m_k$ denoting the masses of the particles in the final state, i.e.~the Standard Model particles and scalars for the annihilation processes, and $X H$ and $X S$ for the semi-annihilation processes), and $s$ stands for the Mandelstam variable $ s = ( p_1 + p_2 )^2 = 2 \left( M^2_X + E_1 E_2 - p_1 p_2 \cos\omega \right)$. Moreover, the prime denotes differentiation with respect to $s/(4M^2_X)$ and $x$ is defined as $x \equiv M_X/T$. In our numerical analysis, we employ the full analytic expressions for the thermally averaged annihilation and semi-annihilation cross sections which can be found in~\cite{Pelaggi:2014wba, Baouche:2021wwa}. 

The annihilation cross section is dominated by the $X X \rightarrow S S$ process. The leading order term is 
%%%%%%%%%%%%%%%%%%%%%%%%%%%
\be 
\left< \sigma v \right>_{\rm ann} \approx \frac{11 g_X^4}{2304 \pi M^2_X} \,.
\ee
%%%%%%%%%%%%%%%%%%%%%%%%%%%
Similarly, the semi-annihilation cross section is dominated by the $X X \rightarrow X S$ process. The leading order term is
%%%%%%%%%%%%%%%%%%%%%%%%%%%
\be 
\left< \sigma v \right>_{\rm semi} \approx \frac{3 g_X^4}{128 \pi M_X^2} \,.
\ee
%%%%%%%%%%%%%%%%%%%%%%%%%%%
One can see that the semi-annihilation processes dominate since $\left< \sigma v \right>_{\rm semi} \sim 5 \left< \sigma v \right>_{\rm ann}$.

Solving the Boltzmann equation, we obtain the dark matter relic abundance given by
%%%%%%%%%%%%%%%%%%%%%%%%%%%%
\be 
% \Omega_{X} h^2 = 3\times 
\Omega_{X} h^2 = \frac{1.04\times 10^9 \, \GeV^{-1}}{\sqrt{g_*} \, M_{\rm Pl} \, J(x_f)}, \qquad J(x_f) = \int_{x_f}^\infty dx \, \frac{\left< \sigma v \right>_{\rm ann} + 2 \left< \sigma v \right>_{\rm semi}}{x^2}, \label{eq:RelicDensity}
\ee
%%%%%%%%%%%%%%%%%%%%%%%%%%%%
where $M_{\rm Pl} = 1.22 \cdot 10^{19} \ \rm GeV$ and the value of the freeze-out point $x_f = M_X / T_{\rm dec}$ can be obtained iteratively~\cite{Kolb:1990vq}:
%%%%%%%%%%%%%%%%%%%%%%%
\be
x_f = \ln \left\{ 0.038 \, \frac{3 M_X M_{\rm Pl}}{\sqrt{g_*(T_{\rm dec}) x_f}} \Big[ \, c \, (c + 2) \left< \sigma v \right>_{\rm ann} + 2 \, c \, ( c + 1 ) \left< \sigma v \right>_{\rm semi} \, \Big] \right\} . \label{eq:fzpoint}
\ee
%%%%%%%%%%%%%%%%%%%%%%%
For the constant $c$ we use $c=1/2$~\cite{Griest:1990kh} and we find typical values between $x_f \approx 25-26$ for the DM mass range that we consider in our numerical analysis. Finally, for $M_X \gg M_S$ and small mixing between the scalars, the correct relic abundance ($\Omega_{\rm DM} h^2 = 0.120 \pm 0.001$~\cite{Planck:2018vyg}) is reproduced if 
%%%%%%%%%%%%%%%%%%%%%%%%%%%%%%%%%%%%%
\be 
g_X \approx 0.9 \times \sqrt{\frac{M_X}{1 \ \rm TeV}} \quad {\rm and} \quad w \approx 2.2 \ \rm TeV \times \sqrt{\frac{M_X}{1 \ \rm TeV}}\,. 
\ee
%%%%%%%%%%%%%%%%%%%%%%%%%%%%%%%%%%%%%
In this limit, the relic abundance fixes the gauge coupling $g_X$ and the parameter space becomes one-dimensional, with the only free parameter being $M_X$.

\begin{figure}
    %\centering
    \center
\includegraphics[width=.8\textwidth]{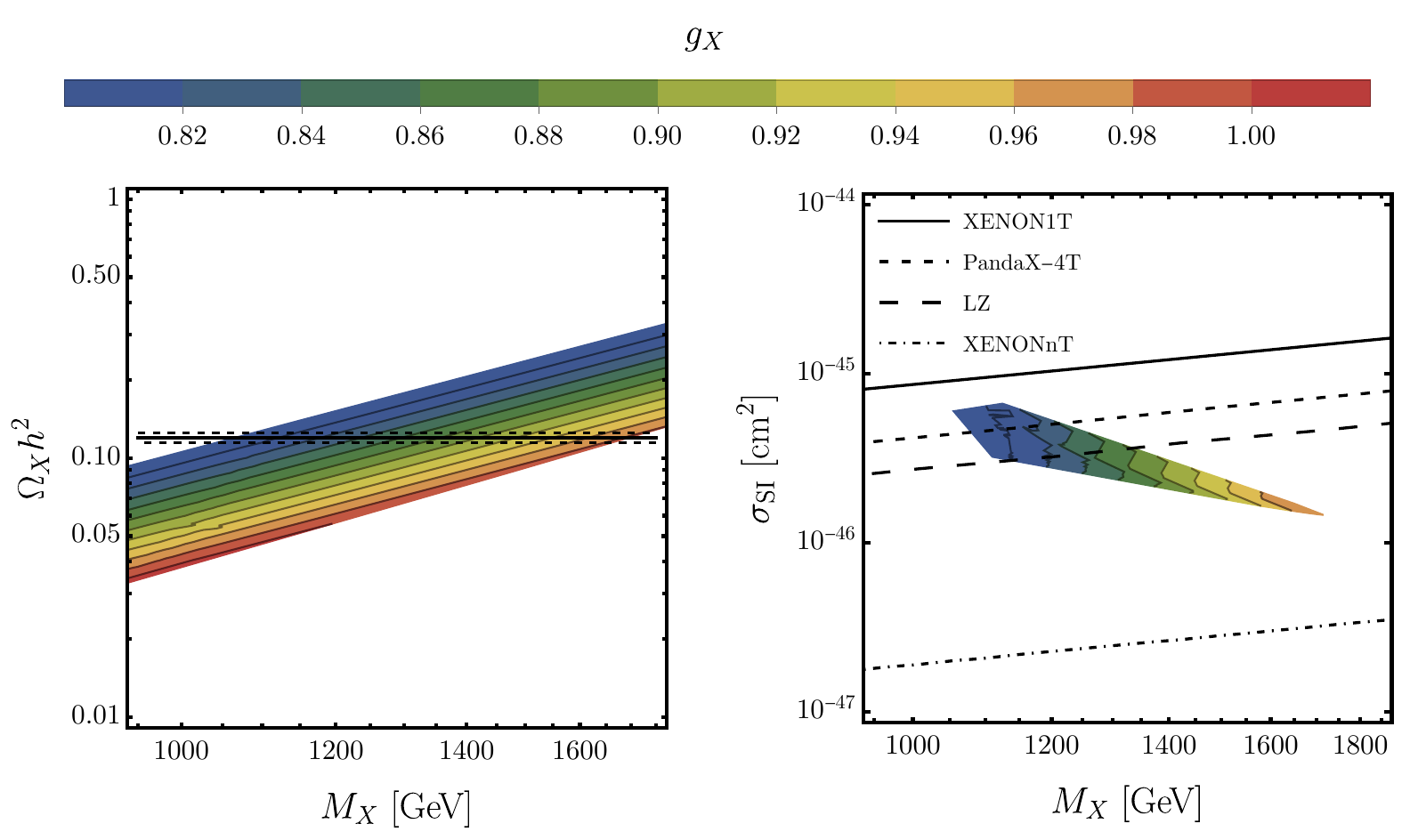}
    \caption{Left: Dark matter relic abundance $\Omega_X h^2$ with colour changing according to the value of the gauge coupling $g_X$. The black lines correspond to the measured value $\Omega_{\rm DM} h^2 = 0.120 \pm 5 \sigma$. Right: The spin-independent dark matter-nucleon cross section. The coloured region corresponds to points that reproduce the measured relic abundance within $5 \sigma$. The lines represent the exclusion limit from the XENON1T 2018~\cite{XENON:2018voc} (solid), PandaX-4T 2021~\cite{PandaX-4T:2021bab} (dashed), LZ 2022~\cite{LZ:2022ufs} (large dashed) and the scheduled XENONnT~\cite{XENON:2020kmp} (dot dashed) experiments.}
    \label{fig:relicplot}
\end{figure}

In the left panel of figure~\ref{fig:relicplot} we show the value of the relic density in terms of DM mass $M_X$ (in logarithmic scales), with the contours indicating values of the gauge coupling $g_X$ in the range found in section~\ref{sec:PT-and-GW}. Furthermore, the black line shows the measured DM abundance value $\Omega_{\rm DM} h^2 = 0.12$. We see that the measured value is reproduced if $M_X \in \left[ 1050, \ 1700 \right] \ \rm GeV$ and $g_X \in \left[ 0.75, \ 1 \right]$.
%%%%%%%%%%%%%%%%%%%%%%%%%%%%%%%%%%%%%
%%%%%%%%%%%%%%%%%%%%%%%%%%%%%%%%%%%%
\subsection{DM direct detection}

In this model, the DM candidate $X$ can in principle scatter off nucleons through the $t$-channel exchange of scalar bosons $h_i = \left\{ H , \, S \right\}$. %The relevant Feynman diagram is presented in Fig.~\ref{fig:DD_Feyn}.
%%%%%%%%%%%%%%%%%%%%%%%%%%%%%%%%%%%%%
%\begin{figure}
%    \centering
%    \includegraphics[width=0.33\textwidth]{plots/DD_Feyn.pdf}
%    \caption{Feynman diagram...}
%    \label{fig:DD_Feyn}
%\end{figure}
%%%%%%%%%%%%%%%%%%%%%%%%%%%%%%%%%%%%
The spin-independent (SI) cross section has the form 
%%%%%%%%%%%%%%%%%%%%%%%%%%%%%%%%%%%%
\be 
\sigma_{\rm SI} = \frac{g^4_X f^2 m^4_N w^2}{64 \pi \left( m_N + M_X \right)^2 v^2} \left( \frac{1}{M^2_H} - \frac{1}{M^2_S} \right)^2 \,,
\label{eq:DD_SI}
\ee
%%%%%%%%%%%%%%%%%%%%%%%%%%%%%%%%%%%%
where $m_N = 0.939 \ \rm GeV$ is the nucleon mass and $f = 0.3$ is the nucleon form factor. The SI cross section is shown in the right panel of figure~\ref{fig:relicplot} for the values of $M_X$ and $g_X$ that also reproduce the measured relic density within $5 \sigma$. One can see that values of $M_X \lesssim 1150 \ \rm GeV$ and $g_X \lesssim 0.82$ are now excluded due to the latest constraints from the LZ experiment~\cite{LZ:2022ufs}. Nevertheless, for $M_X \gtrsim 1150 \ \rm GeV$ and $g_X \gtrsim 0.82$ there is still some parameter space that is allowed and could be probed by the upcoming XENONnT experiment~\cite{XENON:2020kmp}.
%%%%%%%%%%%%%%%%%%%%%%%%%%%%%%%%%%%%%

To sum up, due to the percolation criterion (eq.~\eqref{eq:percolation-crit}), which constrains the parameter space of the first-order PT of interest to $M_X$ below approximately $10^6\g$, %excludes $M_X$ masses above $\sim 10^6 \ \rm \GeV$ 
and to the fact that $\Gamma_\varphi > H(T_p)$ in the considered %rest of the 
DM mass range, we find $T_r > T_{\rm dec}$ for all parameter points. This means that the supercool DM population gets diluted away, the sub-thermal population reaches thermal equilibrium again, and consequently, the relic abundance is produced as in the standard freezeout scenario. The measured DM relic abundance value and the direct detection experiments significantly constrain the parameter space but, nevertheless, we find a small region that is allowed by direct detection experiments and could be tested in the near future.
%XXXXXXXXXXXXXXXXXXXXXXXXXXXXXXXXXXXXXXXXXXXXXXXXX
\section{Combined results for gravitational waves and dark matter\label{sec:results}}
%XXXXXXXXXXXXXXXXXXXXXXXXXXXXXXXXXXXXXXXXXXXXXXXXX

In this section, we analyse the observability of the gravitational-wave signals produced during the phase transition and combine these results with the DM data to see whether our DM candidate could be tested via GW (see also~\cite{Baldes:2018, Marfatia:2020}).

To assess the observability of a signal we compute the signal-to-noise (SNR) ratio for the detectors that have the best potential of observing the predicted signal, i.e.\ LISA and AEDGE. We calculate the SNR using the usual formula \cite{Caprini_2020, Robson_2019}:

\begin{align} 
\mbox{SNR} = \sqrt{
\mathcal{T} 
\int_{f_{\rm{min}}}^{f_{\rm{max}}}
\dd f \left[ \frac{h^2\Omega_{\rm{GW}}(f)}{h^2\Omega_{\rm{Sens}}(f)} \right]^2
},
\end{align}
where $\mathcal{T}$ is the duration of collecting data and $h^2\Omega_{\rm{Sens}}(f)$ is the sensitivity curve of a given detector.
For calculations we have used data collecting durations as $\mathcal{T}_{\rm{LISA}}$ = 75 \% $\cdot$ 4 years \cite{Caprini_2020}  and $\mathcal{T}_{\rm{AEDGE}}$ = 3 years \cite{AEDGE:2019nxb}.
We will assume that a signal could be observed if $\mathrm{SNR}>10$, which is the usual criterion. 

\begin{figure}[h!t]
\center
\includegraphics[width=.4\textwidth]{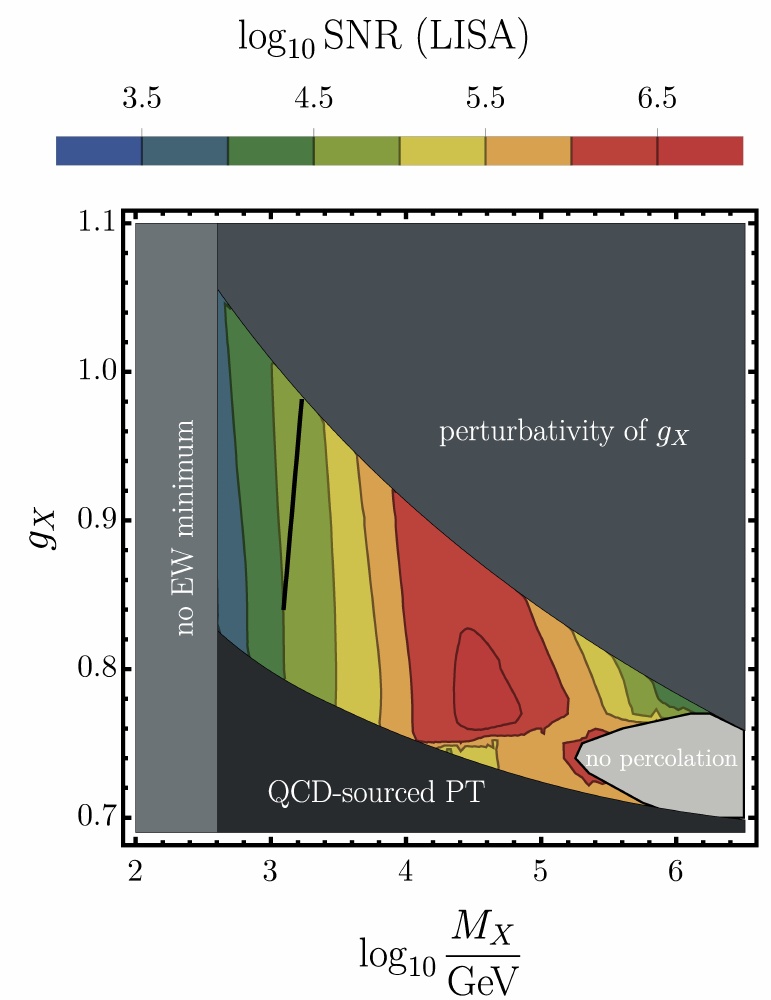}\hspace{20pt}
\includegraphics[width=.4\textwidth]{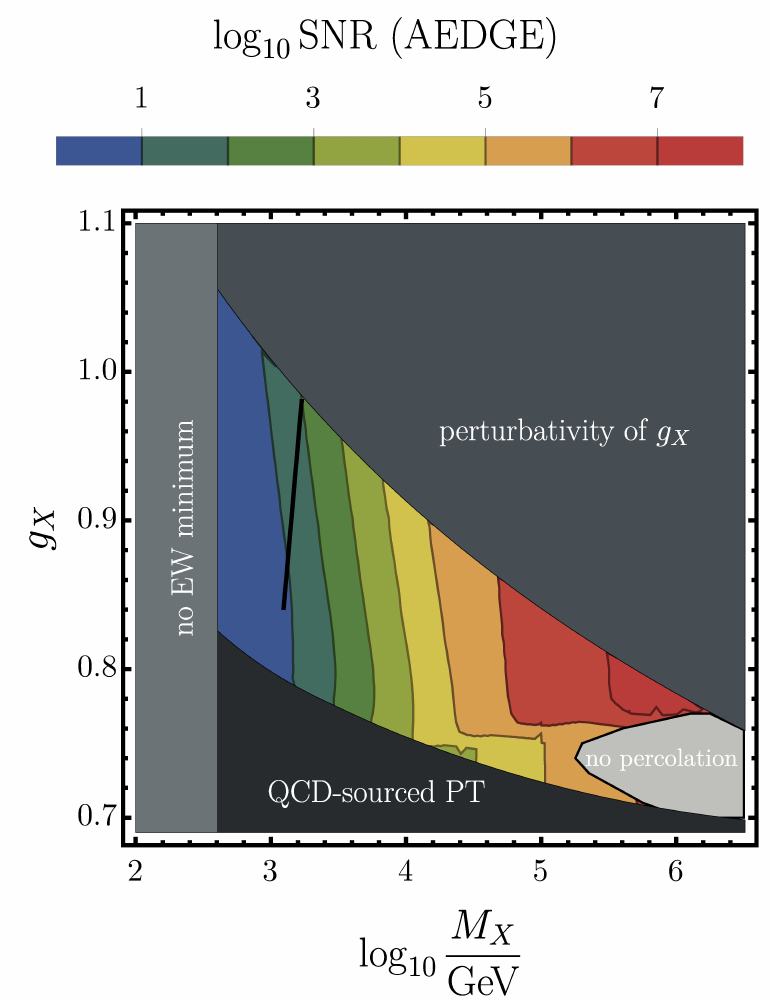}
\caption{Results for the signal-to-noise ratio for LISA (left panel) and AEDGE (right panel) for the predicted GW signal. The black line corresponds to the points that reproduce the measured DM relic abundance and also evade the DM direct detection experimental constraints.\label{fig:SNR}}
\end{figure}

For LIGO and ET we do not find observational prospects (the SNR is small within the whole allowed parameter space), therefore we do not discuss them further. The results are presented in figure~\ref{fig:SNR}. Superimposed is a curve indicating where in the parameter space the correct DM relic density is reproduced and the DM direct detection constraints are satisfied (solid black). Strikingly, the SNR for LISA for the predicted signal is above the observability threshold within the whole parameter space, and almost whole in the case of AEDGE \footnote{For AEDGE we made a conservative assumption cutting the sensitivity at the frequency of 0.01\,Hz.}.
This means that a first-order phase transition sourced by tunnelling of a scalar field in the SU(2)cSM model should be thoroughly testable by LISA and AEDGE. Moreover, in case of not observing a signal consistent with the expectations for the first-order phase transitions this scenario could be falsified.

\begin{figure}[h!t]
\center
\includegraphics[width=.75\textwidth]{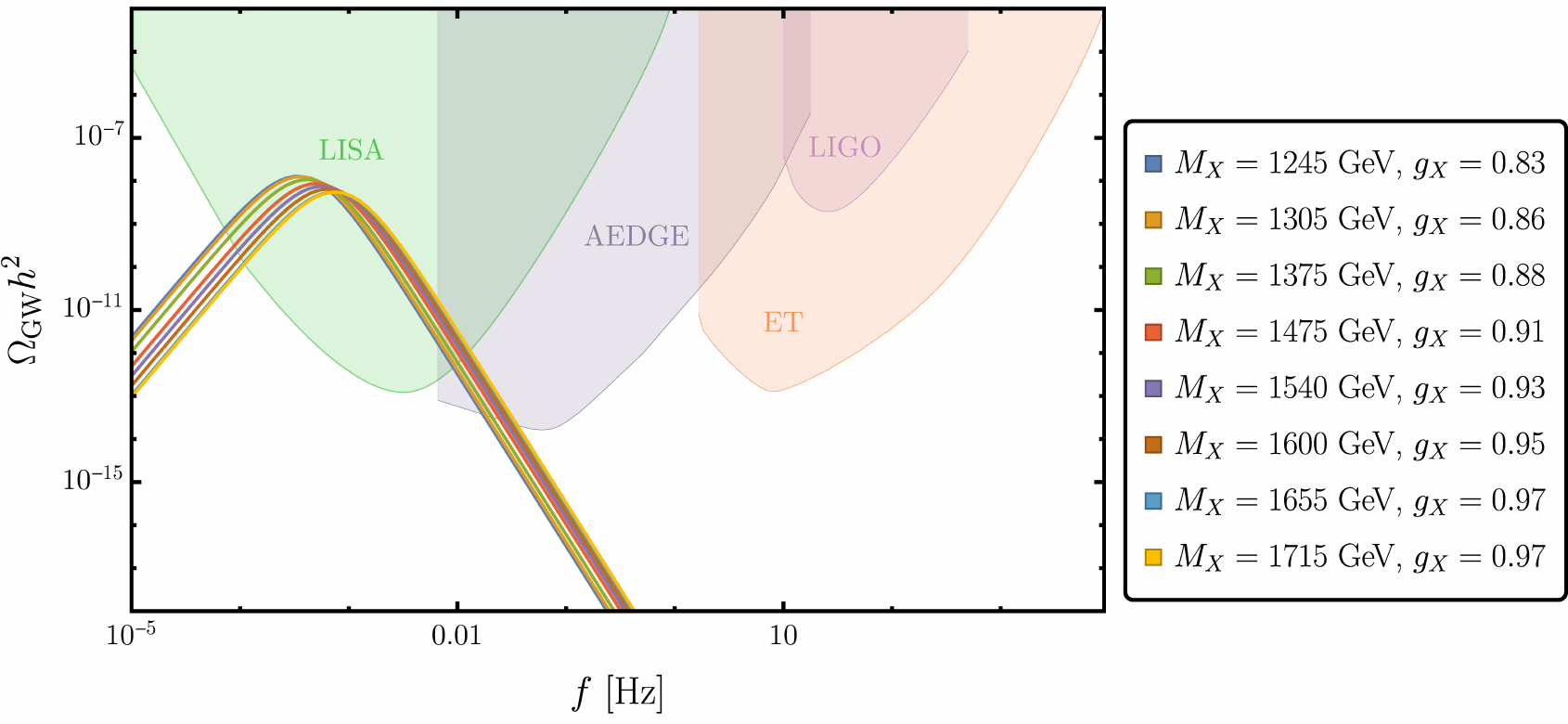}
\caption{Predictions for spectra of gravitational waves together with integrated sensitivity curves for LISA, AEDGE, ET and LIGO for the points in the parameter space where DM relic abundance is saturated.\label{fig:spectra-DM}}
\end{figure}

The correct DM relic abundance and non-exclusion by direct detection experiments (solid black line in figure~\ref{fig:SNR}) is located in the region of relatively weaker signal. It is still well observable with LISA and AEDGE. %\comment{\sout{(this will depend on how good the sensitivity will be for lower frequencies)}}. 
The GW signal in the region where the correct abundance is reproduced is sourced entirely by sound waves. Examples of spectra for points along the black line in figure~\ref{fig:SNR} are shown in figure~\ref{fig:spectra-DM}.
%

%XXXXXXXXXXXXXXXXXXXXX
%~~~~~~~~~~~~~~~~~~~~~~~~~~~~~~~~~~~~~~~~~
\section{Renormalisation-scale dependence\label{sec:scale-dep}}
%~~~~~~~~~~~~~~~~~~~~~~~~~~~~~~~~~~~~~~~~~
%XXXXXXXXXXXXXXXXXXXXX

It is well known that the fixed-loop effective potential depends on the choice of the renormalisation scale. It has also been pointed out that in general, the PT parameters computed using the effective potential and, as a consequence, the predicted GW signals do depend on the choice of the RG scale~\cite{Croon:2020, Athron:2022}. This is even more pronounced in models with classical scaling symmetry due to the presence of vastly different energy scales and so is in the SU(2)cSM model considered in this work. In the main part of this article we used the RG-improved potential in order to suppress the scale dependence of the results. In this section, we will show how the results would change if we performed the computations at a fixed scale.

\begin{figure}[h!t]
\center
\includegraphics[height=.33\textwidth]{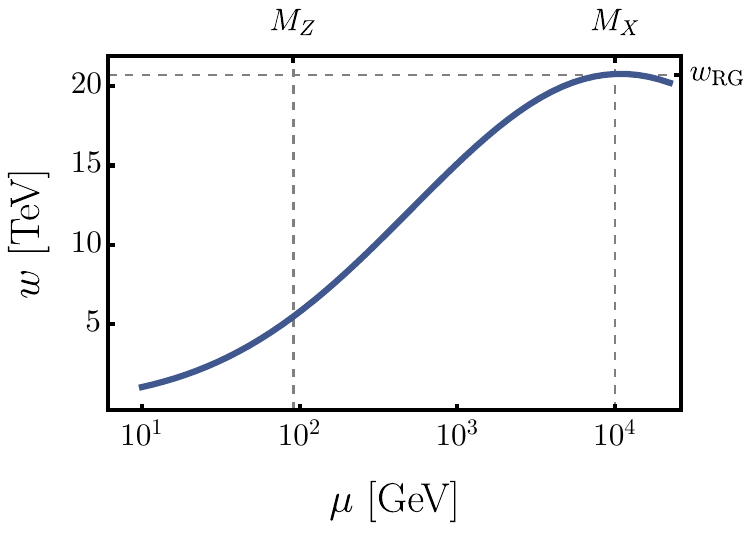}\hspace{2pt}
\includegraphics[height=.33\textwidth]{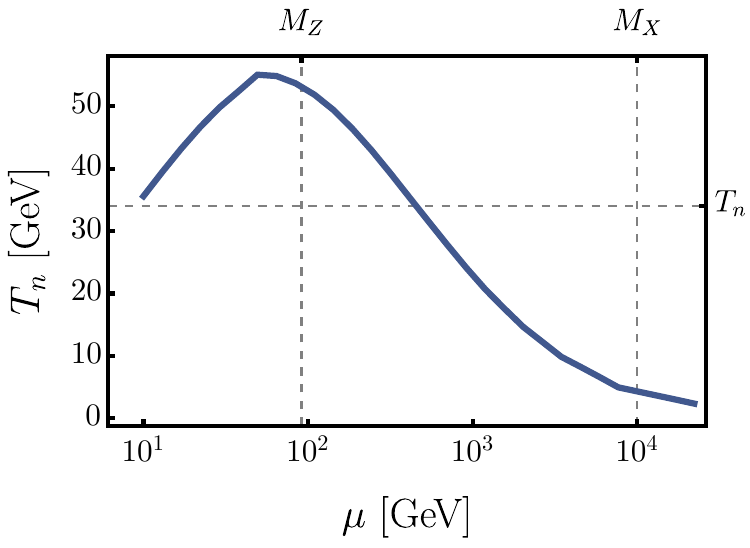}
\caption{RG scale dependence of the location of the minimum in the $\f$ direction (left panel) and of the nucleation temperature (right panel). Vertical lines indicate characteristic energy scales, while horizontal the values of $w$ and $T_n$, respectively, computed using the RG-improved potential. For a benchmark point with $g_X=0.9$, $M_X=10\,\mathrm{TeV}$.\label{fig:scale-dep}}
\end{figure}

To start, in figure~\ref{fig:scale-dep} (left panel) we show the dependence of the VEV of $\f$ on the choice of the scale $\mu$ for a specific benchmark point with $g_X=0.9$ and $M_X=10$\,TeV. To obtain the results we fix the parameters $g_X$ and $M_X$ at $\mu=M_X$. Then we use the one-loop potential with running couplings and fields to move to another scale, at which we perform the computations. The dependence on the scale is dramatic, especially when scales below the electroweak scale are considered. The VEV varies from approximately 1\,TeV at $\mu=10\g$ to about 20\,TeV at $\mu=M_X$. If one recalls that the energy released during the phase transition scales as $w^4$, it is clear that the results for the PT and GW have to depend on the renormalisation scale. Moreover, one cannot simply reject the scales below the EW scale from consideration, since these are the scales around which the tunnelling takes place, see figure~\ref{fig:VatTnuc}. The right panel of figure~\ref{fig:scale-dep} shows the dependence of the nucleation temperature on the RG scale -- $T_n$ varies between around $1\g$ when computed around $M_X$ (and goes down to approximately 0.1\,GeV for higher scales) and $77\g$ slightly below $M_Z$.

\begin{figure}[h!t]
\center
\includegraphics[width=.4\textwidth]{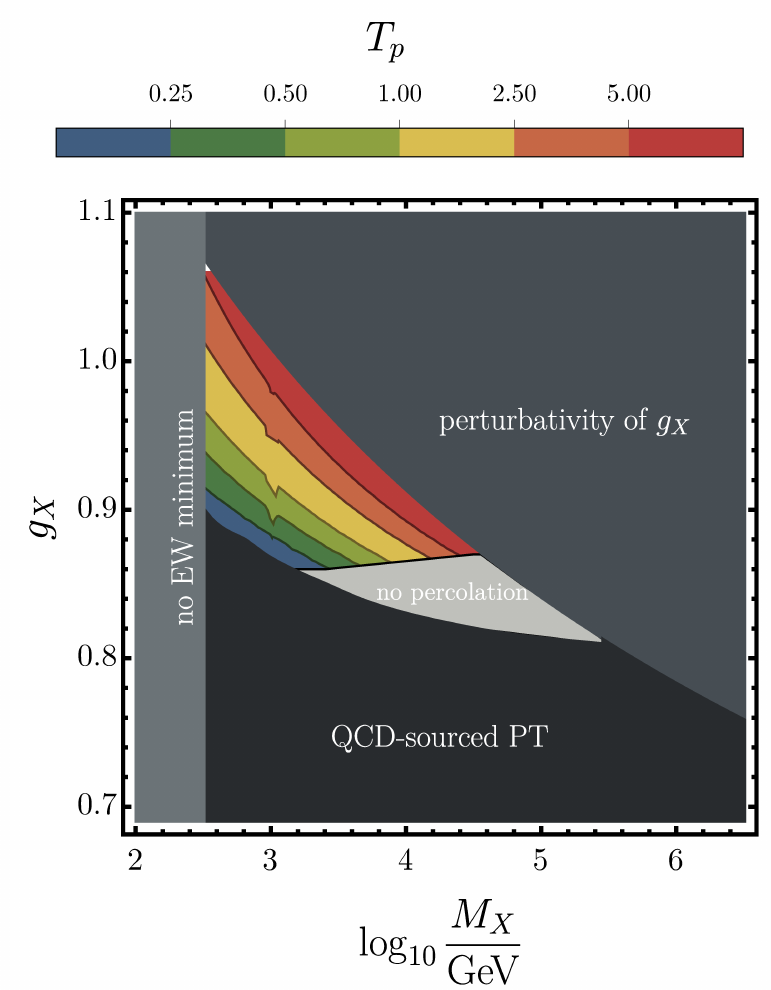}\hspace{20pt}
\includegraphics[width=.4\textwidth]{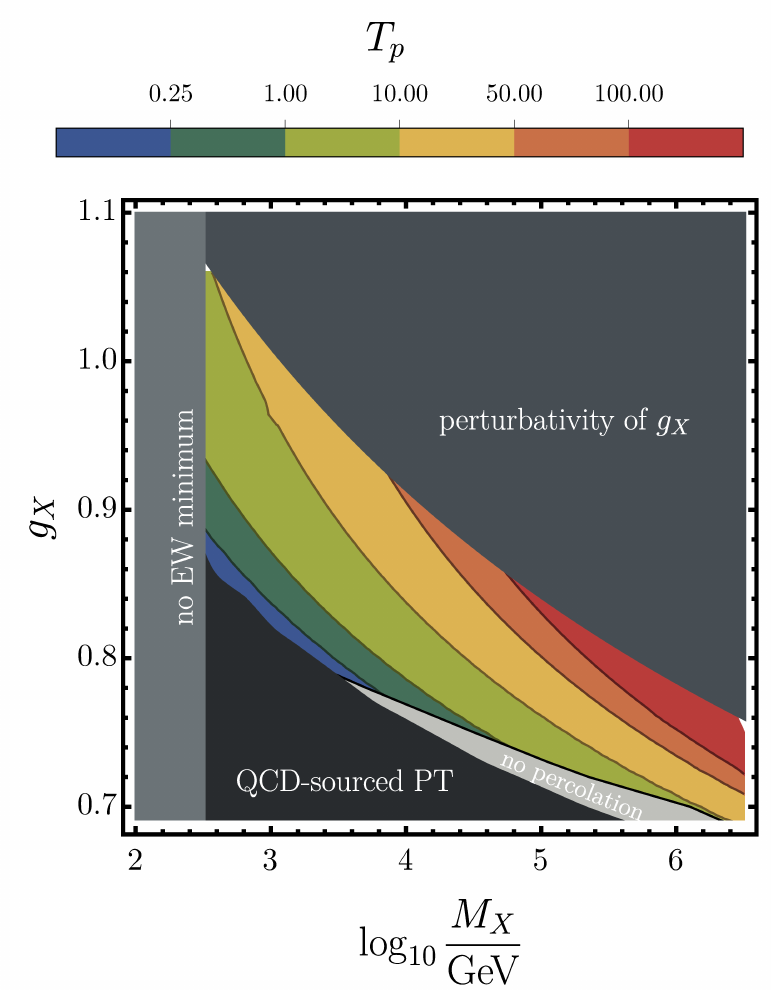}\\
\includegraphics[width=.4\textwidth]{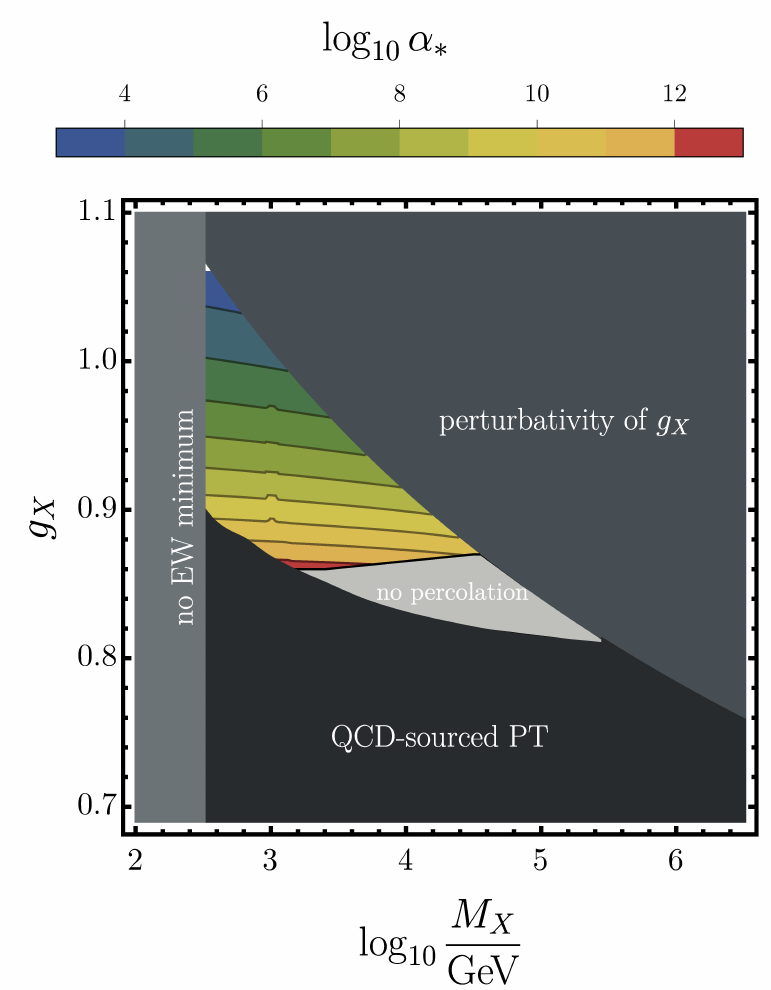}\hspace{20pt}
\includegraphics[width=.4\textwidth]{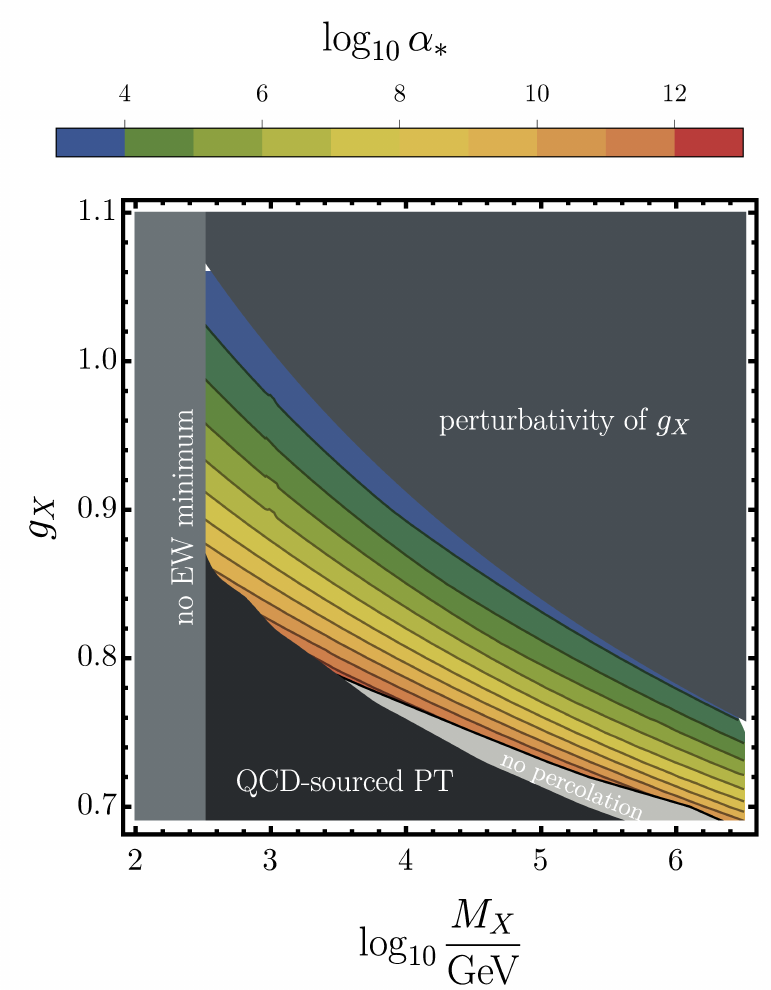}%\\
\caption{Results of the scan with fixed renormalisation scale, $\mu=M_X$ (left column), $\mu=M_Z$ (right column) for selected parameters characterising the phase transition: percolation temperature $T_p$ (upper row), $\log_{10}\alpha_*$ (lower row).\label{fig:scan-at-fixed-mu}}
\end{figure}

Next, motivated by these results, we perform scans of the parameter space, analogous to what has been discussed in section~\ref{sec:results}, but at fixed $\mu$. This will tell us how our understanding of the parameter space and observability of the GW signal depends on the renormalisation scale. Figure~\ref{fig:scan-at-fixed-mu} shows the results for the key PT parameters ($T_p$ -- upper panel and  $\log_{10} \alpha_*$ -- lower panel) computed at different scales ($\mu=M_X$ (left), $\mu=M_Z$ (right)) together with the constraints on the parameter space (for the explanation of these constraints see section~\ref{sec:PT}). The upper row, showing the percolation temperature, indicates a striking dependence on the renormalisation scale, as already suggested by the plot for the nucleation temperature in figure~\ref{fig:scale-dep}. This has further implications, since for $T_p\lesssim 0.1\g$ the PT is believed to be sourced by the QCD effects, which changes the nature and properties of the PT. In this work we focus on the PT sourced by the tunnelling, therefore the considered parameter space changes dramatically as the renormalisation scale is changed. Also the answer to a basic question -- whether or not the PT completes via percolation of bubbles of the true vacuum -- is altered by the change of the renormalisation scale as can be seen by examining the percolation criterion (light-grey shaded region). The strength of the transition is significantly modified as compared to the RG-improved case (compare with figures~\ref{fig:alpha}--\ref{fig:kappa-sw}) -- the strongest transitions with the largest $\alpha_*$ are absent and thus there are no valid points with GW sourced by bubble collisions. We do not display $\kappa_{\textrm{sw}}$ as it is equal to 1 throughout the allowed parameter space.  All the results discussed above show that the change of the scale at which computations are performed not only changes the results quantitatively, by shifting the values of the characteristic parameters of the phase transition, but it also significantly modifies them qualitatively -- by modifying the character of the phase transition, the very fact of its completion and the dominant source of the GW signal.

Comparing the results of the fixed-$\mu$ scans with the results of the scan with RG-improved potential one can see that the high-field-values-related behaviour, related to the location of the VEV and the energy released during the PT is better captured by the scan at the high scale of $M_X$. On the contrary, the scan at the EW scale predicts higher percolation temperatures, not limiting the parameter space as severely as the high-energy scan, being closer (but still quite far from) the RG-improved result. This shows that indeed, RG-improvement links various scales in one consistent description.

One could doubt the sensibility of comparing results obtained at such vastly differing scales. However, these scales (and even lower, as the location of the barrier) are inherently present in the considered model, and, moreover, they are relevant to different aspects of the PT. We cannot thus escape from facing the issue of scale dependence. Our answer to this issue in this work is the RG improvement of the effective potential. There are indications, however, that it might not be enough in the finite temperature setting~\cite{Gould:2021}, thus a next step in improving the analysis could include higher-order corrections and dimensional reduction, which is advocated to be a remedy for the scale dependence in PT computations (for a recent review see ref.~\cite{Schicho:2021gca}, for a recent work combining dimensional reduction in PT computations and a DM analysis see ref.~\cite{Biondini:2022ggt}). It relies, however, on the high-temperature expansion, which does not hold in general in the vicinity of the PT in conformal models, thus its usage in this context is not straightforward and deserves a separate analysis.
%XXXXXXXXXXXXXXXXXXXXXXXXXX
\section{Summary and conclusions\label{sec:summary}}
%XXXXXXXXXXXXXXXXXXXXXXXXXX

In the present work, we studied a model endowed with classical scale invariance, a dark SU(2)$_X$ gauge group and a scalar doublet of this group. This model provides a dynamical mechanism of generating all the mass scales via radiative symmetry breaking, while featuring only two free parameters. Moreover, it provides dark matter candidates -- the three gauge bosons of the SU(2)$_X$ group which are degenerate in mass -- stabilised by an intrinsic $\mathbb{Z}_2 \times \mathbb{Z}_2'$ symmetry. Like other models with scaling symmetry, the studied model exhibits strong supercooling which results in the generation of observable gravitational-wave signal.

Motivated by these attractive features we performed an analysis of the phase transition, gravitational wave generation and dark matter relic abundance, updating and extending the existing results~\cite{Hambye:2013, Carone:2013, Khoze:2014, Pelaggi:2014wba, Karam:2015, Plascencia:2016, Chataignier:2018, Hambye:2018, Baldes:2018, Prokopec:2018, Marfatia:2020}. The analysis features the key ingredients: 
\begin{itemize}
\item careful analysis of the potential in the light of radiative symmetry breaking, in particular a consistent expansion in the powers of couplings;
\item using renormalisation-group improved potential which includes all the leading order terms;
\item using RG-running in order to move between various relevant scales: the electroweak scale for scalar mass generation, the scale of the mass of the new scalar for its decay during reheating;
\item careful analysis of the supercooled phase transition, following recent developments, in particular imposing the percolation criterion which proved crucial for phenomenological predictions;
\item analysis of dark matter relic abundance in the light of the updated picture of the phase transition; %which excluded the supercool scenario (at least when the phase transition is sourced by the tunnelling, not QCD);
\item analysis of gravitational-wave spectra using most recent results from simulations;
\item using fixed-scale potential, in addition to the renormalisation-group-improved one, in order to study the scale dependence of the results.
\end{itemize}

The first and foremost result of our analysis is that within the SU(2)cSM model the gravitational wave signal sourced by a first-order phase transition associated with the SU(2)$_X$ and electroweak symmetry breaking is strong and observable for the whole allowed parameter space. This is an important conclusion, since it allows to falsify this scenario in case of negative LISA results. 

Second, we exclude the supercool dark matter scenario within the region where the phase transition proceeds via nucleation and percolation of bubbles of the true vacuum. It is a result of a combination of two reasons: we include the percolation condition, eq.~\eqref{eq:percolation-crit}, which allows to verify that a strongly supercooled phase transition indeed completes via percolation of bubbles and strongly constrains the parameter space relevant for our analysis. Moreover, we improve on the computation of the decay rate of the scalar field $\f$, which controls the reheating rate, which pushes the onset of inefficient reheating towards higher $M_X$, beyond the region of interest.

Third, we find the parameter space in which the correct relic dark matter abundance is predicted. It is produced via the standard freeze-out mechanism in the region with relatively low $M_X$ and large $g_X$. It is the region where the phase transition is relatively weak (compared with other regions of the parameter space), yet the gravitational-wave signal should be well observable with LISA. This parameter space is further reduced due to the recent direct detection constraints.

Moreover, in the present work we focused on the issue of scale dependence of the predictions. Our approach to reducing this dependence was to implement the renormalisation-group improvement procedure, respecting the power counting of couplings to include all the relevant terms. For comparison, we present results of computations performed at fixed scale, where the dependence on the renormalisation scale is significant. It is important to note that with the change of the scale the predictions do not only change quantitatively, they can change qualitatively. For example, for computations performed at a fixed scale (both $\mu=M_X$ and $M_Z$)  gravitational waves sourced by bubble collisions are not present. At the same time, with RG improvement we see a substantial region where bubble collisions are efficient in producing an observable signal.

A possible issue to be studied further is to carefully verify the scaling of various couplings in the temperature-dependent setting, remembering that the high-temperature expansion is in general not valid in presence of radiative-symmetry breaking. Then one could check whether with the one-loop (and daisy-resummed) terms included in this work the scale dependence is indeed cancelled at the considered order %strictly preserved 
or whether some higher-loop terms are needed. Another way of improving the accuracy of the predictions would be with the use of dimensional reduction, which is not straightforward in the case of radiative symmetry breaking. These issues will be a subject of a forthcoming work.

To sum up, the classically scale-invariant model with an extra SU(2) symmetry remains a valid theoretical framework for describing dark matter and gravitational-wave signal produced during a first-order phase transition in the early Universe. It will be tested experimentally by LISA and other gravitational-wave detectors. The predictions, however, are sensitive to the theoretical procedures implemented. Therefore, it is crucial to improve our understanding of theoretical pitfalls affecting the predictions. The present work is a step in this direction.

%-------------------------------------------------------------------------------
\acknowledgments{
%-------------------------------------------------------------------------------
We would like to thank Kristjan Kannike, Wojciech Kotlarski, Luca Marzola, Tania Robens, Martti Raidal and Rui Santos for useful discussions. We are indebted to Marek Lewicki for numerous discussions, clarifications and sharing data for the SNR plots. We are grateful to Jo\~{a}o Viana for his computation of Higgs decay width using \texttt{hdecay} and IT hints. We would also like to thank Matti Heikinheimo, Tomislav Prokopec, Tommi Tenkanen, Kimmo Tuominen and Ville Vaskonen for collaboration in the early stages of this work. AK was supported by the Estonian Research Council grants MOBTT5, MOBTT86, PSG761 and by the EU through the European Regional Development Fund CoE program TK133 ``The Dark Side of the Universe". The work of B\'{S} and MK is supported by the National Science Centre, Poland, through the SONATA project number 2018/31/D/ST2/03302.
}

\begin{appendix}
\section{Scalar contributions to the effective potential\label{app:scalars}}

In this appendix, we discuss the scalar contributions to the effective potential. This is an extension of the discussion of section~\ref{sec:scaling-of-couplings}.

With the knowledge of how various contributions scale, we can assess which terms in the one-loop correction to the effective potential are relevant for the analysis of radiative symmetry breaking. The SM Goldstone masses in principle are of the same order as other SM contributions since $\la \sim g^2$. However, their mass at the leading-order minimum vanishes (see eq.~\eqref{eq:tree-level-min}) so we can expect that at the slightly loop-corrected VEV of eq.~\eqref{eq:condition-lambda2} their contribution will still be negligible. Since we use the two-dimensional zero-temperature potential only for the determination of the VEVs, masses and the mixing angle (the tunnelling is studied only along the direction of $\f$), we do not need to worry about the Goldstone contributions away from the minimum and thus we can neglect their contributions. The Goldstones belonging to the new SU(2)$_X$ group have their masses suppressed by the small $\lb,\lc$ couplings, however they are enhanced by the VEV of the $\f$ field, see eq.~\eqref{eq:X-Goldstone-masses}. Since there is no tree-level minimum along the $\f$ direction, they are far from vanishing at the loop-generated minimum. In fact, they are negative at the EW scale, where we evaluate the physical masses, since both $\lb$ and $\lc$ are negative, see figure~\ref{fig:scalar-couplings}. Thus, their contribution will be neglected, since it will only contribute to the imaginary part of the effective potential.

In order to assess the contribution of the $H$ and $S$ scalars to the effective potential, one can neglect the contributions to eq.~\eqref{eq:M-plus-minus} suppressed by a product of the small $\lb,\lc$ couplings and the SM Higgs VEV. Then, the two tree-level masses read
\begin{align}
    &M_{+}^{2} =3 \lc \f^2,  \label{eq:tree-level-scalar-masses}\\
    &M_{-}^{2} = 3\la h^2 +\frac{1}{2}\lb \f^2, \label{eq:tree-level-scalar-masses-Higgs}
\end{align}
for $3\la v^2-3\lb w^2+\frac{1}{2}\lb w^2<0$ (which determines the sign of the expression under the square root in eq.~\eqref{eq:M-plus-minus} with the approximations adopted here). The expressions for $M_+$ and $M_-$ are reversed if the opposite condition holds.

It is clear that the expression of eq.~\eqref{eq:tree-level-scalar-masses-Higgs} corresponds to the tree-level approximation to the Higgs mass. We will fix the parameters such as to reproduce the observed Higgs mass $M_H=125\g$ at one-loop level, where the corrections along the Higgs direction are not too big, therefore the tree-level Higgs mass evaluated at the minimum of the potential is close to $M_H$. Thus, it is small compared to e.g.\ the top contribution due to a small number of degrees of freedom and it will be neglected in the computations. The other mass eigenvalue, which corresponds to tree-level $S$ mass, at the electroweak scale is negative, which is not surprising as at tree level the potential does not develop a minimum. Therefore, the $S$ contribution to the one-loop effective potential can be neglected (it will contribute to the imaginary part).
%~~~~~~~~~~~~~~~~~~~~~~~~~~~~~~~~~~~~~~~~~~~~~~~~~~~~~
\section{Self energies \label{app:self-energies}}
%~~~~~~~~~~~~~~~~~~~~~~~~~~~~~~~~~~~~~~~~~~~~~~~~~~~~~

In this appendix, we summarise the results for the self-energies of scalar particles in the SU(2)cSM model.\footnote{A very detailed computation of the loop functions used here can be found in an appendix of ref.~\cite{Swiezewska:2016}, while a general derivation of two-loop self-energies can be found in ref.~\cite{Martin:2003it, Martin:2003qz}.} The contributing diagrams can be found in figure~\ref{fig:self-energy}. The results will be given in terms of the Passarino--Veltman functions~\cite{Passarino:1978} $a$ and $b_0$,
\begin{align}
a(m) &= \frac{m^2}{(4\pi)^2}\left(-\frac{2}{\epsilon}  +\gamma_E- \log (4\pi )+\log \frac{m^2}{\mu^2} -1 \right),\\
b_0(p^2,m_1, m_2) &= \frac{1}{(4\pi)^2}\left(-\frac{2}{\epsilon}  +\gamma_E - \log (4\pi) + \int_0^1 dx \log \frac{\Delta}{\mu^2}\right),
\end{align}
where $\Delta = -x(1-x) p^2 + x m_1^2 +(1-x) m_2^2$. We also introduce non-standard functions $a^b$ and $b^b$ which will be useful in computing bosonic loops,
\begin{align}
a^b(m) &= a(m)+\frac{2}{3}\frac{m^2}{(4\pi)^2},\\
b_0^b(p^2,m_1, m_2) &= b_0(p^2, m_1,m_2)+\frac{1}{2(4\pi)^2}.
\end{align}

Below, we list contributions from the diagrams of figure~\ref{fig:self-energy}. The first two diagrams containing only scalars contribute as $\lambda^2\sim g^8$, therefore we neglect them. The remaining ones read as follows:
\begin{align}
    (C) &\equiv -i \Sigma_{ff}= i 3 Y_f^2 \left[2a(M_f)+(-p^2+4M_f^2)b_0(p^2,M_f,M_f)\right],\\
    (D) &\equiv -i \Sigma_{sv}= -i g_V^2 \frac{C_V}{4M_V^2}\Big[ M_V^2 a\left(M_S\right) + \left(-p^2-M_V^2+M_S^2\right)a(M_V)-(p^2-M_S^2)^2b_0\left(p^2,0,M_S\right)\nonumber\\*
&\quad+\left(p^4+M_S^4+M_V^4-2p^2M_V^2 - 2M_S^2M_V^2-2p^2M_S^2\right) b_0(p^2, M_V,M_S)\Big],\\
    (E) &\equiv -i \Sigma_{vv}= -i g_V^2\frac{C_V}{8M_V^2} \Big[ 2M_V^2a(M_V) +p^4 b_0(p^2, 0, 0)-2\left(p^2-M_V^2\right)^2 b_0(p^2,M_V,0)\nonumber\\*
    &\quad +16M_V^4b_0^b(p^2, M_V, M_V)+\left(p^4-4p^2M_V^2-4M_V^4\right)b_0(p^2, M_V, M_V)\Big].\\
    (F) &\equiv -i \Sigma_{v}= -i g_V^2 C_V \frac{3}{4}  a^b(M_V),
\end{align}
where $C_V$ accounts for symmetry factors ($C_Z=1$ for $Z$ propagating in the loop, $C_{W}=2$ and $C_X=3$), $g_V^2$ represents respective couplings of particles that can propagate in the loop, $g_W^2=g_2^2$, $g_Z^2=g_2^2+g_Y^2$ and $g_X^2$ is the SU(2)$_X$ gauge coupling. With these definitions, we can write down the respective self-energies,
\begin{align}
    \Sigma_{hh}&=\Sigma_{W}+\Sigma_{WW}+\Sigma_{WG}+\Sigma_{Z}+\Sigma_{ZZ}+\Sigma_{ZG}+\Sigma_{tt},\\
    \Sigma_{\f\f}&=\Sigma_{X}+\Sigma_{XX}+\Sigma_{XG_X},
\end{align}
where $G_i$ denote respective Goldstone bosons and the $\pm$ indices are implicitly understood.
%XXXXXXXXXXXXXXXXXXXXX
%~~~~~~~~~~~~~~~~~~~~~~~~~~~~~~~~~~~~~~~~~
\section{Numerical procedure for determining the values of the parameters\label{app:numerical-procedure}}
%~~~~~~~~~~~~~~~~~~~~~~~~~~~~~~~~~~~~~~~~~
%XXXXXXXXXXXXXXXXXXXXX

In this appendix, we describe the details of the numerical procedure used to determine the values of the parameters. For this purpose, we use the one-loop zero-temperature effective potential. The reference scale is set to $\mu=M_Z$ and we make use of the RG equations to evolve the couplings and fields between different scales. The values of the constants are taken from~\cite{ParticleDataGroup:2020ssz, Workman:2022ynf}. The input constants are: the $W$ and $Z$ boson masses $M_W$ and $M_Z$, the top quark mass $M_t$, the Higgs mass $M_H$ and the Fermi constant from which we derive the Higgs VEV. We also use the two-loop-matched values of $g_2$, $g_Y$ and $y_t$ from ref.~\cite{Buttazzo:2013} to plug into the running couplings.

Since the model possesses at least two different scales (at zero $T$), related to the electroweak scale (or $v$) and to the VEV of the new scalar field ($w$), we pay special attention to respecting these scales.

The procedure reads as follows:
\begin{enumerate}
    \item We choose the values of the input parameters, $M_X$ and $g_X$. We assume the tree-level relation for the $X$ mass $M_X=\frac{1}{2}g_X w$ so we can compute the value of the $\f$ VEV, $w$. We assume that $M_X$ corresponds to the physical mass of the $X$ bosons.\footnote{Of course one can compute loop corrections to the $X$ mass and solve the gap equation to obtain a better approximation for this mass. However, since for the time being this mass is not accessible experimentally, we believe that this simplistic choice is sufficient.} The values of $g_X$ and $w$ are treated as evaluated at the scale $\mu=M_X$.
    
    \item We use the minimisation condition along the $\f$ direction, eq.~\eqref{eq:min-phi}, evaluated at $\mu=M_X$ to evaluate $\lc$. This gives us a simple relation
    \begin{equation}
        \lc=\frac{3}{256\pi^2}g_X^4.
    \end{equation}
    
    \item The $g_X$ and $\lambda_3$ couplings are evolved using their RG running (with $\lb$ term neglected in the latter case as being numerically small due to the smallness of $\lb$) and evaluated at the reference scale $\mu=M_Z$.
    
    \item The value of $\lb$  as a function of $\lambda_1$ (at $\mu=M_Z$) is found from eq.~\eqref{eq:condition-lambda2}.
    
    \item The value of $\la$ is computed from the requirement that the physical Higgs mass is equal to 125$\g$, using eq.~\eqref{eq:gap-Higgs} (with a ``$+$'' or ``$-$'' subscript, depending on the region of the parameter space). The evaluation is performed at $\mu=M_Z$, therefore the vacuum expectation value of $\f$ at $\mu=M_Z$ is needed. It is found using eq.~\eqref{eq:min-phi} evaluated at $\mu=M_Z$. 
    
    \item The mass of the $S$ scalar mass-eigenstate is computed by solving iteratively the gap equation~\eqref{eq:gap-scalar}.
    
    \item As discussed in section~\ref{sec:masses-and-mixing}, the mixing between the scalars is evaluated by demanding that the off-diagonal terms of the mass matrix evaluated at $p^2=0$ and in the mass-eigenbasis are zero.
\end{enumerate}

This procedure allows us to determine the values of all the scalar couplings, masses and mixing from two input parameters $g_X$ and $M_X$.
%XXXXXXXXXXXXXXXXXXXXX
%~~~~~~~~~~~~~~~~~~~~~~~~~~~~~~~~~~~~~~~~~
\section{Discussion of various approximations of the energy transfer rate in reheating \label{app:reheating}}
%~~~~~~~~~~~~~~~~~~~~~~~~~~~~~~~~~~~~~~~~~
%XXXXXXXXXXXXXXXXXXXXX

As mentioned in section~\ref{sec:PT}, in the literature there are different approaches to computing the energy transfer rate relevant for reheating after the phase transition. In this appendix we compare some of these approaches to the method we use in section~\ref{sec:PT}.

\begin{figure}
    \centering
    \includegraphics[width=.4\textwidth]{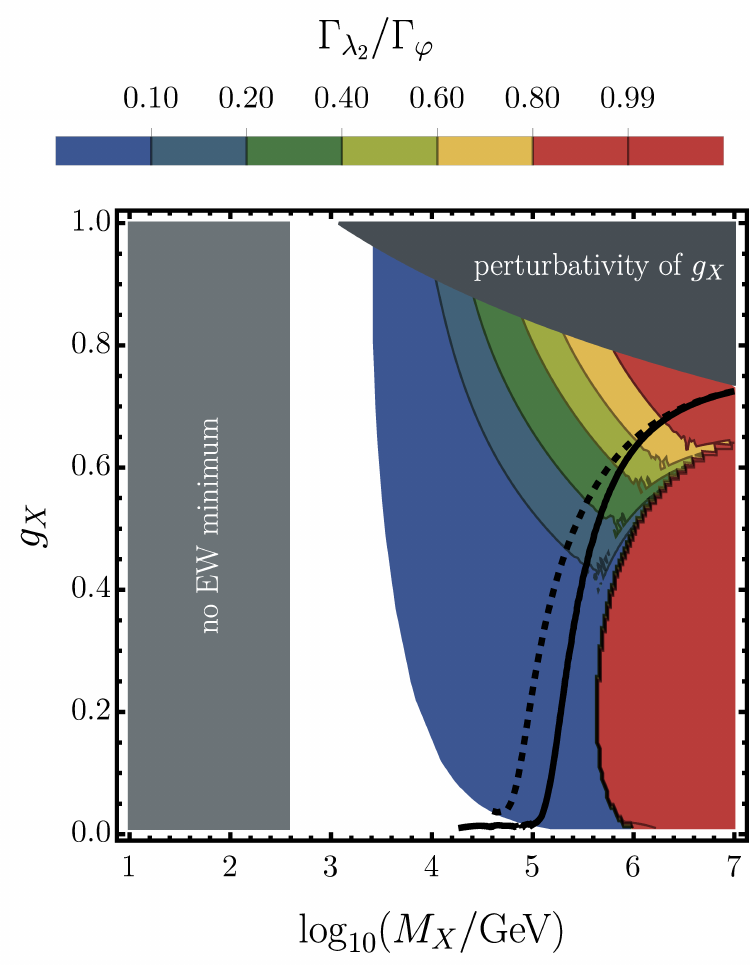}\hspace{2pt}
     \includegraphics[width=.4\textwidth]{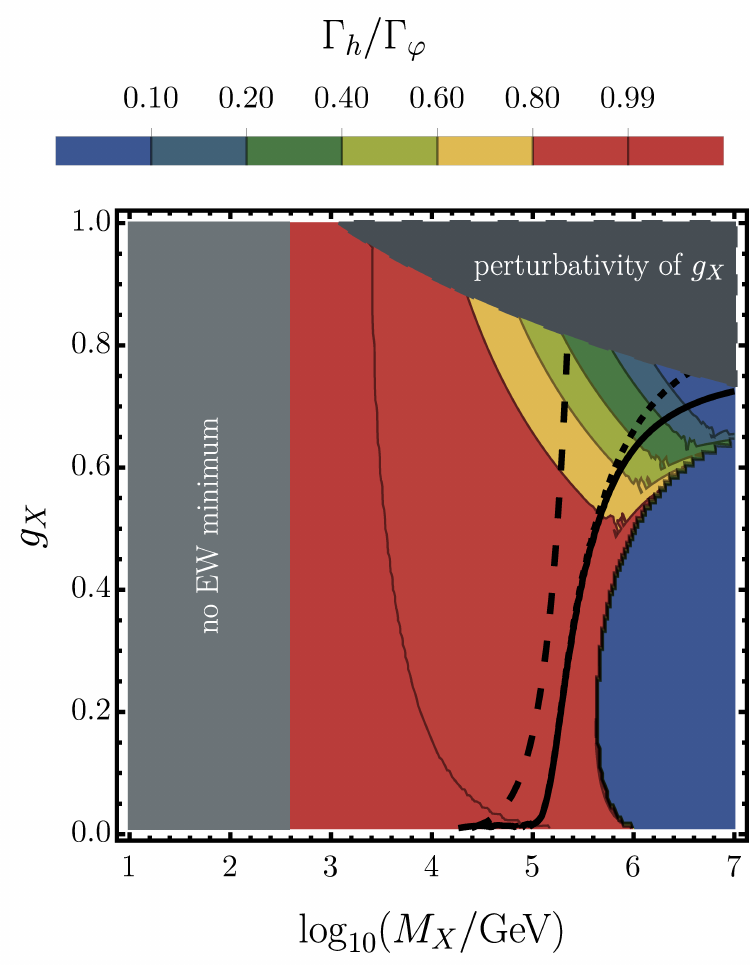}
    \caption{Left panel: The ratio of $\Gamma_{\lb}$ to $\Gamma_{\f}$ of eq.~\eqref{eq:gamma-phi} computed using solely the $S\to HH$ decay, $\Gamma_{\lb}/H_*=1$ line (black, short-dashed), $\Gamma_{\f}/H_*=1$ line (black, solid). Right panel: The ratio of $\Gamma_h$ to $\Gamma_{\f}$ of eq.~\eqref{eq:gamma-phi}, $\Gamma_{h}/H_*=1$ line computed using our methods (black, short-dashed),  $\Gamma_{h}/H_*=1$ line computed using the approach of ref.~\cite{Marfatia:2020} (black, long-dashed) (see the main text for details).}
    \label{fig:dec-rate}
\end{figure}

A common approach, see e.g.\ ref.~\cite{Ellis:2020}, is to assume that $\Gamma_{\f}$ is approximately given by the decay rate $\f\to hh$ assuming that $\f$ and $h$ have the masses equal to $M_S$ and $M_H$, we denote this quantity as $\Gamma_{\lb}$. We have already discussed how it differs from our approach and figure~\ref{fig:dec-rate} shows the values of the ratio $\Gamma_{\lb}/\Gamma_{\f}$ (left panel). The white region corresponds to $M_S<M_H$ where $\Gamma_{\lb}$ is not defined. In the rest of the parameter space the ratio varies almost from 0 to 1, even though the mixing between the scalars is very small everywhere. The red region in the right part of the plot corresponds to $\Gamma_{\lb}/\Gamma_{\f}$ equal practically to 1, as in this regime the mixing becomes numerically vanishing. We have also checked where $\Gamma_{\f}$ and $\Gamma_{\lambda_2}$ become equal to the Hubble rate, which would mean that the reheating is inefficient and one should use formula~\eqref{eq:Treh-general} rather than~\eqref{eq:Treh=TV}. This region is indicated by the solid black line as obtained with $\Gamma_{\f}$ and the dashed line with $\Gamma_{\lb}$. The difference is not huge but it is clear that using $\Gamma_{\f}$ delays the appearance of the matter domination period to larger $M_X$.

In the approach of ref.~\cite{Marfatia:2020} it was assumed that the component of $\Gamma_{\f}$ related to the SM-like decay dominates $\Gamma_{\f}$ and the $S\to HH$ decay was omitted, the resulting width is denoted as $\Gamma_h$. The ratio of $\Gamma_h$ to $\Gamma_{\f}$ is presented in the right panel of figure~\ref{fig:dec-rate}. It is clear from figure~\ref{fig:dec-rate} that approximating $\Gamma_{\f}$ by $\Gamma_h$ is incorrect for large masses. It  results in an underestimation of the energy transfer rate, especially in the large $M_X$ regime which results in finding matter-domination for lower $M_X$ and also in lower reheating temperatures (since for $\Gamma_h<H_*$ eq.~\eqref{eq:Treh-general} has to be used).  The short-dashed line in figure~\ref{fig:dec-rate} represents $\Gamma_h/H_*=1$ line computed with our methods including the determination of the mixing angle and running of the couplings and VEVs. In the approach of ref.~\cite{Marfatia:2020} the mixing was approximated as $\sin^2(v/w)$ (which is only valid for $M_S\ll M_H$~\cite{Hambye:2018}) and the SM rate taken as approximately 4\ MeV. The $\Gamma_h/H_*=1$ line in this approach is represented by the long-dashed line. Clearly, reheating temperatures using this approach must be underestimated.
%XXXXXXXXXXXXXXXXXXXXX
%~~~~~~~~~~~~~~~~~~~~~~~~~~~~~~~~~~~~~~~~~
\section{Gravitational-wave signal according to a recent simulation \label{app:new-spactra}}
%~~~~~~~~~~~~~~~~~~~~~~~~~~~~~~~~~~~~~~~~~
%XXXXXXXXXXXXXXXXXXXXX

\begin{figure}[h!t]
\center
\includegraphics[width=.48\textwidth]{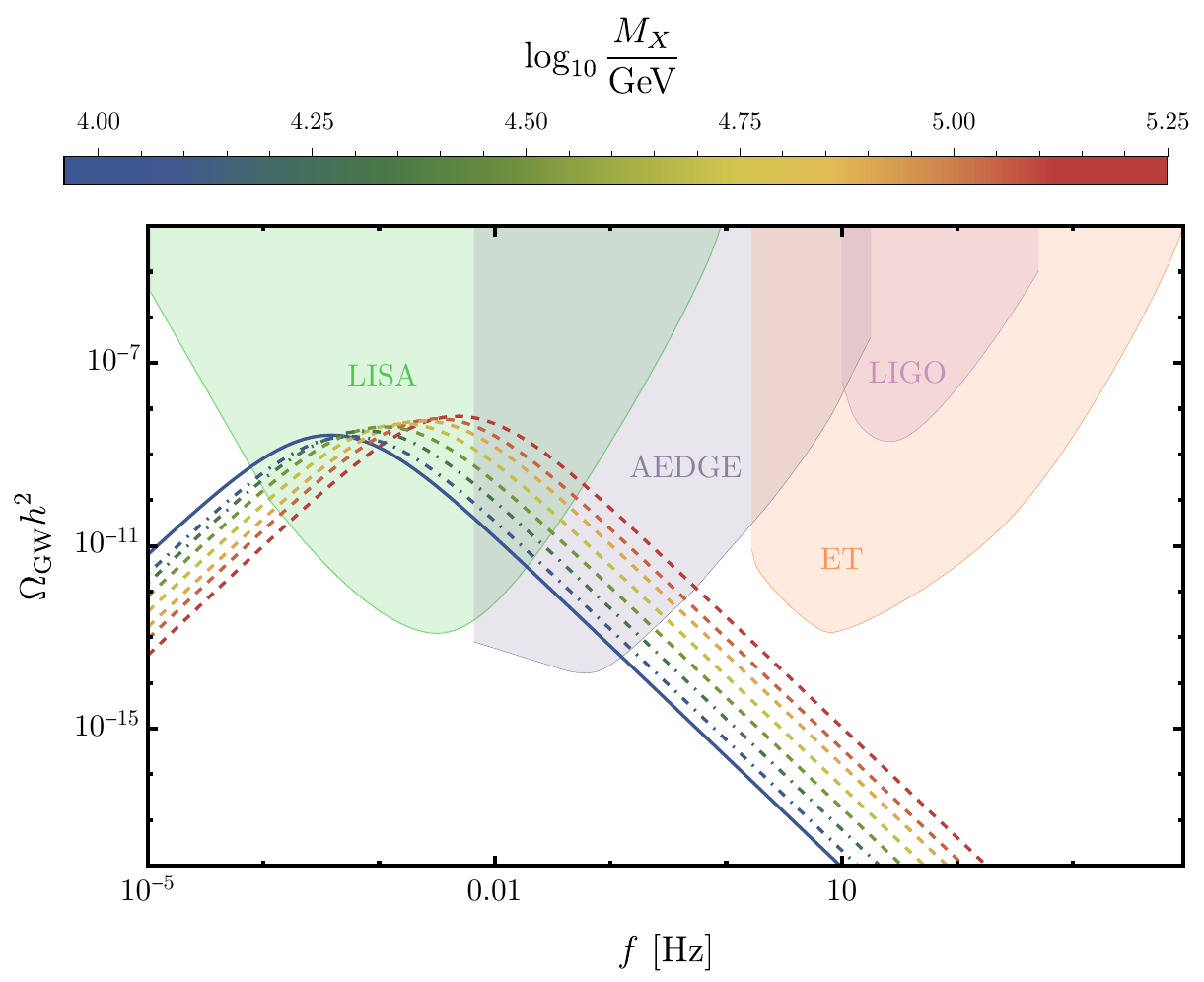}\hspace{2pt}
\includegraphics[width=.48\textwidth]{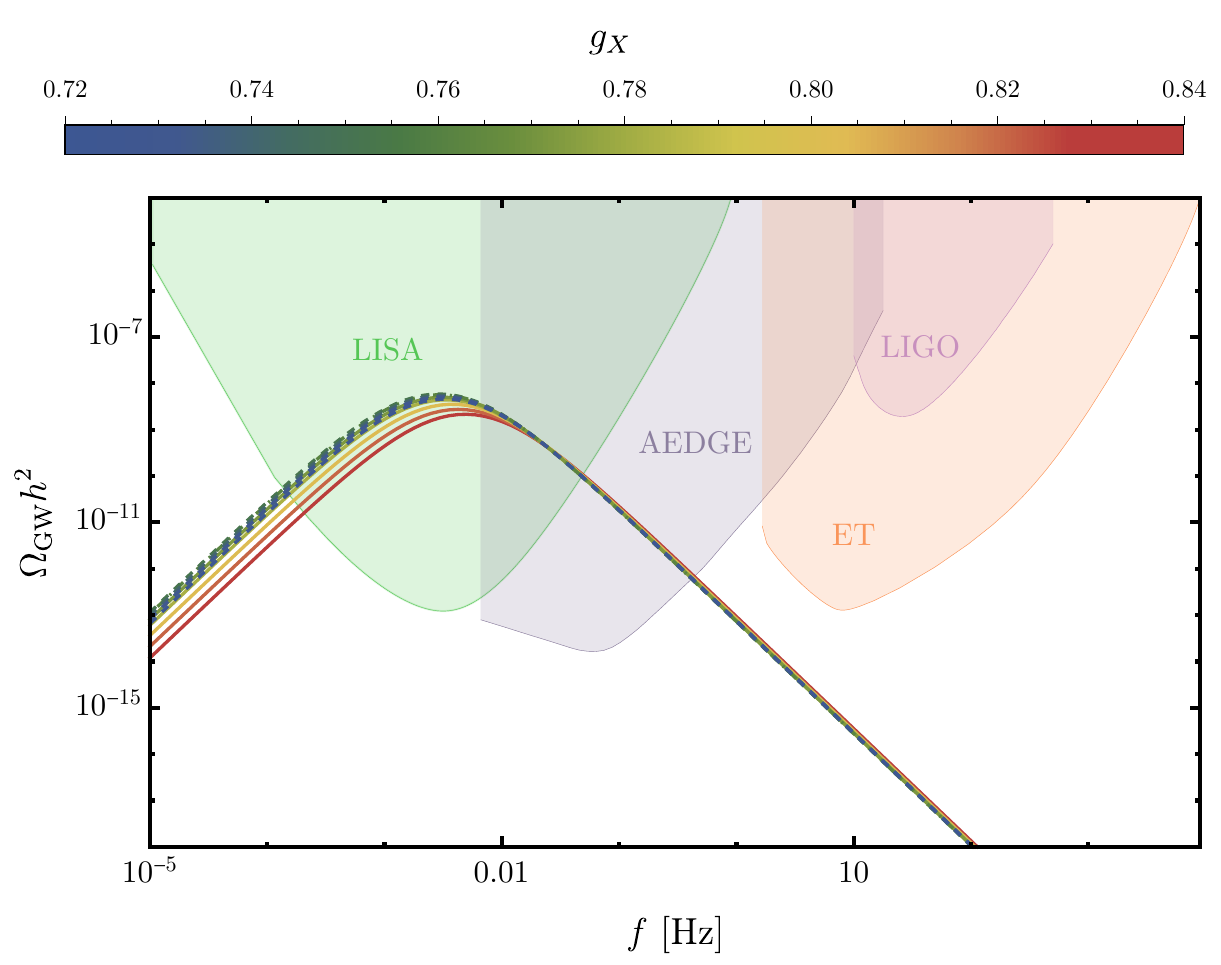}
\caption{Predictions for spectra of gravitational waves according to ref.~\cite{Lewicki:2022pdb} together with integrated sensitivity curves for LISA, AEDGE, ET and LIGO. Left: for a fixed value of $g_X = 0.76$ and varying value of $M_X$ (colour-coded). Right: for a fixed value of $M_X = 100$\,TeV and varying value of $g_X$ (colour-coded).  \label{fig:spectra-gx-MX-lewicki}}
\end{figure}

In this appendix, we present results for the predicted GW spectra computed according to ref.~\cite{Lewicki:2022pdb}, assuming that the fluid shell that follows behind the bubble continues propagating at the speed of light after the collision. The results of the simulation of ref.~\cite{Lewicki:2022pdb} read as follows:
%2.93 is the (8pi)^1/3
\be
\Omega_{\text{col}}(f) = \qty(\frac{R_*H_*}{5})^2 
\qty( \frac{\kappa(R_{\text{eff}}) \alpha }{1+\alpha} )^2 S_{\textrm{strong}}(f), \label{eq:spectrum-new}
\ee
where
\be
\kappa(R)= \frac{1}{1+\frac{R}{R_{\text{eq}}}},
\quad%
S_{\textrm{strong}} = 31.67  \qty[ 2.42 \qty(\frac{f}{ f_{\textrm{strong}} })^{-0.591} + 2.41 \qty(\frac{f}{ f_{\textrm{strong}} })^{0.593} ]^{-4.08}. \label{eq:spectrum-new-2}
\ee
The so-called effective bubble radius is given by $R_{\text{eff}} = 4.81(5/R_*)$, while the peak frequency is $f_{\textrm{strong}} = 0.12(5/R_*H_*)^{-1}$. In this approach, the spectra generated via bubble collisions and sound waves are indistinguishable. Figure~\ref{fig:spectra-gx-MX-lewicki} shows spectra computed according to eqs.~\eqref{eq:spectrum-new}--\eqref{eq:spectrum-new-2}. They can be compared with the spectra of figure~\ref{fig:spectra-gx-MX-caprini}. The shape is slightly modified but the main feature of being within experimental reach is unchanged.

\begin{figure}[H]%[h!t]
\center
\includegraphics[width=.4\textwidth]{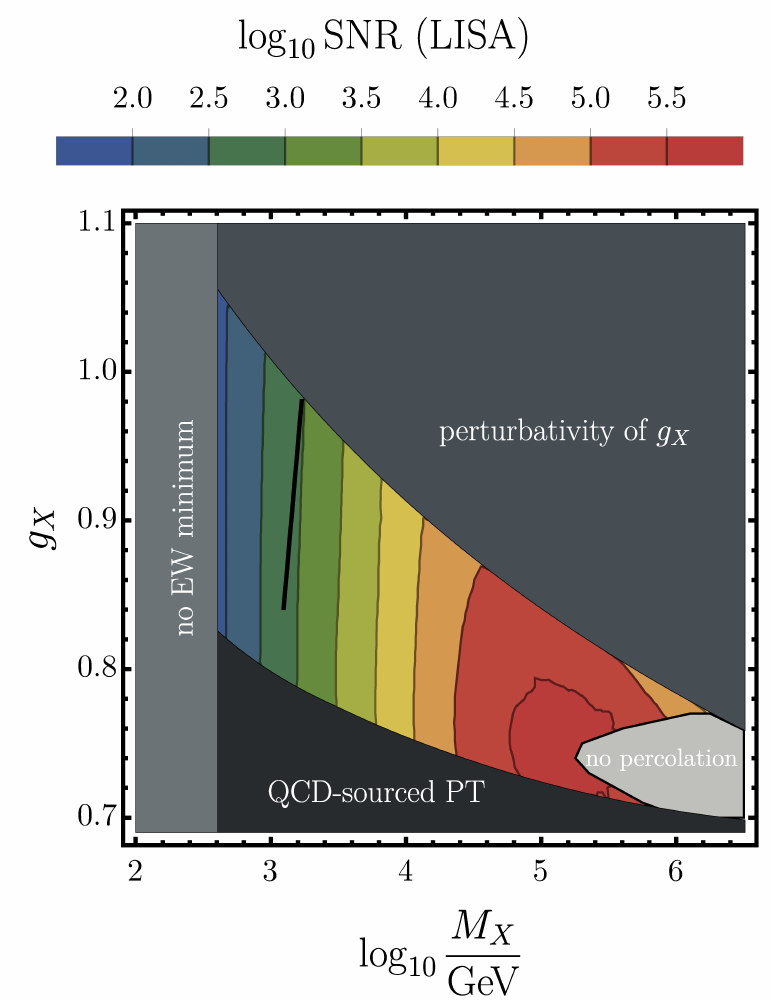}\hspace{20pt}
\includegraphics[width=.4\textwidth]{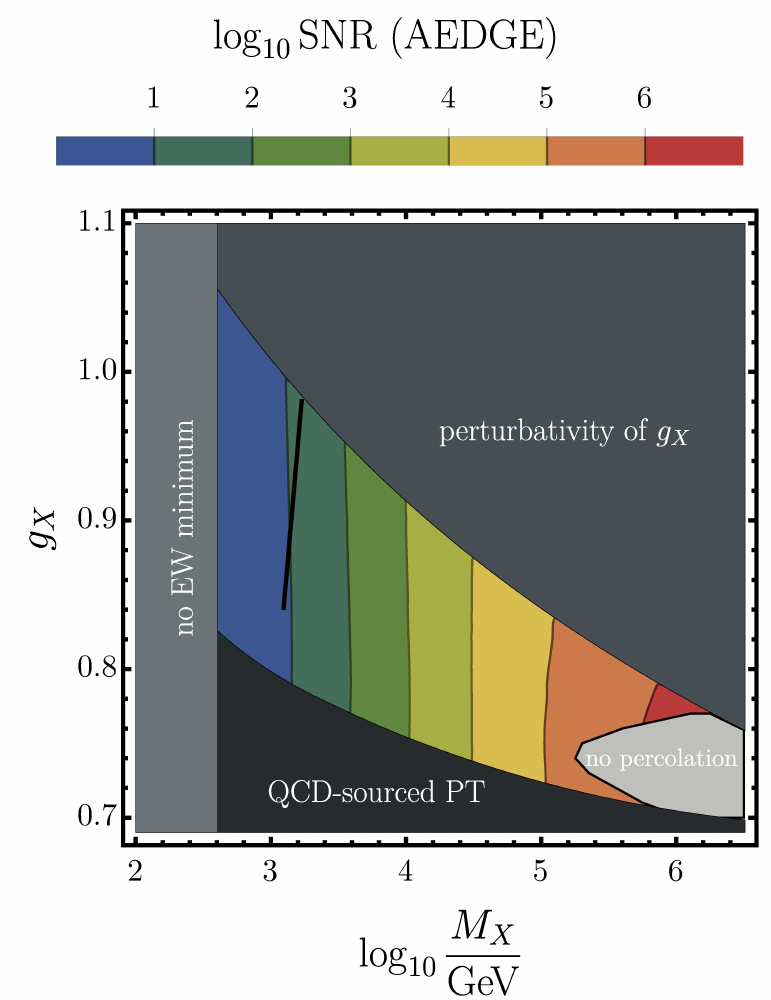}
\caption{Results for the signal-to-noise ratio for LISA (left panel) and AEDGE (right panel) for the GW signal according to ref.~\cite{Lewicki:2022pdb}. The black line corresponds to the points that reproduce the measured DM relic abundance and also evade the DM direct detection experimental constraints.\label{fig:SNR-new}}
\end{figure}

In figure \ref{fig:SNR-new} we show the SNR plots obtained in this approach. They resemble the previous results, as they again show the capability of detection by LISA for the whole of parameter space, but now for the smaller part in the case of AEDGE. The visible differences are a direct consequence of collision-only shapes of the spectra, i.e.\ we do not see any abrupt changes in the values of the contours. 

\end{appendix}

\bibliography{conformal-bib+GW}

\end{document}